\documentclass[prd,aps,nofootinbib,showpacs,10pt]{revtex4}
\usepackage{amsmath,graphicx,color,epsfig,float,textcomp}
\usepackage{epstopdf}

\newcommand{\myslash}[1]{/\kern-0.57 em {#1}}

\begin{document}
\title{Weak Decays of Doubly Heavy Baryons: ${\cal B}_{cc}\to {\cal B} D^{(*)}$}
\author{Run-Hui Li\footnote{E-mail: lirh@imu.edu.cn}, Juan-Juan Hou, Bei He, Ya-Ru Wang}
\affiliation{ School of Physical Science and Technology, Inner Mongolia University, Hohhot 010021, China }
\begin{abstract}
The discovery of $\Xi_{cc}^{++}$ has inspired new interest in studying doubly heavy baryons. In this study, the weak decays of a doubly charmed baryon ${\cal B}_{cc}$ to a light baryon ${\cal B}$ and a charm meson $D^{(*)}$ (either a pseudoscalar or a vector one) are calculated. Following our previous work, we calculate the short distance contributions under the factorization hypothesis, whereas the long distance contributions are modeled as the final state interactions which are calculated with the one particle exchange model. We find that the ${\cal B}_{cc}\to {\cal B} D^{*}$ decays' branching ratios are obviously larger, as they receive contributions of more polarization states. Among the decays that we investigate, the following have the largest branching fractions: ${\cal BR}(\Xi_{cc}^{++}\rightarrow\Sigma^{+}D^{*+}) \in [0.46\%, 3.33\%]$ estimated with $\tau_{\Xi_{cc}^{++}}=256$ fs; ${\cal BR}(\Xi_{cc}^{+}\rightarrow\Lambda D^{*+}) \in  [0.38\%, 2.63\%]$ and ${\cal BR}(\Xi_{cc}^{+}\rightarrow\Sigma^{0} D^{*+}) \in  [0.45\%, 3.16\%]$ with $\tau_{\Xi_{cc}^+}=45$ fs; and ${\cal BR}(\Omega_{cc}^{+}\rightarrow\Xi^{0} D^{*+}) \in [0.27\%, 1.03\%]$, ${\cal BR}(\Omega_{cc}^{+}\rightarrow\Xi^{0} D^{+}) \in [0.07\%, 0.44\%]$ and  ${\cal BR}(\Omega_{cc}^{+}\rightarrow\Sigma^{0} D^{*+}) \in [0.06\%, 0.45\%]$ with $\tau_{\Omega_{cc}^+}=75$ fs. By comparing the decay widths of pure color commensurate channels with those of pure bow-tie ones, we find that the bow-tie mechanism plays an important role in charm decays.
\end{abstract}
 \maketitle
\section{Introduction}\label{sec:introduction}
Studies on doubly heavy baryons that contain two heavy constituent quarks ($c$ or $b$ quark) have been conducted for a long time. They are predicted by the quark model and allowed by the quantum chrodynamics theory. Physicists believe in their existence, even though they have not been found in experiments. The SELEX collaboration announced the discovery of $\Xi_{cc}^+$ in 2002 and 2005\cite{Mattson:2002vu,Ocherashvili:2004hi}. However, the reported production is large and  the lifetime that was measured is very long. Their results do not agree with the theoretical predictions and have not been confirmed by other experiments. In 2017 the LHCb collaboration declared the discovery of $\Xi_{cc}^{++}$ via $\Xi_{cc}^{++}\to\Lambda_c^+ K^-\pi^+\pi^+$ with $m_{\Xi_{cc}^{++}}=3.621$ GeV\cite{Aaij:2017ueg}. In 2018 the LHCb collaboration measured its lifetime as $256$ fs\cite{{Aaij:2018wzf}} and confirmed the discovery via $\Xi_{cc}^{++}\to \Xi_c^+\pi^+$\cite{{Aaij:2018gfl}}. $\Xi_{cc}^{++}$ is the first doubly heavy baryon to be discovered in experiments with properties that agree with the theoretical expectations. Its discovery is meaningful to the study of the hadron spectrum and baryon decays. Physicists have already conducted substantial research on the spectrum of doubly heavy baryons. However, the determination of a proper framework to study their weak decays is a very challenging task. Numerous studies have been conducted on this topic \cite{Ebert:2004ck,Ebert:2005ip,Albertus:2006wb,Albertus:2012jt,Wang:2017mqp,Hu:2017dzi,Zhao:2018mrg,Xing:2018lre,Wang:2017jow,Wang:2018wfj,Shen:2016hyv,Li:2018lxi,Li:2012cfa,Cheng:2020wmk,Gutsche:2019wgu,Gutsche:2018msz,Gutsche:2017hux,Faessler:2009xn,Faessler:2001mr}, and the form factors as well as the semileptonic decays of a doubly heavy baryon to a singly heavy baryon have been studied under various frameworks. However, few systematic methods are available to deal with even two body nonleptonic decays, which is essential for guiding new particle discoveries, understanding the dynamics of strong interactions, and testing the standard model precisely.

In 2017 we applied final state interactions (FSIs) to baryon decays at the charm scale to estimate the branching fractions of two body nonleptonic weak decays of doubly charmed baryons \cite{Yu:2017zst}. We suggested two golden discovery channels of $\Xi_{cc}^{++}$, which, as mentioned above, were adopted by the LHCb collaboration and aided in the discover of the $\Xi_{cc}^{++}$ particle. The discovery inspired the research of doubly heavy baryons and further questions are proposed: which are the golden discovery channels of the other doubly charmed baryons and what else can we find in the decays of doubly charmed baryons? To answer these questions, further research on the weak decays of doubly heavy baryons is required. In our previous work, we calculated the decays of a doubly charmed baryon to a singly charmed baryon and a light vector meson. We also investigated the possibility of these decays as potential discovery channels~\cite{Jiang:2018oak}. After discovering $\Xi_{cc}^{++}$, measuring its lifetime, and confirming the discovery with another decay, the LHCb collaboration also focused on the weak decays of $\Xi_{cc}^{++}$ with a charm meson in the final state~\cite{Aaij:2019dsx}. Motivated by these theoretical questions and experimental efforts, we study the two body nonleptonic decays of a doubly charmed baryon ${\cal B}_{cc}\to {\cal B} D^{(*)}$, where ${\cal B}_{cc}$ represents a doubly charmed baryon, ${\cal B}$ denotes a light baryon, and $D^{(*)}$ is either a pseudoscalar or vector charm meson.

There are many interesting physics to be explored in baryon decays. For example, the $CP$ violations have already been observed in $K$, $B$, and $D$ meson decays but have not been observed in baryon decays. Theoretical progress in this topic is slow because it is challenging to calculate the dynamics. No systematic factorization method has been established thus far, even for two body nonleptonic decays. In general the contributions in two body nonleptonic baryon decays can be topologically classified into several types: $T$, $C$, $E$, and $B$~\cite{Leibovich:2003tw}. In $b$ baryon decays the $E$ and $B$ contributions are numerically small\cite{Lu:2009cm}. In the charm sector, the situation differs, and the $E$ and $B$ contributions may become important~\cite{Gutsche:2019iac}.   The study  of these decays will aid in understanding the dynamics of baryon decays at the charm scale.

The remainder of this paper is organized as follows. In Section \ref{sec:analytic} the phenomenological framework is introduced, the contributions in these decays are discussed, and the analytical expressions are presented. Section \ref{sec:results} presents several inputs, tables of our results and discussions. A summary is provided in Section \ref{sec:summary}. Owing to space limitations, we list all of the expressions of the amplitudes in Appendix \ref{app:amps}, whereas the strong couplings are presented in Appendix \ref{app:stcouplings}.

\section{Theoretical Framework and Analytical Calculations}\label{sec:analytic}
\subsection{Theoretical Framework}\label{ssec:hamiltonian}
In our previous work, we extended the model of FSIs to baryon decays~\cite{Yu:2017zst,Jiang:2018oak} and suggested the discovery channels for $\Xi_{cc}^{++}$ successfully. We also found a misunderstanding of FSIs in certain earlier reports, and interested readers are referred to our upcoming paper. To begin with, we briefly introduce this framework by following the ideas proposed in Ref.~\cite{Cheng:FSIB}. Suppose that the weak Hamiltonian is in the form ${\cal H}_W=\lambda_i Q_i$, where $\lambda_i$ are the combinations of quark mixing matrix elements and $Q_i$ are time reversal invariant weak operators. The amplitude of ${\cal B}_{cc}\to i$ can be decomposed as
\begin{equation}
\langle i;\mbox{out} | Q | {\cal B}_{cc}; \mbox{in}\rangle ^*=\sum_j S_{ji}^* \langle j;\mbox{out}| Q | {\cal B}_{cc}; \mbox{in}\rangle, \label{eq:FSIfactorization}
\end{equation}
where $S_{ji}\equiv \langle i;\mbox{out}|j;\mbox{in}\rangle$ is the strong interaction $S$ matrix element. Using the unitarity of the $S$-matrix and $S=1+iT$,  one can obtain an identity related to the optical theorem:
\begin{equation}
2 {\cal A}bs\,\langle i;\mbox{out} | Q | {\cal B}_{cc}; \mbox{in}\rangle=\sum_j T^*_{ji} \langle j;\mbox{out}| Q | {\cal B}_{cc}; \mbox{in}\rangle.
\end{equation}
Specifically, the absorptive part in the amplitude of the ${\cal B}_{cc}\to {\cal B} {D^{(*)}}$ decay can be obtained as
\begin{eqnarray}
{\cal A}bs\, {\cal M}({\cal B}_{cc}\to {\cal B} {D^{(*)}})& =&\frac{1}{2} \sum_j \left(\prod_{k=1}^j \int \frac{{\rm d}^3 q_k}{(2\pi)^3 2 E_k}\right)(2\pi)^4 \delta^4(p_{\cal B}+p_{D^{(*)}}-\sum_{k=1}^j q_k)\nonumber\\
&&\times {\cal M}(p_{{\cal B}_{cc}}\to \{q_k\}) T^*(p_{\cal B}p_{D^{(*)}}\to \{q_k\}).\label{eq:OPabs}
\end{eqnarray}
Eqs.(\ref{eq:FSIfactorization}) and (\ref{eq:OPabs}) indicate that the decay process can be divided into two steps. The first is the generation of an intermediate state under weak interactions, which is dominated by short distance dynamics, and the second is the subsequent formation of a final state through the strong interactions among intermediate particles. In principle, all the possible intermediate states should be considered. However, based on the argument that the $2$-body $\rightleftharpoons$ $n$-body rescattering is negligible~\cite{FSI:2arguement}, we only need to consider the intermediate states with two particles.

The weak decays ${\cal B}_{cc}\to {\cal B} {D^{(*)}}$ are induced by the charged current $c \to s/d$. For charm decays induced by the flavor changing neutral current (FCNC) with a quark loop effect, cancellation occurs between the $d$ and $s$ quark loop contributions; therefore, the FCNC contributions can safely be ignored. The low energy effective Hamiltonian with a charged current is given by
 \begin{eqnarray}
 {\cal H}_{eff} = \frac{G_{F}}{\sqrt{2}}
     \sum\limits_{q=d,s} V^{*}_{cq} V_{uD} \big[
     C_{1}({\mu}) O^{q}_{1}({\mu})
  +  C_{2}({\mu}) O^{q}_{2}({\mu})\Big] + \mbox{h.c.} ,
 \label{eq:hamiltonian}
\end{eqnarray}
with
\begin{eqnarray}
  O^{q}_{1}=({\bar{u}}_{\alpha}D_{\beta} )_{V-A}
               ({\bar{q}}_{\beta} c_{\alpha})_{V-A},
    \ \ \ \
   O^{q}_{2}=({\bar{u}}_{\alpha}D_{\alpha})_{V-A}
               ({\bar{q}}_{\beta} c_{\beta} )_{V-A},
    \label{eq:operators}
\end{eqnarray}
where $D=s,d$, $V_{cq}$ and
$V_{uD}$ are the Cabibbo-Kobayashi-Maskawa (CKM) matrix elements whose values are used from the CKMfitter Group \cite{CKM}, $C_{1/2}(\mu)$ represents the Wilson coefficients, the Fermi constant is $G_F=1.166\times 10^{-5}\mbox{ GeV}^{-2}$, and $O^q_{1/2}$ are the local four-quark operators in which $\alpha$ and $\beta$ are color indices.
\begin{figure}[htp]
\begin{center}
\centerline{\includegraphics[scale=0.7]{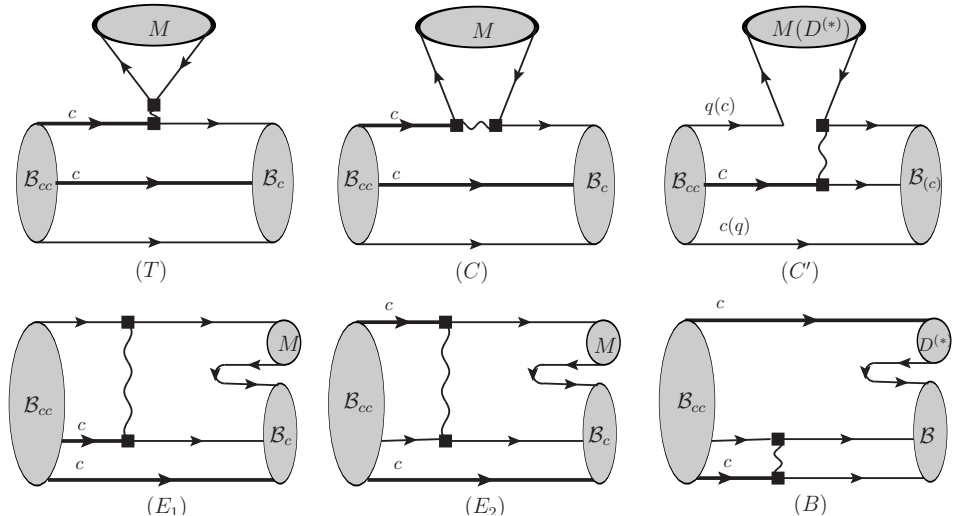}}
\end{center}
\caption{Topological diagrams of ${\cal B}_{cc}\to{\cal B}_{(c)} M(D^{(*)})$ at tree level. ${\cal B}_{(c)}$ denotes a light baryon or singly charmed baryon. $M$ is a light meson. The thick lines represent $c$ quarks and the wavy lines represent $W$ bosons}
\label{fig:topos}
\end{figure}

The contributions induced by the above Hamiltonian in two body nonleptonic decays of ${\cal B}_{cc}$ can be classified into eight topological diagrams, which are depicted in Fig.~\ref{fig:topos}. The external $W$ emission contribution is denoted by the symbol $T$. The internal $W$ emission contributions can be classified as two types. In the $C$ diagram, the two constituent quarks of the meson are all obtained from the weak vertex. In the $C^\prime$ diagram one constituent quark of the meson is obtained from the initial state baryon. The diagrams in the second line of Fig. \ref{fig:topos} are all $W$ exchange diagrams. In $E_1$, the light quark, which is obtained from the $c$ quark by emitting a $W$ boson, is picked up by the final state baryon, whereas in $E_2$, it is picked up by the final state meson. In the bow-tie diagram (denoted by $B$), both light quarks generated in the weak interaction are picked up by the final state baryon. The strong interactions in Fig. \ref{fig:topos} , both short and long distance, are included although they are not drawn.

\begin{figure}[htp]
\begin{center}
\includegraphics[scale=0.5]{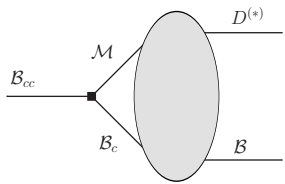}
\end{center}
\caption{Decay process indicated at hadron level. The black square is a weak vertex at which the intermediate state ${\cal B}_c {\cal M}$ is generated. The gray ellipse represents the strong interactions between intermediate particles, which is realized by exchanging hadrons}
\label{fig:hadronl}
\end{figure}

The short distance strong interactions are associated with the weak vertex, and this part of the contribution occurs at a high energy scale so that the perturbative calculation is still valid. Drawing on the experience of studying $b$ baryon decays~\cite{Leibovich:2003tw,Lu:2009cm}, one can observe that the $W$ exchange contribution can be safely neglected at a short distance. The situation differs when one considers the long distance contributions, which are thought to be dominating because of the low energy release and the $W$ exchange mechanism may become important~\cite{FSIs}. A decay process of ${\cal B}_{cc} \to {\cal B}D^{(*)}$ can be divided into two steps: a ${\cal B}_{cc}$ baryon first decays to ${\cal B}_c {\cal M}$ and then to ${\cal B}D^{(*)}$ via long distance interactions. The former step occurs at a short distance; therefore, the $W$ exchange contribution can be omitted. The long distance part, which is essentially nonperturbative, is difficult to calculate. In this work, we model it as the FSIs and perform the calculation at the hadron level. In this model, the long distance dynamics are realized by exchanging hadron-state particles (depicted in Fig. \ref{fig:hadronl}). Now we arrive at the step for calculating the amplitude in detail.

\subsection{Calculation of weak vertices}\label{ssec:SDcontribution}
As stated in the previous subsection, the first step in obtaining the amplitude is to calculate the weak production of an intermediate state. To avoid double counting, this part of the contribution is short distance dynamics in principle. At the hadron level this part is represented by a weak vertex. At a short distance the $W$ exchange mechanism can safely be neglected. Therefore, the weak vertex can be calculated reliably with the factorization hypothesis. Given the Hamiltonian in Eqs. (\ref{eq:hamiltonian}) and (\ref{eq:operators}), the $T$ diagram in the factorization hypothesis is factorized as follows:
\begin{equation}
{\cal A}({\cal B}_{cc}\to {\cal B}_c {\cal M})=\frac{G_{F}}{\sqrt{2}}
     \sum\limits_{q=d,s} V^{*}_{cq} V_{uD} \left(C_2+C_1/N_C\right) \langle{\cal M}|({\bar{u}}_{\alpha}D_{\alpha})_{V-A}|0\rangle
              \langle{\cal B}_c| ({\bar{q}}_{\beta} c_{\beta} )_{V-A}|{\cal B}_{cc}\rangle
\end{equation}
with $N_c=3$.
The weak transition of ${\cal B}_{cc}$ to a spin-$1/2$ singly charmed baryon ${\cal B}_c$ is parameterized as
\begin{eqnarray}
\langle {\cal B}_c(p^\prime,s_z^\prime)| (V-A)_\mu |{\cal B}_{cc}(p,s_z)\rangle
=&& \bar u(p^\prime,s^\prime_z)\left[ \gamma_\mu f_1(q^2) + i\sigma_{\mu\nu}\frac{q^\nu}{M} f_2(q^2) +\frac{q^\mu}{M} f_3(q^2) \right] u(p,s_z) \nonumber\\
&&- \bar u(p^\prime,s^\prime_z)\left[ \gamma_\mu g_1(q^2) + i\sigma_{\mu\nu}\frac{q^\nu}{M} g_2(q^2) +\frac{q^\mu}{M} g_3(q^2) \right] \gamma_5 u(p,s_z),\label{eq:ff}
\end{eqnarray}
where $q=p-p^\prime$, $M$ is the mass of ${\cal B}_{cc}$, and $f_i$ and $g_i$ are the form factors.

The expressions of $f_i$ and $g_i$, which can be obtained with the aid of certain quark models or sum rules, are used as inputs here. In this study, we adopt the results calculated under the light-front quark model in Ref.~\cite{Wang:2017mqp}.

The decay constants of pseudoscalar and vector mesons are respectively defined as
\begin{equation}
\langle 0|A_\mu|P(q)\rangle=if_Pq_\mu \, ,
\label{eq:pdc}
\end{equation}
and
\begin{equation}
\langle 0|V_\mu|V(q)\rangle=f_Vm_V\epsilon_\mu \, ,
\label{eq:vdc}
\end{equation}
where the subscripts ``$P$" and ``$V$" correspond to a pseudoscalar and vector meson, respectively.
Combining Eqs. (\ref{eq:ff}), (\ref{eq:pdc}) and (\ref{eq:vdc}), the weak vertex of ${\cal B}_{cc}\to{\cal B}_c P$ is expressed as
\begin{eqnarray}
W_T({\cal B}_{cc}\to{\cal B}_c P)=i\frac{G_F}{\sqrt{2}} V^*_{cq} V_{uD} a_1 f_P
\bar u(p^\prime,s^\prime_z)\left[(M-M^\prime) f_1(m_P^2)+ (M+M^\prime) g_1(m_P^2) \gamma_5 \right] u(p,s_z)\, .
\end{eqnarray}
The $C$ diagram can be calculated via its relation to the $T$ diagram under Fierz transformation:
\begin{eqnarray}
W_C({\cal B}_{cc}\to{\cal B}_c P)=i\frac{G_F}{\sqrt{2}} V^*_{cq} V_{uD} a_2 f_P
\bar u(p^\prime,s^\prime_z)\left[(M-M^\prime) f_1(m_P^2)+ (M+M^\prime) g_1(m_P^2) \gamma_5 \right] u(p,s_z)\, .\label{eq:B2BP}
\end{eqnarray}
In the above equations, $a_1=C_2+C_1/N_C$ and $a_2=C_1+C_2/N_C$ are the combinations of Wilson coefficients. In this work, the decays are under the charm scale, so we use $a_1(m_c)$ and $a_2(m_c)$ in Ref.~\cite{Ali:2007ff}. $M^\prime$ is the mass of ${\cal B}_c$. We omit the terms with $f_3$ and $g_3$ in Eq. (\ref{eq:B2BP}), because they are suppressed by $m_P^2/M^2$.

For ${\cal B}_{cc}\to{\cal B}_c V$ we obtain
\begin{eqnarray}
W_T({\cal B}_{cc}\to{\cal B}_c V)=&&\frac{G_F}{\sqrt{2}} V^*_{cq} V_{uD} a_1 f_V \epsilon^*_\mu
\bar u(p^\prime,s^\prime_z)\left[\left(f_1(m_V^2)-\frac{M+M^\prime}{M}f_2(m_V^2)\right)\gamma^\mu +\frac{2}{M}f_2(m_V^2)p^{\prime\mu} \right. \nonumber\\
&&-\left.\left(g_1(m_V^2)+\frac{M-M^\prime}{M}g_2(m_V^2)\right)\gamma^\mu\gamma_5 -\frac{2}{M}g_2(m_V^2)p^{\prime\mu}\gamma_5\right] u(p,s_z)\, ,\nonumber\\
W_C({\cal B}_{cc}\to{\cal B}_c V)=&&\frac{G_F}{\sqrt{2}} V^*_{cq} V_{uD} a_2 f_V \epsilon^*_\mu
\bar u(p^\prime,s^\prime_z)\left[\left(f_1(m_V^2)-\frac{M+M^\prime}{M}f_2(m_V^2)\right)\gamma^\mu +\frac{2}{M}f_2(m_V^2)p^{\prime\mu}\right. \nonumber\\
&&-\left.\left(g_1(m_V^2)+\frac{M-M^\prime}{M}g_2(m_V^2)\right)\gamma^\mu\gamma_5 -\frac{2}{M}g_2(m_V^2)p^{\prime\mu}\gamma_5\right] u(p,s_z)\, .
\label{eq:B2BV}
\end{eqnarray}

\subsection{Rescattering at long distance}\label{ssec:LDcontribution}
The rescattering between the intermediate particles is nonperturbative dynamics by nature and very difficult to calculate. In this work we employ the framework of FSIs and perform the calculation with the one-particle-exchange model at the hadron level~\cite{FSIs,Cheng:FSIB,Lu:2005mx}. In the following, we use $\Omega_{cc}^{+}\to\Xi^{0} D_s^+$ as an example to demonstrate the detailed process of our calculation. This decay can proceed as $\Omega_{cc}^{+}\to\Omega_c^0(K^+/K^{*+})\to\Xi^{0} D_s^+$ , $\Omega_{cc}^{+}\to(\Xi_c^+/\Xi_c^{\prime +})(\phi/\eta_1/\eta_8)\to\Xi^{0} D_s^+$, and $\Omega_{cc}^{+}\to(\Xi_c^0/\Xi_c^{\prime 0})(\pi^+/\rho^+)\to\Xi^{0} D_s^+$. The first one is induced by $c\to s u \bar s$ at the quark level and the latter two ones are induced by $c \to d u \bar d$, which indicates that it is a singly CKM suppressed decay. The intermediate states $\Omega_c^0(K^+/K^{*+})$ and $(\Xi_c^0/\Xi_c^{\prime 0})(\pi^+/\rho^+)$ are generated via the $T$ diagram and $(\Xi_c^+/\Xi_c^{\prime +})(\phi/\eta_1/\eta_8)$ originates from the $C$ mechanism.

\begin{figure}[htb]
\begin{center}
\hspace{-0.5cm}
\begin{minipage}{0.25\linewidth}
\centerline{\includegraphics[scale=0.4]{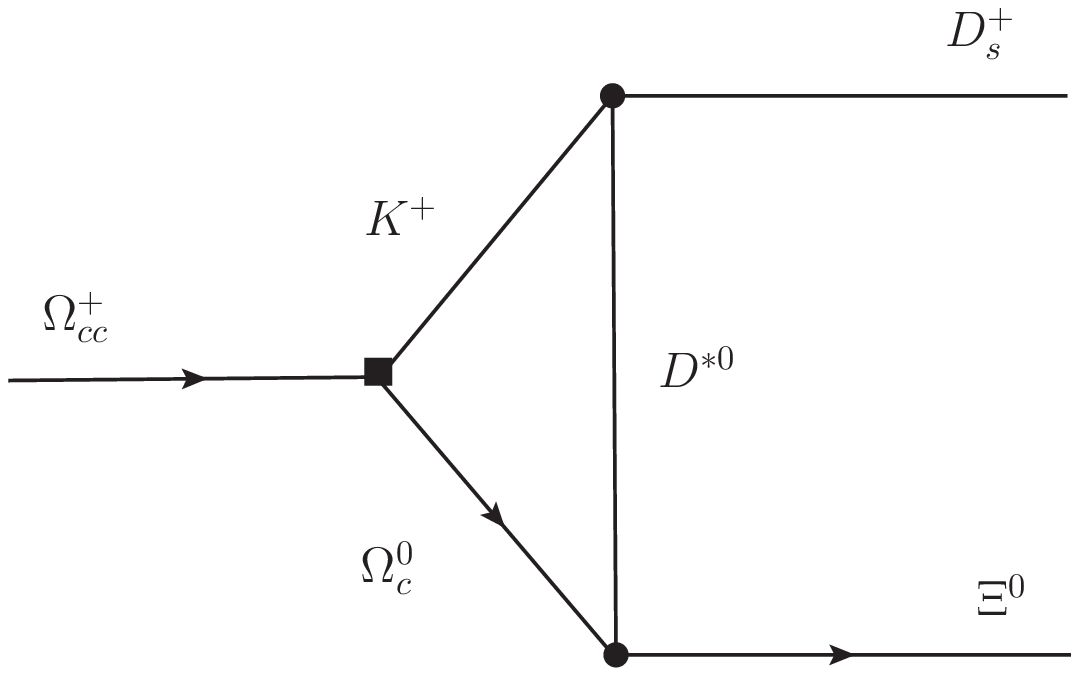}}
(a)
\centerline{\includegraphics[scale=0.4]{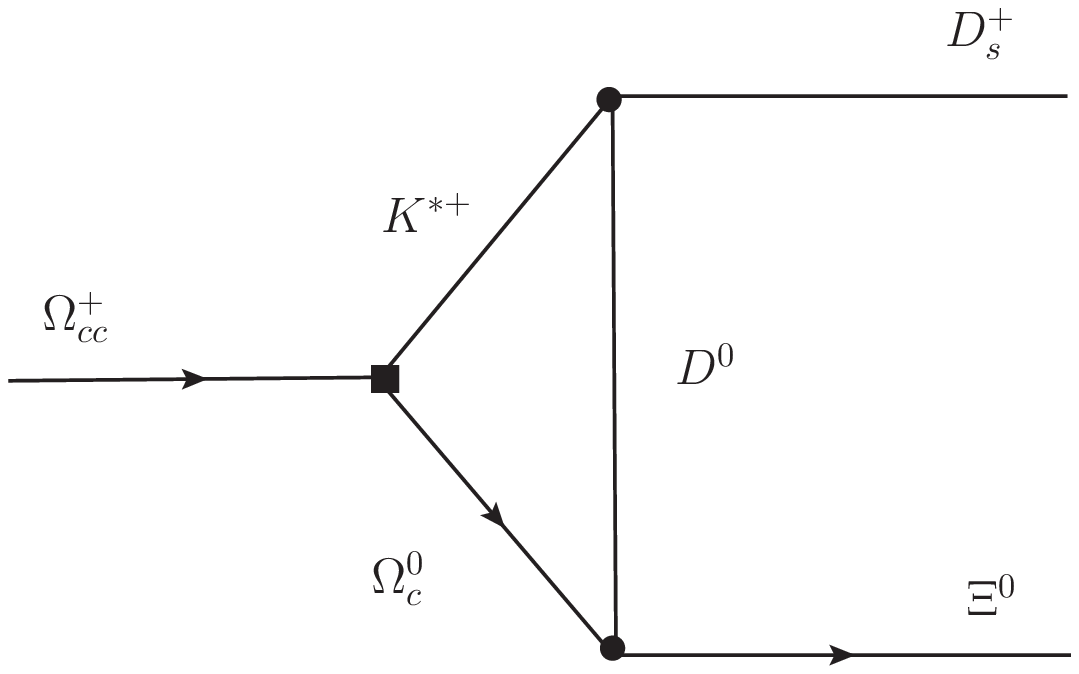}}
(d)
\end{minipage}
\begin{minipage}{0.25\linewidth}
\centerline{\includegraphics[scale=0.4]{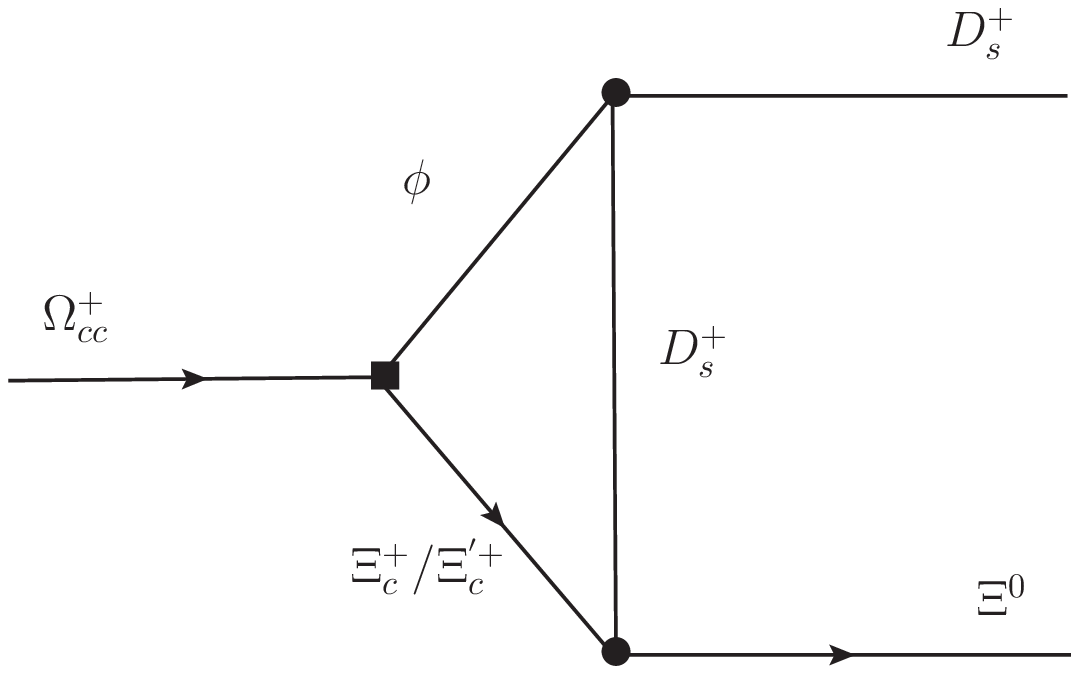}}
(b)
\centerline{\includegraphics[scale=0.4]{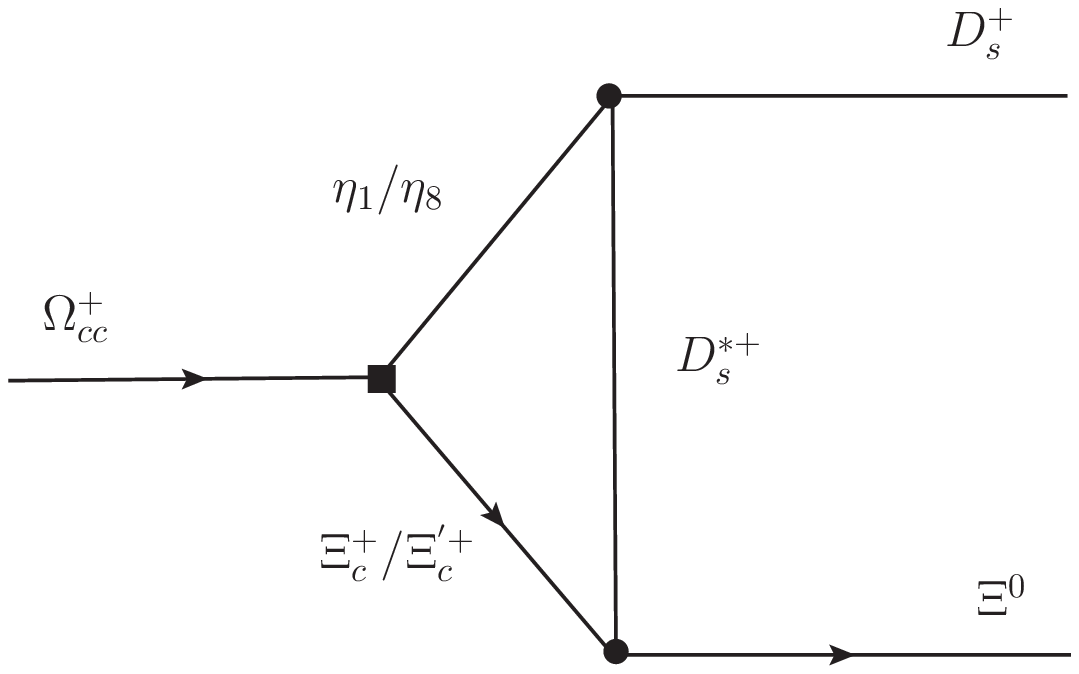}}
(e)
\end{minipage}
\begin{minipage}{0.25\linewidth}
\centerline{\includegraphics[scale=0.4]{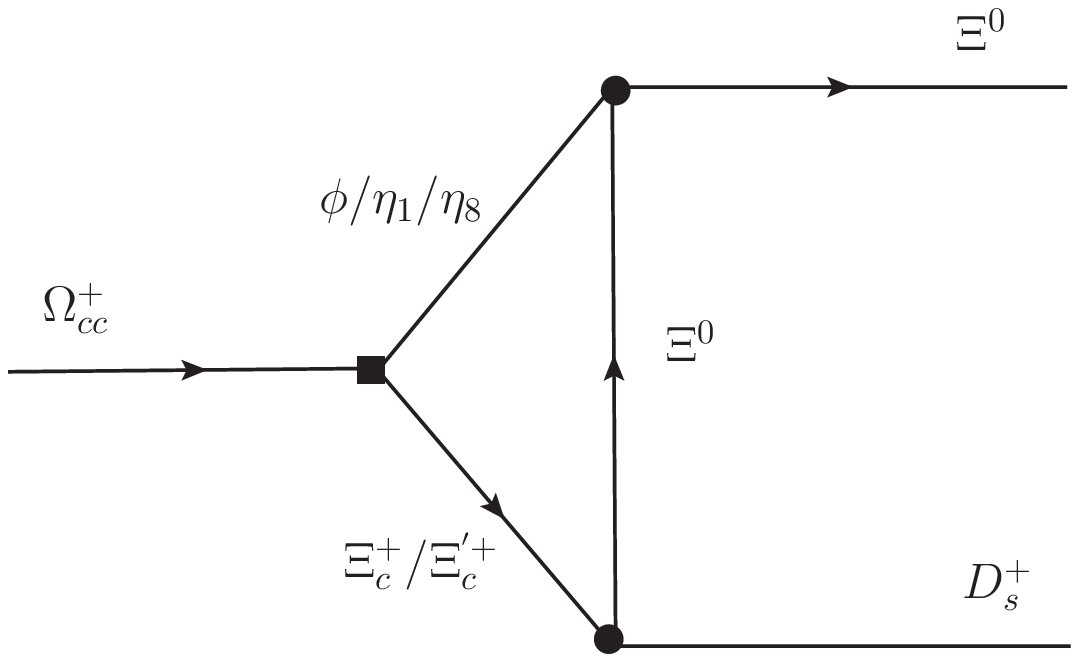}}
(c)
\centerline{\includegraphics[scale=0.4]{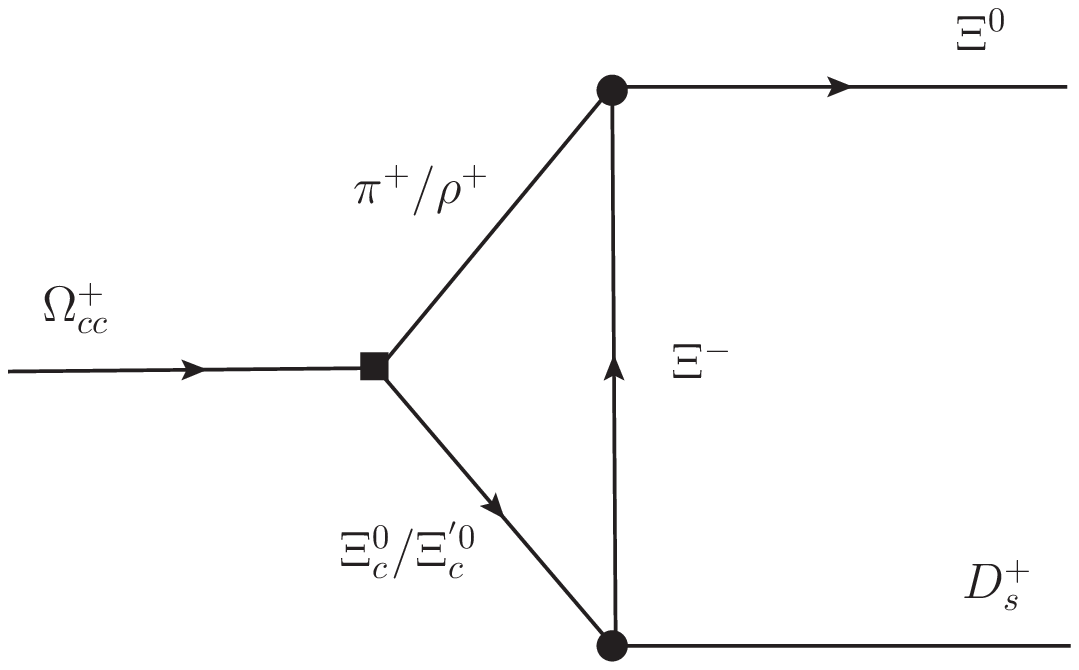}}
(f)
\end{minipage}\\
\begin{minipage}{0.25\linewidth}
\centerline{\includegraphics[scale=0.4]{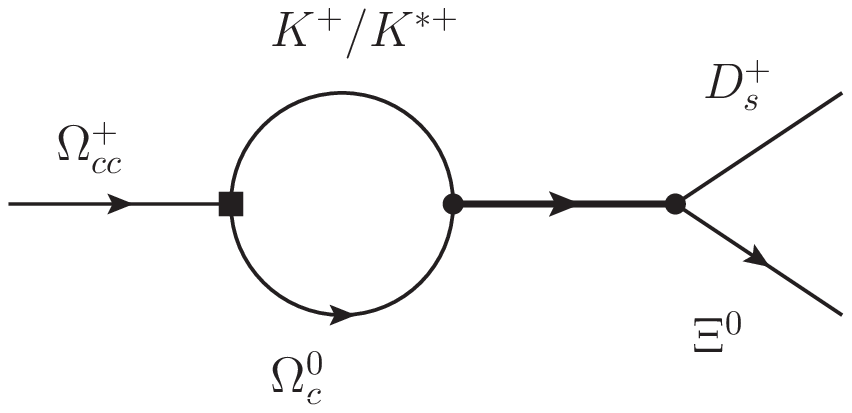}}
(g)
\end{minipage}
\begin{minipage}{0.25\linewidth}
\centerline{\includegraphics[scale=0.4]{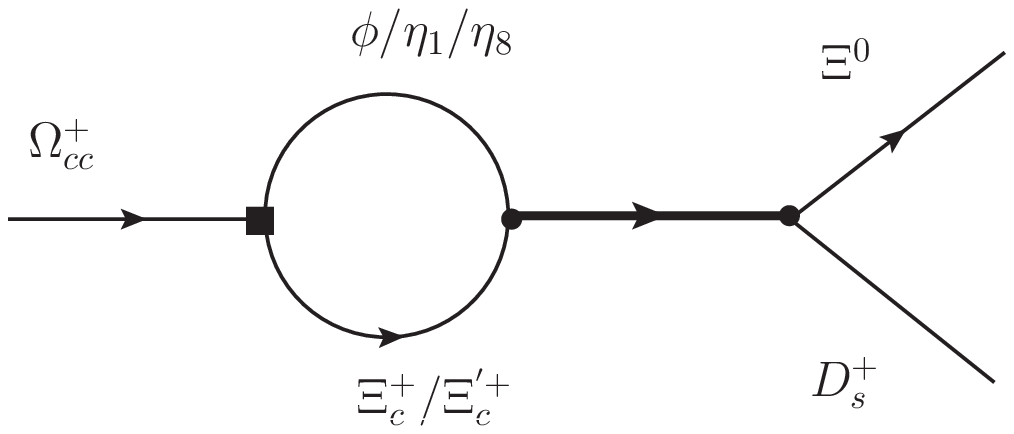}}
(h)
\end{minipage}
\begin{minipage}{0.25\linewidth}
\centerline{\includegraphics[scale=0.4]{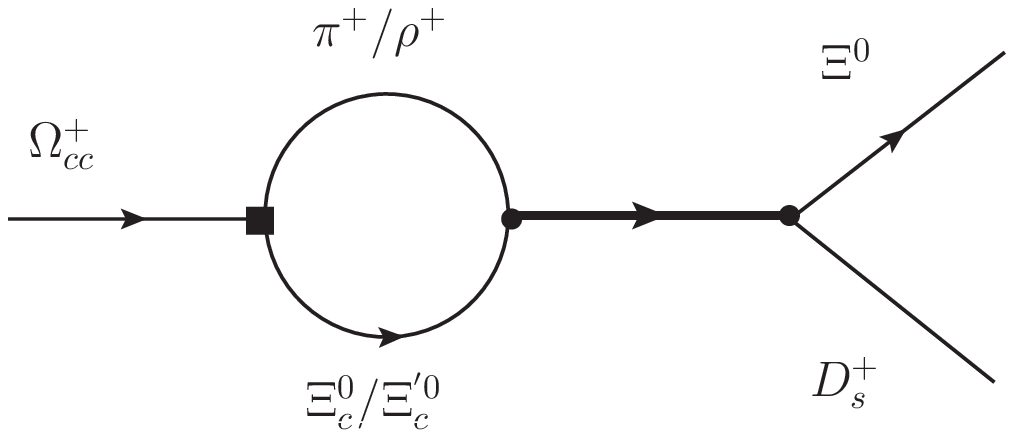}}
(i)
\end{minipage}
\end{center}
\caption{Leading FSI contributions to $\Omega_{cc}^{+}\to\Xi^{0} D_s^+$ manifested at hadron level. The black squares denote weak vertices and dots represent strong vertices. Each thick line in diagrams (g), (h), and (i) denotes a resonant structure. Diagrams (a), (d), and (g) are induced by the rescattering between $\Omega_c^0$ and $ K^+/K^{*+}$, diagrams (b), (c), (e), and (h) by the rescattering between $\Xi_c^+/\Xi_c^{\prime +}$ and $\phi/\eta_1/\eta_8$, and diagram (f) by $\Xi_c^0/\Xi_c^{\prime 0}$ and $\pi^+/\rho^+$}
\label{fig:FSIq}
\end{figure}

As mentioned previously and depicted in Fig.~\ref{fig:hadronl}, the long distance contributions are calculated at the hadron level. The calculation is performed with the chiral Lagrangian. One can draw all of the leading diagrams according to the perturbation theory with only one particle exchanged as in Fig. \ref{fig:FSIq}. The Lagrangian used in this study is botained from Refs. \cite{Yan:1992gz,Casalbuoni:1996pg,Meissner:1987ge,Li:2012bt}. The readers can refer to Ref. \cite{Yu:2017zst} for specific expressions.

The three diagrams of the $s$ channel, presented in Figs. \ref{fig:FSIq}(g), (h) and (i), make a sizeable contribution only when the mass of each resonant state is quite close to the mass of the mother particle $\Omega_{cc}^{+}$.  Among the discovered singly charmed baryons, even the heaviest one is approximately $500$ MeV lighter than $\Xi_{cc}^{++}$. Therefore, these contributions are supposed to be suppressed by the off-shell effect. As a result, we neglect these contributions in our calculation and only consider the $t$ channel contributions, which are the typical triangle diagrams depicted in Figs. \ref{fig:FSIq}(a)-(f). Eq.~(\ref{eq:OPabs}) is employed to calculate the absorptive part of these diagrams. In principle, the amplitude of the diagram can be obtained via the dispersion relation
\begin{equation}
A(m_1^2)=\frac{1}{\pi}\int_s^\infty \frac{{\cal A}bs\,A(s^\prime)}{s^\prime - m_1^2-i\epsilon} \mbox{d}s^\prime.
\end{equation}
As opposed to QCD sum rules, our calculation is performed in the physical kinematics region, where a singularity exists in the above integration. In this study we follow the scheme adopted by Hai-Yang Cheng, Chun-Khiang Chua, and Amarjit Soni in Ref.~\cite{Cheng:FSIB}. Only the absorptive part of the amplitude is maintained for order estimation. An additional phenomenological factor is associated with the exchanged particle to account for its off-shell effect and to make the theoretical framework consistent. The expression of this factor provided in the following subsection.

\subsection{Analytical expressions for diagrams}\label{ssec:expressions}
\begin{figure}[htp]
\begin{center}
\begin{minipage}{0.25\linewidth}
\centerline{\includegraphics[scale=1]{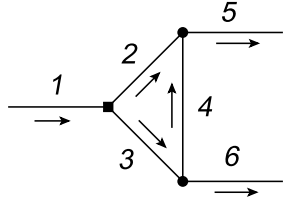}}
\end{minipage}
\end{center}
\caption{Numbers assigned to lines in triangle diagram. The arrows define the momentum directions in our calculation}
\label{fig:number}
\end{figure}

We derive the analytical expressions of the amplitudes by combining the discussions in Sections~\ref{ssec:SDcontribution} and \ref{ssec:LDcontribution} in this subsection.  To simplify the subscripts we assign numbers to the particles in a triangle diagram as illustrated in Fig.~\ref{fig:number}, in which the momentum flows are also defined. We use $M_{a/b/c/d/e/f}(P2;P3;P4)$ to denote the amplitude of such a triangle diagram. The subscripts ``$a/b/c/d/e/f$" correspond to Figs.~\ref{fig:FSIq}(a)-(f), whereas $P2$, $P3$ and $P4$ denote the particles at positions $2$, $3$, and $4$, respectively.

Specifically, the absorptive part of Fig.~\ref{fig:FSIq}(a) is given by $P2= K^+$, $P3=\Omega_c^0$, and $P4= D^{*0}$:
\begin{eqnarray}
{\cal A}bs\,M_{a}(K^{+};\Omega_{c}^{0};D^{*0})&=&
\int\frac{|\vec{p_2}|sin\theta d\theta
d\varphi}{32\pi^{2}m_{\Omega_{cc}^{+}}}
\frac{G_{F}}{\sqrt{2}}V_{cs}^{*}V_{us}a_{1}f_{K^+}
\frac{F^{2}(t,m_{D^{*0}})}{t-m_{D^{*0}}^{2}+im_{D^{*0}}\Gamma_{D^{*0}}}g_{D^{*0}D_s^{+}K^+}p_{2\alpha}\nonumber\\
  &&\times\overline{u}(p_{6},s^{\prime}_{z})
\left[f_{1\Omega_{c}^{0}\Xi^{0}D^{*0}}\gamma_{\mu}(-g^{\mu\alpha}+\frac{p_{4}^{\mu}p_{4}^{\alpha}}{m_{D^{*0}}^{2}})
+\frac{f_{2\Omega_{c}^{0}\Xi^{0}D^{*0}}}{m_{\Omega_{c}^{0}} +m_{\Xi^{0}}}\sigma_{\mu\nu}ip_{4}^{\mu}
(-g^{\nu\alpha}+\frac{p_{4}^{\nu}p_{4}^{\alpha}}{m_{D^{*0}}^{2}})\right]\nonumber\\
 &&\times(\myslash{p_{3}}+m_{\Omega_{c}^{0}})
  \left[(m_{\Omega_{cc}^{+}}-m_{\Omega_{c}^{0}})f_{1}(m^{2}_{K^+})
+(m_{\Omega_{cc}^{+}}+m_{\Omega_{c}^{0}})g_{1}(m^{2}_{k^+})\gamma_{5}\right]
u(p_{1},s_{z}).
\label{eq:a2PXP}
\end{eqnarray}
In the calculation, summations over the polarization states of the intermediate and exchanged particles need to be performed. For example, in Fig.~\ref{fig:FSIq}(a), one needs to sum over the polarization states of $\Omega_{c}^{0}$ and $D^{*0}$. In Eq.~(\ref{eq:a2PXP}), $\theta$ and $\phi$ are the polar and azimuthal angles of $\vec{p_3}$ in the spherical coordinate system, respectively, whereas $g_{D^{*0}D_s^{+}K^+}$, $f_{1\Omega_{c}^{0}\Xi^{0}D^{*0}}$, and $f_{2\Omega_{c}^{0}\Xi^{0}D^{*0}}$ are strong coupling constants. Furthermore, $P2$ and $P3$ are set to be on-shell. To account for the off-shell effect and to make the theoretical framework self-consistent, a Breit-Wigner structure and a form factor $F(t,m)$ are associated with the exchanged particle. The form factor $F(t,m)$ is parameterized as \cite{Cheng:FSIB}

\begin{equation}
F(t,m)=\left(\frac{\Lambda^2-m^2}{\Lambda^2-t}\right)^n,
\label{eq:Ffactor}
\end{equation}
which is normalized to $1$ at $t=m^2$, whereas $m$ is the mass of the exchanged particle. The cutoff $\Lambda$ is given as
\begin{equation}
\Lambda= m + \eta \Lambda_{\rm QCD}
\end{equation}
with $\Lambda_{\rm QCD}=330\,{\rm MeV}$. The phenomenological parameter $\eta$ depends on all of the particles at the strong vertex. Because numerous strong vertices appear in the calculation, a huge amount of experimental data are required to determine these parameters individually. In our calculation, we set $\eta=1.5$ and vary it from $1$ to $2$ for the error estimations. In Eq.~(\ref{eq:Ffactor}) $n$ is another phenomenological parameter that needs to be extracted from experimental data. Owing to a lack of experimental data, we draw on the experience of Ref. \cite{Cheng:FSIB} and set it to $1$.

Similarly the absorptive part of Fig.~\ref{fig:FSIq}(d) is given as
\begin{eqnarray}
{\cal A}bs\,M_{d}(K^{*+};\Omega_{c}^{0};D^{0})&=&
-i\int\frac{|\vec{p_2}|sin\theta d\theta
d\varphi}{32\pi^{2}m_{\Omega_{cc}^{+}}}
\frac{G_{F}}{\sqrt{2}}V_{cs}^{*}V_{us}a_{1}f_{K^{*+}}
\frac{F^{2}(t,m_{D^{0}})}{t-m_{D^{0}}^{2}+im_{D^{0}}\Gamma_{D^{0}}}
g_{\Omega_c^{0}\Xi^{0}D^{0}}
g_{D_s^{+}D^{0}K^{*+}}(p_{5\alpha}+p_{4\alpha})\nonumber\\
  &&\times\overline{u}(p_{6},s^{\prime}_{z})\gamma_{5}(\myslash{p_{3}}+m_{\Omega_{c}^{0}})
(-g^{\mu\alpha}+\frac{p_{2}^{\mu}p_{2}^{\alpha}}{m_{K^{*+}}^{2}})\nonumber\\
  &&\times\left[
\left(f_{1}(m^2_{K^{*+}})
-\frac{m_{\Omega_{cc}^{+}}+m_{\Omega_{c}^{0}}}{m_{\Omega_{cc}^{+}}}{f_{2}(m^2_{K^{*+}})}\right)\gamma_{\mu}\right.
+\frac{2}{m_{\Omega_{cc}^+}}{f_{2}(m^2_{K^{*+}})}p_{3\mu}\nonumber\\
&&\left.-\left
(g_{1}(m^2_{K^{*+}})+\frac{m_{\Omega_{cc}^{+}}-m_{\Omega_{c}^{0}}}{m_{\Omega_{cc}^{+}}}g_{2}(m^2_{K^{*+}})\right)
\gamma_{\mu}\gamma_{5}\right.
\left.-\frac{2}{m_{\Omega_{cc}^+}}g_{2}(m^2_{K^{*+}})p_{3\mu}\gamma_{5}
  \right]
u(p_{1},s_{z}),
\end{eqnarray}
where the spin summation is performed over the polarization states of $\Omega_c^0$ and $K^{*+}$. It should be stressed that certain symbols of strong coupling constants resemble those of weak transition form factors. Readers can distinguish these according to the feature that strong coupling constants have particle names as subscripts. The expressions of the remaining  diagrams in Fig.~\ref{fig:FSIq} are provided in Appendix \ref{app:Fig3diagrams}.
With all of the diagrams calculated, the amplitude of $\Omega_{cc}^{+}\to\Xi^{0} D_s^+$ is given as
\begin{eqnarray}
{\cal A}(\Omega_{cc}^{+}\to\Xi^{0} D_s^+)=i&{\cal A}bs[&
M_{a}(K^{+};\Omega_{c}^{0};D^{*0})+M_{b}(\phi;\Xi_c^+;D_s^{+})+M_{b}(\phi;\Xi_c^{\prime +};D_s^{+})+M_{c}(\phi;\Xi_c^+;\Xi^0)\nonumber\\
&&+M_{c}(\phi;\Xi_c^{\prime +};\Xi^0)+M_{c}(\eta_1;\Xi_c^+;\Xi^0)+M_{c}(\eta_1;\Xi_c^{\prime +};\Xi^0)+M_{c}(\eta_8;\Xi_c^+;\Xi^0)\nonumber\\
&&+M_{c}(\eta_8;\Xi_c^{\prime +};\Xi^0)+M_{d}(K^{*+};\Omega_{c}^{0};D^{0})+M_{e}(\eta_1;\Xi_c^{+}; D_s^{*+})+M_{e}(\eta_1;\Xi_c^{\prime +};D_s^{*+})\nonumber\\
&&+M_{e}(\eta_8;\Xi_c^{+};D_s^{*+})+M_{e}(\eta_8;\Xi_c^{\prime +};D_s^{*+})+M_{f}(\pi^+;\Xi_c^0;\Xi^-)+M_{f}(\pi^+;\Xi_c^{\prime 0};\Xi^-)\nonumber\\
&&+M_{f}(\rho^+;\Xi_c^0;\Xi^-)+M_{f}(\rho^+;\Xi_c^{\prime 0};\Xi^-)].
\end{eqnarray}
The amplitudes of the other decays can be obtained in the same manner. Owing to space limitations, we provide these expressions in Appendix \ref{app:amps}.
\section{Numerical Results and Discussions}\label{sec:results}
The decay width of ${\cal B}_{cc}\to{\cal B}D^{(*)}$ can be calculated at the rest frame of ${\cal B}_{cc}$ by
\begin{eqnarray}
\Gamma({\cal B}_{cc}\to {\cal B}D^{(*)})=\frac{\sqrt{(m_{{\cal B}_{cc}}^2-(m_{\cal B}+m_{D^{(*)}})^2)((m_{{\cal B}_{cc}}^2-(m_{\cal B}-m_{D^{(*)}})^2}}{32\pi m_{{\cal B}_{cc}}^3}\sum_{\rm pol.}|{\cal A}({\cal B}_{cc}\to {\cal B}D^{(*)})|^2,
\end{eqnarray}
where the summations are performed over the polarizations of the initial and final states.Moreover, a factor $1/2$ has already been multiplied to average over the polarizations of the mother particle ${\cal B}_{cc}$.
\begin{table}[!htbp]
  \centering
 \caption{Decay constants of light pseudoscalar and vector mesons obtained from Refs. \cite{PDG,Choi:2015ywa} (in units of MeV). $f_{\eta_8}$ and $f_{\eta_1}$ are calculated using the formulae in Ref. \cite{Feldmann:1998vh}}
 \label{tab:decaycons}
  \begin{tabular}{cccccccc}  
   \hline\hline %
     $f_{\pi}$   &$f_{K}$    &$f_{\eta_8}$   &$f_{\eta_1}$   &$f_{\rho}$   &$f_{K^*}$   &$f_{\omega}$   &$f_{\phi}$\\
   \hline %
     $130$   &$156$     &$163$          &$152$          &$216$        &$217$       &$195$          &$233$    \\
   \hline\hline
   \end{tabular}
\end{table}

The calculation of the short distance contribution requires the decay constants of several pseudoscalar and vector mesons, which are presented in Table \ref{tab:decaycons}. Furthermore, numerous strong couplings are required, most of which are obtained from Refs. \cite{Cheng:FSIB,Aliev:2006xr,Aliev:2009ei,Khodjamirian:2011jp,Azizi:2014bua,Yu:2016pyo,Azizi:2015tya,Ballon-Bayona:2017bwk}. Certain strong couplings that cannot be found in literature are calculated under the $SU(3)_F$ symmetry. The data of the strong couplings are presented in Appendix \ref{app:stcouplings}.

\begin{table}[!htbp]
  \centering
  \caption{Our results for branching ratios of $\Xi_{cc}^{++}\to{\cal B} D^{(*)}$. The terms ``CF", ``SCS", and ``DCS" represent CKM favored, singly CKM suppressed and doubly CKM suppressed processes, respectively. The errors are estimated by varying $\eta$ from $1$ to $2$, and the central values are given at $\eta=1.5$. Topologically, these decays are all classified as the $C^\prime$ diagram}
  \label{tab:brsXiccpp}
  \begin{tabular}{|c|c|c|c|c|c|}  
     \hline %
   Channel   &${\cal BR}(10^{-3})$      &CKM    &Channel    &${\cal BR}(10^{-3})$         &CKM  \\
     \hline %
   $\Xi_{cc}^{++}\rightarrow\Sigma^{+}D^{+}$    &$2.98_{-2.02}^{+3.16}$                     &CF&
   $\Xi_{cc}^{++}\rightarrow\Sigma^{+}D^{*+}$    &$16.06_{-10.50}^{+17.28}$                 &CF   \\
     \hline %
   $\Xi_{cc}^{++}\rightarrow\Sigma^{+}D^{+}_{s}$    &$0.17_{-0.12}^{+0.18}$                 &SCS&
   $\Xi_{cc}^{++}\rightarrow\Sigma^{+}D^{*+}_{s}$    &$2.68_{-1.71}^{+2.64}$                &SCS  \\
     \hline %
   $\Xi_{cc}^{++}\rightarrow p D^{+}$    &$0.16_{-0.11}^{+0.18}$                            &SCS&
   $\Xi_{cc}^{++}\rightarrow p D^{*+}$    &$2.96_{-2.06}^{+3.38}$                           &SCS  \\
     \hline %
   $\Xi_{cc}^{++}\rightarrow p D^{+}_{s}$    &$0.01_{-0.00}^{+0.02}$                        &DCS&
   $\Xi_{cc}^{++}\rightarrow p D^{*+}_{s}$    &$0.11_{-0.07}^{+0.13}$                       &DCS  \\
     \hline %
   \end{tabular}
\end{table}

Now we can obtain the numerical values of the related decays. We use the lifetime $\tau_{\Xi_{cc}^{++}}=256\,{\rm fs}$, which was measured by the LHCb collaboration~\cite{Aaij:2018wzf}, to calculate the branching fractions of the $\Xi_{cc}^{++}$ decays, and our results are presented in table \ref{tab:brsXiccpp}. It can be observed that the branching ratios of the $\Xi_{cc}^{++}\to {\cal B} D^*$ decays tend to be larger than those of the $\Xi_{cc}^{++}\to {\cal B} D$ mode when the quark constituents are the same. This can easily be understood by the fact that $\Xi_{cc}^{++}\to {\cal B} D^*$ decays have more polarization states. Among these decays the CKM favored ones certainly have the largest branching ratios. The branching ratio of $\Xi_{cc}^{++}\rightarrow\Sigma^{+}D^{*+}$ is estimated to reach the percentage level.

\begin{table}[!htbp]
  \centering
 \caption{Results for branching ratios of $\Xi_{cc}^{+}\to{\cal B} D^{(*)}$. The terms ``CF", ``SCS", and ``DCS" represent CKM favored, singly CKM suppressed and doubly CKM suppressed processes, respectively. The errors are estimated by varying $\eta$ from $1$ to $2$, and the central values are given at $\eta=1.5$. $B$ and $C^\prime$ represent the contributions in Fig.~\ref{fig:topos}}
   \label{tab:brsXiccp}
  \begin{tabular}{|c|c|c|c|c|c|c|c|}  
   \hline %
   Channel   &$\Gamma/{\rm GeV}$      &CKM &Contribution    &Channels    &$\Gamma/{\rm GeV}$         &CKM  &Contributions\\
   \hline
  $\Xi_{cc}^{+}\rightarrow\Sigma^{0} D^{+}$    &$(5.93_{-4.05}^{+6.31})*10^{-15}$       &CF    &$C^\prime$ $B$  &
  $\Xi_{cc}^{+}\rightarrow\Lambda D^{*+}$   &$(1.82_{-1.26}^{+2.03})*10^{-13}$          &CF  &$C^\prime$ $B$ \\
   \hline
  $\Xi_{cc}^{+}\rightarrow\Lambda D^{+}$   &$(5.84_{-3.98}^{+6.16})*10^{-15}$           &CF    & $C^\prime$ $B$  &
  $\Xi_{cc}^{+}\rightarrow\Sigma^{0} D^{*+}$  &$(2.17 _{-1.51}^{+2.45})*10^{-13}$       &CF     & $C^\prime$ $B$  \\
  \hline
  $\Xi_{cc}^{+}\rightarrow\Sigma^{+} D^{0}$    &$(1.23 _{-0.77 }^{+1.24 })*10^{-15}$    &CF  &$B$  &
  $\Xi_{cc}^{+}\rightarrow\Sigma^{+} D^{*0}$    &$(6.77 _{-4.46}^{+7.37})*10^{-14}$     &CF     &$B$ \\
  \hline
  $\Xi_{cc}^{+}\rightarrow\Xi^{0} D_{s}^{+}$    &$(4.52 _{-3.49}^{+5.22})*10^{-16}$     &CF  &$B$ &
  $\Xi_{cc}^{+}\rightarrow\Xi^{0} D_{s}^{*+}$    &$(2.52 _{-1.48 }^{+2.23})*10^{-14}$   &CF    &$B$  \\

  \hline
  $\Xi_{cc}^{+}\rightarrow p D^{0}$    &$(1.85 _{-1.27}^{+2.02})*10^{-15}$              &SCS  &$B$ &
  $\Xi_{cc}^{+}\rightarrow\Sigma^{0} D_{s}^{*+}$   &$(1.15_{-0.85}^{+1.41})*10^{-14}$   &SCS     & $C^\prime$ $B$  \\
  \hline
   $\Xi_{cc}^{+}\rightarrow\Lambda  D_{s}^{+}$    &$(3.00_{-2.00}^{+2.93})*10^{-16}$     &SCS   & $C^\prime$ $B$  &
   $\Xi_{cc}^{+}\rightarrow\Lambda  D_{s}^{*+}$    &$(1.58_{-1.09}^{+1.73})*10^{-14}$    &SCS    & $C^\prime$ $B$   \\
  \hline
  $\Xi_{cc}^{+}\rightarrow n D^{+}$    &$(1.59 _{-1.13}^{+1.87})*10^{-16}$              &SCS    & $C^\prime$ $B$  &
  $\Xi_{cc}^{+}\rightarrow n D^{*+}$    &$(1.04 _{-0.87 }^{+1.41})*10^{-15}$            &SCS     & $C^\prime$ $B$  \\
  \hline
  $\Xi_{cc}^{+}\rightarrow\Sigma^{0} D_{s}^{+}$    &$(2.85_{-1.78}^{+2.72})*10^{-16}$   &SCS   & $C^\prime$ $B$  &
  $\Xi_{cc}^{+}\rightarrow p D^{*0}$    &$(9.46 _{-6.49}^{+10.20 })*10^{-15}$           &SCS &$B$ \\
   \hline
  $\Xi_{cc}^{+}\rightarrow n D_{s}^{+}$   &$(3.26 _{-2.41}^{+3.88})*10^{-17}$           &DCS &$C^\prime$&
  $\Xi_{cc}^{+}\rightarrow n D_{s}^{*+}$    &$(1.47 _{-1.00}^{+1.57})*10^{-16}$         &DCS   &$C^\prime$\\
   \hline
   \end{tabular}
   \end{table}

\begin{table}[!htbp]
  \centering
 \caption{The same as table \ref{tab:brsXiccp} but for decay widths of $\Omega_{cc}^{+}\to{\cal B} D^{(*)}$}
  \label{tab:brsOmegacc}
  \begin{tabular}{|c|c|c|c|c|c|c|c|}
   \hline %
   Channel   &$\Gamma/{\rm GeV}$      &CKM   &Contributions   &Channels   &$\Gamma/{\rm GeV}$         &CKM   &Contributions    \\
   \hline
   $\Omega_{cc}^{+}\rightarrow\Xi^{0} D^{+}$    &$(1.88_{-1.25}^{+1.97})*10^{-14}$      &CF &$C^\prime$&
   $\Omega_{cc}^{+}\rightarrow\Xi^{0} D^{*+}$     &$(4.99_{-2.62}^{+4.05})*10^{-14}$     &CF   &$C^\prime$ \\
   \hline
   $\Omega_{cc}^{+}\rightarrow\Sigma^{+} D^{0}$    &$(1.76_{-1.07}^{+1.71})*10^{-15}$    &SCS& $B$ &
   $\Omega_{cc}^{+}\rightarrow\Sigma^{0} D^{*+}$    &$(1.80_{-1.27}^{+2.14})*10^{-14}$      &SCS      & $C^\prime$ $B$   \\
   \hline
   $\Omega_{cc}^{+}\rightarrow\Lambda D^{+}$    &$(1.75_{-1.21}^{+1.98})*10^{-15}$      &SCS     & $C^\prime$ $B$  &
   $\Omega_{cc}^{+}\rightarrow\Lambda D^{*+}$    &$(7.65_{-5.32}^{+8.69})*10^{-15}$     &SCS       & $C^\prime$ $B$  \\
   \hline
   $\Omega_{cc}^{+}\rightarrow\Xi^{0} D_{s}^{+}$     &$(9.93_{-6.84}^{+10.87})*10^{-16}$      &SCS     & $C^\prime$ $B$  &
   $\Omega_{cc}^{+}\rightarrow\Xi^{0} D_{s}^{*+}$     &$(4.26_{-2.81}^{+4.06})*10^{-16}$     &SCS       & $C^\prime$ $B$  \\
   \hline
   $\Omega_{cc}^{+}\rightarrow\Sigma^{0} D^{+}$    &$(2.37_{-1.20}^{+2.14})*10^{-16}$      &SCS     & $C^\prime$ $B$  &
   $\Omega_{cc}^{+}\rightarrow\Sigma^{+} D^{*0}$    &$(6.91_{-4.15}^{+6.24})*10^{-15}$     &SCS       &$B$  \\
    \hline
   $\Omega_{cc}^{+}\rightarrow\Sigma^{0} D_{s}^{+}$     &$(1.17_{-6.57}^{+10.93})*10^{-16}$     &DCS     & $C^\prime$ $B$  &
   $\Omega_{cc}^{+}\rightarrow\Sigma^{0} D_{s}^{*+}$    &$(1.68_{-1.18}^{+1.92})*10^{-16}$     &DCS      & $C^\prime$ $B$  \\
   \hline
   $\Omega_{cc}^{+}\rightarrow p D^{0}$    &$(1.74_{-1.23}^{+2.05})*10^{-17}$     &DCS     &$B$  &
   $\Omega_{cc}^{+}\rightarrow p D^{*0}$     &$(3.83_{-2.74}^{+4.85})*10^{-16}$     &DCS        &  $B$ \\
   \hline
   $\Omega_{cc}^{+}\rightarrow n D^{+}$     &$(4.32_{-3.30}^{+5.56})*10^{-17}$      &DCS     &$B$  &
   $\Omega_{cc}^{+}\rightarrow\Lambda D_{s}^{*+}$    &$(1.06_{-0.75}^{+1.27})*10^{-16}$     &DCS      &$B$   \\
   \hline
   $\Omega_{cc}^{+}\rightarrow\Lambda D_{s}^{+}$     &$(7.00_{-4.85}^{+7.87})*10^{-18}$     &DCS     &$B$  &
   $\Omega_{cc}^{+}\rightarrow n D^{*+}$   &$(2.74_{-2.09}^{+3.29})*10^{-17}$   &DCS      &$B$   \\
   \hline
\end{tabular}
\end{table}
In Tables \ref{tab:brsXiccp} and \ref{tab:brsOmegacc} the decay widths of $\Xi_{cc}^+$ and $\Omega_{cc}^+$ decays instead of the branching ratios, because no experimental data of their lifetimes are available. Among the decays of the same mode, ${\cal B}_{cc}\to {\cal B} D$ or ${\cal B}_{cc}\to {\cal B} D^*$, the CKM favored, singly CKM suppressed, and doubly CKM suppressed decays fall into a hierarchy naturally.

$\Xi_{cc}^{+}\rightarrow\Lambda D^{*+}$ and $\Xi_{cc}^{+}\rightarrow\Sigma^{0} D^{*+}$ possess the largest decay widths among the ${\Xi}_{cc}^+\to {\cal B} D^{(*)}$ decays. When estimated with a recently calculated lifetime of $\tau_{\Xi_{cc}^+}=45~\rm{fs}$ in Ref. \cite{Cheng:2018mwu}, their branching ratios are given by
\begin{eqnarray}
{\cal BR}(\Xi_{cc}^{+}\rightarrow\Lambda D^{*+})&\in& [0.38\%, 2.63\%],\nonumber\\
{\cal BR}(\Xi_{cc}^{+}\rightarrow\Sigma^{0} D^{*+})&\in& [0.45\%, 3.16\%].
\end{eqnarray}

The lifetime of $\Omega_{cc}^+$ is predicted to lie in the range of $75\sim 180$ fs in Ref. \cite{Cheng:2018mwu}. Here, we use the boundary of $75$ fs to estimate the three largest branching fractions in the $\Omega_{cc}^+ \to {\cal B} D^{(*)}$ decays, which are given as
\begin{eqnarray}
\label{eq:OmegaBRs}
 {\cal BR}(\Omega_{cc}^{+}\rightarrow\Xi^{0} D^{*+}) &\in& [0.27\%, 1.03\%],\nonumber\\
 {\cal BR}(\Omega_{cc}^{+}\rightarrow\Xi^{0} D^{+}) &\in& [0.07\%, 0.44\%],\nonumber\\
 {\cal BR}(\Omega_{cc}^{+}\rightarrow\Sigma^{0} D^{*+}) &\in& [0.06\%, 0.45\%].
\end{eqnarray}
We also specify the topological contributions of the decays in the tables. It can be observed that the bow-tie mechanism make a sizeable contribution to the charm decays. Let us take $\Omega_{cc}^+\to \Xi^0 D^+$ and $\Omega_{cc}^+ \to \Sigma^+ D^0$ as an example for clarification. The former decay is a pure color commensurate process, whereas latter one is purely dominated by the bow-tie mechanism. It can be observed from Table~\ref{tab:brsOmegacc} that
\begin{eqnarray}
\frac{\Gamma(\Omega_{cc}^+\to \Xi^0 D^+)}{\Gamma(\Omega_{cc}^+ \to \Sigma^+ D^0)}\sim 10.
\end{eqnarray}
The ratio of their CKM matrix elements is
\begin{eqnarray}
\frac{V_{cs}V^*_{ud}}{V_{cs}V^*_{us}}\sim 4.4.
\end{eqnarray}
Considering that the CKM factors are squared in the calculation of decay widths, It can be found that the CKM factors will cause a difference of approximately $20$ times. This means that the bow-tie mechanism and color commensurate mechanism contribute with the same order.
\section{Summary}\label{sec:summary}
The discovery of $\Xi_{cc}^{++}$ in 2017 has inspired the interest in studying doubly charmed baryons. Among all of the related topics, how to calculate their weak decays is a meaningful and challenging one, which can provide valuable suggestions for experimental research as well as aid in understanding the dynamics of baryon decays. In our previous work, we applied the model of FSIs to baryon decays and realized the estimation of two body nonleptonic decays of charm baryons.

In this study we calculate the decays of a doubly charmed baryon to a light baryon and a charm meson. In the same decay mode, ${\cal B}_{cc} \to {\cal B} D$ or ${\cal B}_{cc} \to {\cal B} D^*$, the CKM favored, singly CKM suppressed and doubly CKM suppressed decays fall into a hierarchy naturally. The ${\cal B}_{cc} \to {\cal B} D^*$ decays tends to have larger branching ratios or decay widths because they have more polarization states. Moreover, $\Xi_{cc}^{++}\rightarrow\Sigma^{+}D^{*+}$ has the largest branching ratio in $\Xi_{cc}^{++} \to {\cal B} D^{(*)}$ decays, which lies in the range of $(0.46\sim 3.33)\%$. The two largest branching ratios in the $\Xi_{cc}^{+} \to {\cal B} D^{(*)}$ mode are ${\cal BR}(\Xi_{cc}^{+}\rightarrow\Lambda D^{*+}) \in  [0.38\%, 2.63\%]$ and ${\cal BR}(\Xi_{cc}^{+}\rightarrow\Sigma^{0} D^{*+}) \in  [0.45\%, 3.16\%]$, which are estimated with $\tau_{\Xi_{cc}^+}=45$ fs. For the $\Omega_{cc}^{+} \to {\cal B} D^{(*)}$ mode ${\cal BR}(\Omega_{cc}^{+}\rightarrow\Xi^{0} D^{*+}) \in [0.27\%, 1.03\%]$, ${\cal BR}(\Omega_{cc}^{+}\rightarrow\Xi^{0} D^{+}) \in [0.07\%, 0.44\%]$, and  ${\cal BR}(\Omega_{cc}^{+}\rightarrow\Sigma^{0} D^{*+}) \in [0.06\%, 0.45\%]$ are the three largest ones, and they are calculated with $\tau_{\Omega_{cc}^+}=75$ fs.

By comparing the decay widths of pure color commensurate processes with those of pure bow-tie processes, we found that the bow-tie mechanism also plays an important role in charm decays.

\section{acknowledgement}
This work is supported by the National Natural Science Foundation of China under the Grant Nos. 11765012 and 12075126. We would like to thank Z.-T. Zou, F.-S. Yu, and F.-K. Guo for helpful discussions.
\appendix

\section{Expressions of Diagrams (c)-(i) in Fig. \ref{fig:FSIq}}
\label{app:Fig3diagrams}
{\scriptsize
\begin{eqnarray}
{\cal A}bs\,M_{b}(\phi;\Xi_c^+;D_s^{+})&=&
i\int\frac{|\vec{p_2}|sin\theta d\theta
d\varphi}{32\pi^{2}m_{\Omega_{cc}^{+}}}
\frac{G_{F}}{\sqrt{2}}V_{cs}^{*}V_{us}a_{2}f_{\phi}
\frac{F^{2}(t,m_{D_s^{+}})}{t-m_{D_s^{+}}^{2}+im_{D_s^{+}}\Gamma_{D_s^{+}}}
g_{\Xi_c^{+}\Xi^0D_s^{+}}
g_{D_s^+D_s^+\phi}(p_{4\alpha}+p_{5\alpha})\nonumber\\
  &&\times\overline{u}(p_{6},s^{\prime}_{z})\gamma_{5}(\myslash{p_{3}}+m_{\Xi_{c}^{+}})
(-g^{\mu\alpha}+\frac{p_{2}^{\mu}p_{2}^{\alpha}}{m_{\phi}^{2}})\nonumber\\
  &&\times\left[
\left(f_{1}(m^2_{\phi})
-\frac{m_{\Omega_{cc}^{+}}+m_{\Xi_{c}^{+}}}{m_{\Omega_{cc}^{+}}}{f_{2}(m^2_{\phi})}\right)\gamma_{\mu}\right.
+\frac{2}{m_{\Omega_{cc}^+}}{f_{2}(m^2_{\phi})}p_{3\mu}\nonumber\\
&&\left.-\left
(g_{1}(m^2_{\phi})+\frac{m_{\Omega_{cc}^{+}}-m_{\Xi_{c}^{+}}}{m_{\Omega_{cc}^{+}}}{g_{2}(m^2_{\phi})}\right)
\gamma_{\mu}\gamma_{5}\right.
\left.-\frac{2}{m_{\Omega_{cc}^+}}{g_{2}(m^2_{\phi})}p_{3\mu}\gamma_{5}
  \right]
u(p_{1},s_{z}).
\end{eqnarray}
\begin{eqnarray}
{\cal A}bs\,M_{b}(\phi;\Xi_c^{\prime +};D_s^{+})&=&
i\int\frac{|\vec{p_2}|sin\theta d\theta
d\varphi}{32\pi^{2}m_{\Omega_{cc}^{+}}}
\frac{G_{F}}{\sqrt{2}}V_{cs}^{*}V_{us}a_{2}f_{\phi}
\frac{F^{2}(t,m_{D_s^+})}{t-m_{D_s^+}^{2}+im_{D_s^{+}}\Gamma_{D_s^{+}}}
g_{\Xi_c^{\prime+}\Xi^0D_s^{+}}
g_{D_s^+D_s^+\phi}(p_{4\alpha}+p_{5\alpha})\nonumber\\
  &&\times\overline{u}(p_{6},s^{\prime}_{z})\gamma_{5}(\myslash{p_{3}}+m_{\Xi_{c}^{\prime +}})
(-g^{\mu\alpha}+\frac{p_{2}^{\mu}p_{2}^{\alpha}}{m_{\phi}^{2}})\nonumber\\
  &&\times\left[
\left(f_{1}(m^2_{\phi})
-\frac{m_{\Omega_{cc}^{+}}+m_{\Xi_{c}^{\prime+}}}{m_{\Omega_{cc}^{+}}}{f_{2}(m^2_{\phi})}\right)\gamma_{\mu}\right.
+\frac{2}{m_{\Omega_{cc}^+}}{f_{2}(m^2_{\phi})}p_{3\mu}\nonumber\\
&&\left.-\left
(g_{1}(m^2_{\phi})+\frac{m_{\Omega_{cc}^{+}}-m_{\Xi_{c}^{\prime+}}}{m_{\Omega_{cc}^{+}}}{g_{2}(m^2_{\phi})}\right)
\gamma_{\mu}\gamma_{5}\right.
\left.-\frac{2}{m_{\Omega_{cc}^+}}{g_{2}(m^2_{\phi})}p_{3\mu}\gamma_{5}
  \right]
u(p_{1},s_{z}).
\end{eqnarray}
\begin{eqnarray}
{\cal A}bs\,M_{e}(\eta_1;\Xi_c^{+}; D_s^{*+})&=&
-\int\frac{|\vec{p_2}|sin\theta d\theta
d\varphi}{32\pi^{2}m_{\Omega_{cc}^{+}}}
\frac{G_{F}}{\sqrt{2}}V_{cs}^{*}V_{us}a_{2}f_{\eta_1}
\frac{F^{2}(t,m_{D_s^{*+}})}{t-m_{D_s^{*+}}^{2}+im_{D_s^{*+}}\Gamma_{D_s^{*+}}}g_{D_s^{*+}D_s^{+}\eta_1}p_{2\alpha}\nonumber\\
  &&\times\overline{u}(p_{6},s^{\prime}_{z})
\left[f_{1\Xi_{c}^{+}\Xi^{0}D_s^{*+}}\gamma_{\mu}(-g^{\mu\alpha}+\frac{p_{4}^{\mu}p_{4}^{\alpha}}{m_{D_s^{*+}}^{2}})
+\frac{f_{2\Xi_{c}^{+}\Xi^{0}D_s^{*+}}}{m_{\Xi_{c}^{+}} +m_{\Xi^{0}}}\sigma_{\mu\nu}ip_{4}^{\mu}
(-g^{\nu\alpha}+\frac{p_{4}^{\nu}p_{4}^{\alpha}}{m_{D_s^{*+}}^{2}})\right]\nonumber\\
 &&\times(\myslash{p_{3}}+m_{\Xi_{c}^{+}})
  \left[(m_{\Omega_{cc}^{+}}-m_{\Xi_{c}^{+}})f_{1}(m^{2}_{\eta_1})
+(m_{\Omega_{cc}^{+}}+m_{\Xi_{c}^{+}})g_{1}(m^{2}_{\eta_1})\gamma_{5}\right]
u(p_{1},s_{z}),
\end{eqnarray}
\begin{eqnarray}
{\cal A}bs\,M_{e}(\eta_1;\Xi_c^{\prime +};D_s^{*+})&=&
-\int\frac{|\vec{p_2}|sin\theta d\theta
d\varphi}{32\pi^{2}m_{\Omega_{cc}^{+}}}
\frac{G_{F}}{\sqrt{2}}V_{cs}^{*}V_{us}a_{2}f_{\eta_1}
\frac{F^{2}(t,m_{D_s^{*+}})}{t-m_{D_s^{*+}}^{2}+im_{D_s^{*+}}\Gamma_{D_s^{*+}}}g_{D_s^{*+}D_s^{+}\eta_1}p_{2\alpha}\nonumber\\
  &&\times\overline{u}(p_{6},s^{\prime}_{z})
\left[f_{1\Xi_{c}^{\prime +}\Xi^{0}D_s^{*+}}\gamma_{\mu}(-g^{\mu\alpha}+\frac{p_{4}^{\mu}p_{4}^{\alpha}}{m_{D_s^{*+}}^{2}})
+\frac{f_{2\Xi_{c}^{\prime +}\Xi^{0}D_s^{*+}}}{m_{\Xi_{c}^{\prime +}} +m_{\Xi^{0}}}\sigma_{\mu\nu}ip_{4}^{\mu}
(-g^{\nu\alpha}+\frac{p_{4}^{\nu}p_{4}^{\alpha}}{m_{D_s^{*+}}^{2}})\right]\nonumber\\
 &&\times(\myslash{p_{3}}+m_{\Xi_{c}^{\prime +}})
  \left[(m_{\Omega_{cc}^{+}}-m_{\Xi_{c}^{\prime +}})f_{1}(m^{2}_{\eta_1})
+(m_{\Omega_{cc}^{+}}+m_{\Xi_{c}^{\prime +}})g_{1}(m^{2}_{\eta_1})\gamma_{5}\right]
u(p_{1},s_{z}),
\end{eqnarray}
\begin{eqnarray}
{\cal A}bs\,M_{e}(\eta_8;\Xi_c^{+};D_s^{*+})&=&
-\int\frac{|\vec{p_2}|sin\theta d\theta
d\varphi}{32\pi^{2}m_{\Omega_{cc}^{+}}}
\frac{G_{F}}{\sqrt{2}}V_{cs}^{*}V_{us}a_{2}f_{\eta_8}
\frac{F^{2}(t,m_{D_s^{*+}})}{t-m_{D_s^{*+}}^{2}+im_{D_s^{*+}}\Gamma_{D_s^{*+}}}g_{D_s^{*+}D_s^{+}\eta_8}p_{2\alpha}\nonumber\\
  &&\times\overline{u}(p_{6},s^{\prime}_{z})
\left[f_{1\Xi_{c}^{+}\Xi^{0}D_s^{*+}}\gamma_{\mu}(-g^{\mu\alpha}+\frac{p_{4}^{\mu}p_{4}^{\alpha}}{m_{D_s^{*+}}^{2}})
+\frac{f_{2\Xi_{c}^{+}\Xi^{0}D_s^{*+}}}{m_{\Xi_{c}^{+}} +m_{\Xi^{0}}}\sigma_{\mu\nu}ip_{4}^{\mu}
(-g^{\nu\alpha}+\frac{p_{4}^{\nu}p_{4}^{\alpha}}{m_{D_s^{*+}}^{2}})\right]\nonumber\\
 &&\times(\myslash{p_{3}}+m_{\Xi_{c}^{+}})
  \left[(m_{\Omega_{cc}^{+}}-m_{\Xi_{c}^{+}})f_{1}(m^{2}_{\eta_8})
+(m_{\Omega_{cc}^{+}}+m_{\Xi_{c}^{+}})g_{1}(m^{2}_{\eta_8})\gamma_{5}\right]
u(p_{1},s_{z}),
\end{eqnarray}
\begin{eqnarray}
{\cal A}bs\,M_{e}(\eta_8;\Xi_c^{\prime +};D_s^{*+})&=&
-\int\frac{|\vec{p_2}|sin\theta d\theta
d\varphi}{32\pi^{2}m_{\Omega_{cc}^{+}}}
\frac{G_{F}}{\sqrt{2}}V_{cs}^{*}V_{us}a_{2}f_{\eta_8}
\frac{F^{2}(t,m_{D_s^{*+}})}{t-m_{D_s^{*+}}^{2}+im_{D_s^{*+}}\Gamma_{D_s^{*+}}}g_{D_s^{*+}D_s^{+}\eta_8}p_{2\alpha}\nonumber\\
  &&\times\overline{u}(p_{6},s^{\prime}_{z})
\left[f_{1\Xi_{c}^{\prime +}\Xi^{0}D_s^{*+}}\gamma_{\mu}(-g^{\mu\alpha}+\frac{p_{4}^{\mu}p_{4}^{\alpha}}{m_{D_s^{*+}}^{2}})
+\frac{f_{2\Xi_{c}^{\prime +}\Xi^{0}D_s^{*+}}}{m_{\Xi_{c}^{\prime +}} +m_{\Xi^{0}}}\sigma_{\mu\nu}ip_{4}^{\mu}
(-g^{\nu\alpha}+\frac{p_{4}^{\nu}p_{4}^{\alpha}}{m_{D_s^{*+}}^{2}})\right]\nonumber\\
 &&\times(\myslash{p_{3}}+m_{\Xi_{c}^{\prime +}})
  \left[(m_{\Omega_{cc}^{+}}-m_{\Xi_{c}^{\prime +}})f_{1}(m^{2}_{\eta_8})
+(m_{\Omega_{cc}^{+}}+m_{\Xi_{c}^{\prime +}})g_{1}(m^{2}_{\eta_8})\gamma_{5}\right]
u(p_{1},s_{z}),
\end{eqnarray}
\begin{eqnarray}
{\cal A}bs\,M_{c}(\phi;\Xi_c^+;\Xi^0)&=&
-i\int\frac{|\vec{p_2}|sin\theta d\theta
d\varphi}{32\pi^{2}m_{\Omega_{cc}^{+}}}
\frac{G_{F}}{\sqrt{2}}V_{cs}^{*}V_{us}a_{2}f_{\phi}
\frac{F^{2}(t,m_{\Xi^{0}})}{t-m_{\Xi^{0}}^{2}+im_{\Xi^{0}}\Gamma_{\Xi^{0}}}g_{\Xi_c^{+}\Xi^0D_s^{+}}\nonumber\\
 &&\times\overline{u}(p_{5},s^{\prime}_{z})
 \left[
f_{1\Xi^0\Xi^0\phi}\gamma_{\mu}
(-g^{\alpha\mu}+\frac{p_{2}^{\alpha}p_{2}^{\mu}}{m_{\phi}^{2}})
+\frac{f_{2\Xi^{0}\Xi^{0}\phi}}{m_{\Xi^{0}}+m_{\Xi^{0}}}\sigma_{\mu\nu}(-ip_{2}^{\mu})
(-g^{\alpha\nu}+\frac{p_{2}^{\alpha}p_{2}^{\nu}}{m_{\phi}^{2}})\right]\nonumber\\
&&\times(\myslash{p_{4}}+m_{\Xi^{0}})\gamma_{5}(\myslash{p_{3}}+m_{\Xi_c^{+}})
\left[
\left(f_{1}(m^2_{\phi})
-\frac{m_{\Omega_{cc}^{+}}+m_{\Xi_{c}^{+}}}{m_{\Omega_{cc}^{+}}}{f_{2}(m^2_{\phi})}\right)\gamma_{\alpha}\right.
+\frac{2}{m_{\Omega_{cc}^+}}{f_{2}(m^2_{\phi})}p_{3\alpha}\nonumber\\&&
\left.-\left
(g_{1}(m^2_{\phi})+\frac{m_{\Omega_{cc}^{+}}-m_{\Xi_{c}^{+}}}{m_{\Omega_{cc}^{+}}}{g_{2}(m^2_{\phi})}\right)
\gamma_{\alpha}\gamma_{5}\right.
\left.-\frac{2}{m_{\Omega_{cc}^+}}{g_{2}(m^2_{\phi})}p_{3\alpha}\gamma_{5}
  \right]
u(p_{1},s_{z}).
\end{eqnarray}
\begin{eqnarray}
{\cal A}bs\,M_{c}(\phi;\Xi_c^{\prime +};\Xi^0)&=&
-i\int\frac{|\vec{p_2}|sin\theta d\theta
d\varphi}{32\pi^{2}m_{\Omega_{cc}^{+}}}
\frac{G_{F}}{\sqrt{2}}V_{cs}^{*}V_{us}a_{2}f_{\phi}
\frac{F^{2}(t,m_{\Xi^{0}})}{t-m_{\Xi^{0}}^{2}+im_{\Xi^{0}}\Gamma_{\Xi^{0}}}g_{\Xi_c^{\prime +}\Xi^0D_s^{+}}\nonumber\\
 &&\times\overline{u}(p_{5},s^{\prime}_{z})
 \left[
f_{1\Xi^0\Xi^0\phi}\gamma_{\mu}
(-g^{\alpha\mu}+\frac{p_{2}^{\alpha}p_{2}^{\mu}}{m_{\phi}^{2}})
+\frac{f_{2\Xi^{0}\Xi^{0}\phi}}{m_{\Xi^{0}}+m_{\Xi^{0}}}\sigma_{\mu\nu}(-ip_{2}^{\mu})
(-g^{\alpha\nu}+\frac{p_{2}^{\alpha}p_{2}^{\nu}}{m_{\phi}^{2}})\right]\nonumber\\
&&\times(\myslash{p_{4}}+m_{\Xi^{0}})\gamma_{5}(\myslash{p_{3}}+m_{\Xi_c^{\prime +}})
\left[
\left(f_{1}(m^2_{\phi})
-\frac{m_{\Omega_{cc}^{+}}+m_{\Xi_{c}^{\prime +}}}{m_{\Omega_{cc}^{+}}}{f_{2}(m^2_{\phi})}\right)\gamma_{\alpha}\right.
+\frac{2}{m_{\Omega_{cc}^+}}{f_{2}(m^2_{\phi})}p_{3\alpha}\nonumber\\&&
\left.-\left
(g_{1}(m^2_{\phi})+\frac{m_{\Omega_{cc}^{+}}-m_{\Xi_{c}^{\prime +}}}{m_{\Omega_{cc}^{+}}}{g_{2}(m^2_{\phi})}\right)
\gamma_{\alpha}\gamma_{5}\right.
\left.-\frac{2}{m_{\Omega_{cc}^+}}{g_{2}(m^2_{\phi})}p_{3\alpha}\gamma_{5}
  \right]
u(p_{1},s_{z}).
\end{eqnarray}
\begin{eqnarray}
{\cal A}bs\,M_{c}(\eta_1;\Xi_c^+;\Xi^0)&=&
i\int\frac{|\vec{p_2}|sin\theta d\theta
d\varphi}{32\pi^{2}m_{\Omega_{cc}^{+}}}
\frac{G_{F}}{\sqrt{2}}V_{cs}^{*}V_{us}a_{2}f_{\eta_1}
\frac{F^{2}(t,m_{\Xi^{0}})}{t-m_{\Xi^{0}}^{2}+im_{\Xi^{0}}\Gamma_{\Xi^{0}}}
g_{\Xi^{0}\Xi^{0}\eta_1}
g_{\Xi_c^{+}\Xi^{0}D_s^+}\nonumber\\
 &&\times\overline{u}(p_{5},s^{\prime}_{z})\gamma_{5}(\myslash{p_{4}}+m_{\Xi^{0}})
\gamma_{5}
(\myslash{p_{3}}+m_{\Xi_c^{+}})
\nonumber\\
 &&\times\left[(m_{\Omega_{cc}^{+}}-m_{\Xi_{c}^{+}})f_{1}(m^{2}_{\eta_1})
+(m_{\Omega_{cc}^{+}}+m_{\Xi_{c}^{+}})g_{1}(m^{2}_{\eta_1})\gamma_{5}\right]u(p_{1},s_{z}).
\end{eqnarray}
\begin{eqnarray}
{\cal A}bs\,M_{c}(\eta_1;\Xi_c^{\prime +};\Xi^0)&=&
i\int\frac{|\vec{p_2}|sin\theta d\theta
d\varphi}{32\pi^{2}m_{\Omega_{cc}^{+}}}
\frac{G_{F}}{\sqrt{2}}V_{cs}^{*}V_{us}a_{2}f_{\eta_1}
\frac{F^{2}(t,m_{\Xi^{0}})}{t-m_{\Xi^{0}}^{2}+im_{\Xi^{0}}\Gamma_{\Xi^{0}}}
g_{\Xi^{0}\Xi^{0}\eta_1}
g_{\Xi_c^{\prime +}\Xi^{0}D_s^+}\nonumber\\
 &&\times\overline{u}(p_{5},s^{\prime}_{z})\gamma_{5}(\myslash{p_{4}}+m_{\Xi^{0}})
\gamma_{5}
(\myslash{p_{3}}+m_{\Xi_c^{\prime +}})
\nonumber\\
 &&\times\left[(m_{\Omega_{cc}^{+}}-m_{\Xi_{c}^{\prime +}})f_{1}(m^{2}_{\eta_1})
+(m_{\Omega_{cc}^{+}}+m_{\Xi_{c}^{\prime +}})g_{1}(m^{2}_{\eta_1})\gamma_{5}\right]u(p_{1},s_{z}).
\end{eqnarray}
\begin{eqnarray}
{\cal A}bs\,M_{c}(\eta_8;\Xi_c^+;\Xi^0)&=&
i\int\frac{|\vec{p_2}|sin\theta d\theta
d\varphi}{32\pi^{2}m_{\Omega_{cc}^{+}}}
\frac{G_{F}}{\sqrt{2}}V_{cs}^{*}V_{us}a_{2}f_{\eta_8}
\frac{F^{2}(t,m_{\Xi^{0}})}{t-m_{\Xi^{0}}^{2}+im_{\Xi^{0}}\Gamma_{\Xi^{0}}}
g_{\Xi^{0}\Xi^{0}\eta_8}
g_{\Xi_c^{+}\Xi^{0}D_s^+}\nonumber\\
 &&\times\overline{u}(p_{5},s^{\prime}_{z})\gamma_{5}(\myslash{p_{4}}+m_{\Xi^{0}})
\gamma_{5}
(\myslash{p_{3}}+m_{\Xi_c^{+}})
\nonumber\\
 &&\times\left[(m_{\Omega_{cc}^{+}}-m_{\Xi_{c}^{+}})f_{1}(m^{2}_{\eta_8})
+(m_{\Omega_{cc}^{+}}+m_{\Xi_{c}^{+}})g_{1}(m^{2}_{\eta_8})\gamma_{5}\right]u(p_{1},s_{z}).
\end{eqnarray}
\begin{eqnarray}
{\cal A}bs\,M_{c}(\eta_8;\Xi_c^{\prime +};\Xi^0)&=&
i\int\frac{|\vec{p_2}|sin\theta d\theta
d\varphi}{32\pi^{2}m_{\Omega_{cc}^{+}}}
\frac{G_{F}}{\sqrt{2}}V_{cs}^{*}V_{us}a_{2}f_{\eta_8}
\frac{F^{2}(t,m_{\Xi^{0}})}{t-m_{\Xi^{0}}^{2}+im_{\Xi^{0}}\Gamma_{\Xi^{0}}}
g_{\Xi^{0}\Xi^{0}\eta_8}
g_{\Xi_c^{\prime +}\Xi^{0}D_s^+}\nonumber\\
 &&\times\overline{u}(p_{5},s^{\prime}_{z})\gamma_{5}(\myslash{p_{4}}+m_{\Xi^{0}})
\gamma_{5}
(\myslash{p_{3}}+m_{\Xi_c^{\prime +}})
\nonumber\\
 &&\times\left[(m_{\Omega_{cc}^{+}}-m_{\Xi_{c}^{\prime +}})f_{1}(m^{2}_{\eta_8})
+(m_{\Omega_{cc}^{+}}+m_{\Xi_{c}^{\prime +}})g_{1}(m^{2}_{\eta_8})\gamma_{5}\right]u(p_{1},s_{z}).
\end{eqnarray}
\begin{eqnarray}
{\cal A}bs\,M_{f}(\pi^+;\Xi_c^0;\Xi^-)&=&
i\int\frac{|\vec{p_2}|sin\theta d\theta
d\varphi}{32\pi^{2}m_{\Omega_{cc}^{+}}}
\frac{G_{F}}{\sqrt{2}}V_{cd}^{*}V_{ud}a_{2}f_{\pi^+}
\frac{F^{2}(t,m_{\Xi^{-}})}{t-m_{\Xi^{-}}^{2}+im_{\Xi^{-}}\Gamma_{\Xi^{-}}}
g_{\Xi^{0}\Xi^{-}\pi^+}
g_{\Xi_c^{0}\Xi^{-}D_s^+}\nonumber\\
 &&\times\overline{u}(p_{5},s^{\prime}_{z})\gamma_{5}(\myslash{p_{4}}+m_{\Xi^{-}})
\gamma_{5}
(\myslash{p_{3}}+m_{\Xi_c^{0}})
\nonumber\\
 &&\times\left[(m_{\Omega_{cc}^{+}}-m_{\Xi_{c}^{0}})f_{1}(m^{2}_{\pi^+})
+(m_{\Omega_{cc}^{+}}+m_{\Xi_{c}^{0}})g_{1}(m^{2}_{\pi^+})\gamma_{5}\right]u(p_{1},s_{z}).
\end{eqnarray}
\begin{eqnarray}
{\cal A}bs\,M_{f}(\pi^+;\Xi_c^{\prime 0};\Xi^-)&=&
i\int\frac{|\vec{p_2}|sin\theta d\theta
d\varphi}{32\pi^{2}m_{\Omega_{cc}^{+}}}
\frac{G_{F}}{\sqrt{2}}V_{cd}^{*}V_{ud}a_{2}f_{\pi^+}
\frac{F^{2}(t,m_{\Xi^{-}})}{t-m_{\Xi^{-}}^{2}+im_{\Xi^{-}}\Gamma_{\Xi^{-}}}
g_{\Xi^{0}\Xi^{-}\pi^+}
g_{\Xi_c^{\prime 0}\Xi^{-}D_s^+}\nonumber\\
 &&\times\overline{u}(p_{5},s^{\prime}_{z})\gamma_{5}(\myslash{p_{4}}+m_{\Xi^{-}})
\gamma_{5}
(\myslash{p_{3}}+m_{\Xi_c^{\prime 0}})
\nonumber\\
 &&\times\left[(m_{\Omega_{cc}^{+}}-m_{\Xi_{c}^{\prime 0}})f_{1}(m^{2}_{\pi^+})
+(m_{\Omega_{cc}^{+}}+m_{\Xi_{c}^{\prime 0}})g_{1}(m^{2}_{\pi^+})\gamma_{5}\right]u(p_{1},s_{z}).
\end{eqnarray}
\begin{eqnarray}
{\cal A}bs\,M_{f}(\rho^+;\Xi_c^0;\Xi^-)&=&
-i\int\frac{|\vec{p_2}|sin\theta d\theta
d\varphi}{32\pi^{2}m_{\Omega_{cc}^{+}}}
\frac{G_{F}}{\sqrt{2}}V_{cd}^{*}V_{ud}a_{2}f_{\rho^+}
\frac{F^{2}(t,m_{\Xi^{-}})}{t-m_{\Xi^{-}}^{2}+im_{\Xi^{-}}\Gamma_{\Xi^{-}}}
g_{\Xi_c^{0}\Xi^-D_s^{+}}\nonumber\\
 &&\times\overline{u}(p_{5},s^{\prime}_{z})
 \left[
f_{1\Xi^0\Xi^-\rho^+}\gamma_{\mu}
+\frac{f_{2\Xi^{0}\Xi^{-}\rho^+}}{m_{\Xi^{0}}+m_{\Xi^{-}}}\sigma_{\mu\nu}(-ip_{2}^{\mu})
\right]\nonumber\\
&&\times(\myslash{p_{4}}+m_{\Xi^{-}})\gamma_{5}(\myslash{p_{3}}+m_{\Xi_c^{0}})
(-g^{\alpha\mu}+\frac{p_{2}^{\alpha}p_{2}^{\mu}}{m_{\rho^+}^{2}})
(-g^{\alpha\nu}+\frac{p_{2}^{\alpha}p_{2}^{\nu}}{m_{\rho^+}^{2}})\nonumber\\
&&\times\left[
\left(f_{1}(m^2_{\rho^+})
-\frac{m_{\Omega_{cc}^{+}}+m_{\Xi_{c}^{0}}}{m_{\Omega_{cc}^{+}}}{f_{2}(m^2_{\rho^+})}\right)\gamma_{\alpha}\right.
+\frac{2}{m_{\Omega_{cc}^+}}{f_{2}(m^2_{\rho^+})}p_{3\alpha}\nonumber\\
&&\left.-\left
(g_{1}(m^2_{\rho^+})+\frac{m_{\Omega_{cc}^{+}}-m_{\Xi_{c}^{0}}}{m_{\Omega_{cc}^{+}}}{g_{2}(m^2_{\rho^+})}\right)
\gamma_{\alpha}\gamma_{5}\right.
\left.-\frac{2}{m_{\Omega_{cc}^+}}{g_{2}(m^2_{\rho^+})}p_{3\alpha}\gamma_{5}
  \right]
u(p_{1},s_{z}).
\end{eqnarray}
\begin{eqnarray}
{\cal A}bs\,M_{f}(\rho^+;\Xi_c^{\prime 0};\Xi^-)&=&
-i\int\frac{|\vec{p_2}|sin\theta d\theta
d\varphi}{32\pi^{2}m_{\Omega_{cc}^{+}}}
\frac{G_{F}}{\sqrt{2}}V_{cd}^{*}V_{ud}a_{2}f_{\rho^+}
\frac{F^{2}(t,m_{\Xi^{-}})}{t-m_{\Xi^{-}}^{2}+im_{\Xi^{-}}\Gamma_{\Xi^{-}}}
g_{\Xi_c^{\prime 0}\Xi^-D_s^{+}}\nonumber\\
 &&\times\overline{u}(p_{5},s^{\prime}_{z})
 \left[
f_{1\Xi^0\Xi^-\rho^+}\gamma_{\mu}
+\frac{f_{2\Xi^{0}\Xi^{-}\rho^+}}{m_{\Xi^{0}}+m_{\Xi^{-}}}\sigma_{\mu\nu}(-ip_{2}^{\mu})
\right]\nonumber\\
&&\times(\myslash{p_{4}}+m_{\Xi^{-}})\gamma_{5}(\myslash{p_{3}}+m_{\Xi_c^{\prime 0}})
(-g^{\alpha\mu}+\frac{p_{2}^{\alpha}p_{2}^{\mu}}{m_{\rho^+}^{2}})
(-g^{\alpha\nu}+\frac{p_{2}^{\alpha}p_{2}^{\nu}}{m_{\rho^+}^{2}})\nonumber\\
&&\times\left[
\left(f_{1}(m^2_{\rho^+})
-\frac{m_{\Omega_{cc}^{+}}+m_{\Xi_{c}^{\prime 0}}}{m_{\Omega_{cc}^{+}}}{f_{2}(m^2_{\rho^+})}\right)\gamma_{\alpha}\right.
+\frac{2}{m_{\Omega_{cc}^+}}{f_{2}(m^2_{\rho^+})}p_{3\alpha}\nonumber\\
&&\left.-\left
(g_{1}(m^2_{\rho^+})+\frac{m_{\Omega_{cc}^{+}}-m_{\Xi_{c}^{\prime 0}}}{m_{\Omega_{cc}^{+}}}{g_{2}(m^2_{\rho^+})}\right)
\gamma_{\alpha}\gamma_{5}\right.
\left.-\frac{2}{m_{\Omega_{cc}^+}}{g_{2}(m^2_{\rho^+})}p_{3\alpha}\gamma_{5}
  \right]
u(p_{1},s_{z}).
\end{eqnarray}
}

\section{Expressions of Amplitudes}
\label{app:amps}
The expressions of the amplitudes for all of the ${\cal B}_{cc}\to{\cal B} D^{(*)}$ decays are presented in this section. To make the expressions simpler, we define a function ${\cal M}(P1,P2,P3,P4,P5,P6)$ to represent the absorptive part of the triangle diagram depicted in Fig. \ref{fig:number}. The absorptive part in Eq. (\ref{eq:a2PXP}) is related to this function as follows:
{\scriptsize
\begin{equation}
{\cal A}bs\,M_{a}(K^+;\Omega_c^0;D^{*0})={\cal M}(\Omega_{cc}^{+}, K^+,\Omega_c^0,D^{*0},\Xi^0,D_s^{+}).
\end{equation}}
The amplitudes of all ${\cal B}_{cc}\to{\cal B} D^{(*)}$ decays are calculated as follows using this function:

{\scriptsize
\begin{eqnarray}
{\cal A}(\Xi_{cc}^{++}\to\Sigma^{+}D^{+})
&=&i [
    {\cal M}(\Xi_{cc}^{++}, \pi^+,       \Xi_c^+,           D^{*0},      D^+,        \Sigma^{+})
  + {\cal M}(\Xi_{cc}^{++}, \pi^+,       \Xi_c^{\prime +},  D^{*0},      D^+,        \Sigma^{+})
  + {\cal M}(\Xi_{cc}^{++}, \rho^+,      \Xi_c^+,           D^{0},       D^+,        \Sigma^{+})   \nonumber\\
&&+ {\cal M}(\Xi_{cc}^{++}, \rho^+,      \Xi_c^{\prime +},  D^0,         D^+,        \Sigma^{+})
  + {\cal M}(\Xi_{cc}^{++}, \bar K^{0},  \Sigma_c^{++},     D_s^{*+},    D^+,        \Sigma^{+})
  + {\cal M}(\Xi_{cc}^{++}, \bar K^{*0}, \Sigma_c^{++},     D_s^{+},     D^+,        \Sigma^{+})   \nonumber\\
&&+ {\cal M}(\Xi_{cc}^{++}, \pi^+,       \Xi_c^+,          \Sigma^0,    \Sigma^{+},   D^+  )
  + {\cal M}(\Xi_{cc}^{++}, \pi^+,       \Xi_c^+,          \Lambda,     \Sigma^{+},   D^+  )
  + {\cal M}(\Xi_{cc}^{++}, \pi^+,       \Xi_c^{\prime +}, \Sigma^0,    \Sigma^{+},   D^+  )       \nonumber\\
&&+ {\cal M}(\Xi_{cc}^{++}, \pi^+,       \Xi_c^{\prime +}, \Lambda,     \Sigma^{+},   D^+  )
  + {\cal M}(\Xi_{cc}^{++}, \rho^+,      \Xi_c^+,          \Sigma^0,    \Sigma^{+},   D^+  )
  + {\cal M}(\Xi_{cc}^{++}, \rho^+,      \Xi_c^+,          \Lambda,     \Sigma^{+},   D^+  )       \nonumber\\
&&+ {\cal M}(\Xi_{cc}^{++}, \rho^+,      \Xi_c^{\prime +}, \Sigma^0,    \Sigma^{+},   D^+  )
  + {\cal M}(\Xi_{cc}^{++}, \rho^+,      \Xi_c^{\prime +}, \Lambda,     \Sigma^{+},   D^+  )
  + {\cal M}(\Xi_{cc}^{++}, \bar K^{0},  \Sigma_c^{++},     p,          \Sigma^{+},   D^+  )       \nonumber\\
&&+ {\cal M}(\Xi_{cc}^{++}, \bar K^{*0}, \Sigma_c^{++},     p,          \Sigma^{+},   D^+  )
],
\end{eqnarray}
\begin{eqnarray}
{\cal A}(\Xi_{cc}^{++}\to\Sigma^{+}D^{*+})
&=&i [
    {\cal M}(\Xi_{cc}^{++}, \pi^+,       \Xi_c^+,           D^{0},       D^{*+},        \Sigma^{+})
  + {\cal M}(\Xi_{cc}^{++}, \pi^+,       \Xi_c^{\prime +},  D^{0},       D^{*+},        \Sigma^{+})
  + {\cal M}(\Xi_{cc}^{++}, \rho^+,      \Xi_c^+,           D^{*0},      D^{*+},        \Sigma^{+})   \nonumber\\
&&+ {\cal M}(\Xi_{cc}^{++}, \rho^+,      \Xi_c^{\prime +},  D^{*0},      D^{*+},        \Sigma^{+})
  + {\cal M}(\Xi_{cc}^{++}, \bar K^{0},  \Sigma_c^{++},     D_s^{+},     D^{*+},        \Sigma^{+})
  + {\cal M}(\Xi_{cc}^{++}, \bar K^{*0}, \Sigma_c^{++},     D_s^{*+},    D^{*+},        \Sigma^{+})   \nonumber\\
&&+ {\cal M}(\Xi_{cc}^{++}, \pi^+,       \Xi_c^+,          \Sigma^0,    \Sigma^{+},   D^{*+}  )
  + {\cal M}(\Xi_{cc}^{++}, \pi^+,       \Xi_c^+,          \Lambda,     \Sigma^{+},   D^{*+}  )
  + {\cal M}(\Xi_{cc}^{++}, \pi^+,       \Xi_c^{\prime +}, \Sigma^0,    \Sigma^{+},   D^{*+}  )       \nonumber\\
&&+ {\cal M}(\Xi_{cc}^{++}, \pi^+,       \Xi_c^{\prime +}, \Lambda,     \Sigma^{+},   D^{*+}  )
  + {\cal M}(\Xi_{cc}^{++}, \rho^+,      \Xi_c^+,          \Sigma^0,    \Sigma^{+},   D^{*+}  )
  + {\cal M}(\Xi_{cc}^{++}, \rho^+,      \Xi_c^+,          \Lambda,     \Sigma^{+},   D^{*+}  )       \nonumber\\
&&+ {\cal M}(\Xi_{cc}^{++}, \rho^+,      \Xi_c^{\prime +}, \Sigma^0,    \Sigma^{+},   D^{*+}  )
  + {\cal M}(\Xi_{cc}^{++}, \rho^+,      \Xi_c^{\prime +}, \Lambda,     \Sigma^{+},   D^{*+}  )
  + {\cal M}(\Xi_{cc}^{++}, \bar K^{0},  \Sigma_c^{++},     p,          \Sigma^{+},   D^{*+}  )       \nonumber\\
&&+ {\cal M}(\Xi_{cc}^{++}, \bar K^{*0}, \Sigma_c^{++},     p,          \Sigma^{+},   D^{*+}  )
],
\end{eqnarray}

\begin{eqnarray}
{\cal A}(\Xi_{cc}^{++}\rightarrow\Sigma^{+}D^{+}_{s})
&=&i [
    {\cal M}(\Xi_{cc}^{++},   K^+,        \Xi_c^+,           D^{*0},      D_s^{+},        \Sigma^{+})
  + {\cal M}(\Xi_{cc}^{++},   k^+,        \Xi_c^{\prime +},  D^{*0},      D_s^{+},        \Sigma^{+})
  + {\cal M}(\Xi_{cc}^{++},   K^{*+},     \Xi_c^+,           D^{0},       D_s^{+},        \Sigma^{+})   \nonumber\\
&&+ {\cal M}(\Xi_{cc}^{++},   K^{*+},     \Xi_c^{\prime +},  D^{0},       D_s^{+},        \Sigma^{+})
  + {\cal M}(\Xi_{cc}^{++},  \phi,        \Sigma_c^{++},     D_s^{+},     D_s^{+},        \Sigma^{+})
  + {\cal M}(\Xi_{cc}^{++},  \eta_1,      \Sigma_c^{++},     D_s^{*+},    D_s^{+},        \Sigma^{+})   \nonumber\\
&&+ {\cal M}(\Xi_{cc}^{++},  \eta_8,      \Sigma_c^{++},     D_s^{*+},    D_s^{+},        \Sigma^{+})
  + {\cal M}(\Xi_{cc}^{++},   K^+,        \Xi_c^+,          \Xi^0,       \Sigma^{+},       D_s^{+}  )
  + {\cal M}(\Xi_{cc}^{++},   K^+,        \Xi_c^{\prime +}, \Xi^0,       \Sigma^{+},       D_s^{+}  )   \nonumber\\
&&+ {\cal M}(\Xi_{cc}^{++},   K^{*+},     \Xi_c^+,          \Xi^0,       \Sigma^{+},       D_s^{+}  )
  + {\cal M}(\Xi_{cc}^{++},   K^{*+},     \Xi_c^{\prime +}, \Xi^0,       \Sigma^{+},       D_s^{+}  )
  + {\cal M}(\Xi_{cc}^{++},  \phi,        \Sigma_c^{++},    \Sigma^{+},  \Sigma^{+},       D_s^{+}  )   \nonumber\\
&&+ {\cal M}(\Xi_{cc}^{++},  \eta_1,      \Sigma_c^{++},    \Sigma^{+},  \Sigma^{+},       D_s^{+}  )
  + {\cal M}(\Xi_{cc}^{++},  \eta_8,      \Sigma_c^{++},    \Sigma^{+},  \Sigma^{+},       D_s^{+}  )
  + {\cal M}(\Xi_{cc}^{++},  \pi^+,       \Lambda_c^{+},    \Sigma^{0},  \Sigma^{+},       D_s^{+}  )   \nonumber\\
&&+ {\cal M}(\Xi_{cc}^{++},  \pi^+,       \Lambda_c^{+},    \Lambda,     \Sigma^{+},       D_s^{+}  )
  + {\cal M}(\Xi_{cc}^{++},  \pi^+,       \Sigma_c^{+},     \Sigma^{0},  \Sigma^{+},       D_s^{+}  )
  + {\cal M}(\Xi_{cc}^{++},  \pi^+,       \Sigma_c^{+},     \Lambda,     \Sigma^{+},       D_s^{+}  )   \nonumber\\
&&+ {\cal M}(\Xi_{cc}^{++},  \rho^+,      \Lambda_c^{+},    \Sigma^{0},  \Sigma^{+},       D_s^{+}  )
  + {\cal M}(\Xi_{cc}^{++},  \rho^+,      \Lambda_c^{+},    \Lambda,     \Sigma^{+},       D_s^{+}  )
  + {\cal M}(\Xi_{cc}^{++},  \rho^+,      \Sigma_c^{+},     \Sigma^{0},  \Sigma^{+},       D_s^{+}  )   \nonumber\\
&&+ {\cal M}(\Xi_{cc}^{++},  \rho^+,      \Sigma_c^{+},     \Lambda,     \Sigma^{+},       D_s^{+}  )
],
\end{eqnarray}

\begin{eqnarray}
{\cal A}(\Xi_{cc}^{++}\rightarrow\Sigma^{+}D^{*+}_{s})
&=&i [
    {\cal M}(\Xi_{cc}^{++},   K^+,        \Xi_c^+,           D^{0},       D_s^{*+},        \Sigma^{+})
  + {\cal M}(\Xi_{cc}^{++},   K^+,        \Xi_c^{\prime +},  D^{0},       D_s^{*+},        \Sigma^{+})
  + {\cal M}(\Xi_{cc}^{++},   K^{*+},     \Xi_c^+,           D^{*0},      D_s^{*+},        \Sigma^{+})   \nonumber\\
&&+ {\cal M}(\Xi_{cc}^{++},   K^{*+},     \Xi_c^{\prime +},  D^{*0},      D_s^{*+},        \Sigma^{+})
  + {\cal M}(\Xi_{cc}^{++},  \phi,        \Sigma_c^{++},     D_s^{*+},    D_s^{*+},        \Sigma^{+})
  + {\cal M}(\Xi_{cc}^{++},  \eta_1,      \Sigma_c^{++},     D_s^{+},     D_s^{*+},        \Sigma^{+})   \nonumber\\
&&+ {\cal M}(\Xi_{cc}^{++},  \eta_8,      \Sigma_c^{++},     D_s^{+},     D_s^{*+},        \Sigma^{+})
  + {\cal M}(\Xi_{cc}^{++},   K^+,        \Xi_c^+,          \Xi^0,       \Sigma^{+},        D_s^{*+}  )
  + {\cal M}(\Xi_{cc}^{++},   K^+,        \Xi_c^{\prime +}, \Xi^0,       \Sigma^{+},        D_s^{*+}  )  \nonumber\\
&&+ {\cal M}(\Xi_{cc}^{++},  K^{*+},      \Xi_c^+,          \Xi^0,       \Sigma^{+},        D_s^{*+}  )
  + {\cal M}(\Xi_{cc}^{++},  K^{*+},      \Xi_c^{\prime +}, \Xi^0,       \Sigma^{+},        D_s^{*+}  )
  + {\cal M}(\Xi_{cc}^{++},  \phi,        \Sigma_c^{++},    \Sigma^{+},  \Sigma^{+},        D_s^{*+}  )  \nonumber\\
&&+ {\cal M}(\Xi_{cc}^{++},  \eta_1,      \Sigma_c^{++},    \Sigma^{+},  \Sigma^{+},        D_s^{*+}  )
  + {\cal M}(\Xi_{cc}^{++},  \eta_8,      \Sigma_c^{++},    \Sigma^{+},  \Sigma^{+},        D_s^{*+}  )
  + {\cal M}(\Xi_{cc}^{++},  \pi^+,       \Lambda_c^{+},    \Sigma^{0},  \Sigma^{+},        D_s^{*+}  )  \nonumber\\
&&+ {\cal M}(\Xi_{cc}^{++},  \pi^+,       \Lambda_c^{+},    \Lambda,     \Sigma^{+},        D_s^{*+}  )
  + {\cal M}(\Xi_{cc}^{++},  \pi^+,       \Sigma_c^{+},     \Sigma^{0},  \Sigma^{+},        D_s^{*+}  )
  + {\cal M}(\Xi_{cc}^{++},  \pi^+,       \Sigma_c^{+},     \Lambda,     \Sigma^{+},        D_s^{*+}  )  \nonumber\\
&&+ {\cal M}(\Xi_{cc}^{++},  \rho^+,      \Lambda_c^{+},    \Sigma^{0},  \Sigma^{+},        D_s^{*+}  )
  + {\cal M}(\Xi_{cc}^{++},  \rho^+,      \Lambda_c^{+},    \Lambda,     \Sigma^{+},        D_s^{*+}  )
  + {\cal M}(\Xi_{cc}^{++},  \rho^+,      \Sigma_c^{+},     \Sigma^{0},  \Sigma^{+},        D_s^{*+}  )  \nonumber\\
&&+ {\cal M}(\Xi_{cc}^{++},  \rho^+,      \Sigma_c^{+},     \Lambda,     \Sigma^{+},        D_s^{*+}  )
],
\end{eqnarray}

\begin{eqnarray}
{\cal A}(\Xi_{cc}^{++}\rightarrow p D^{+})
&=&i [
    {\cal M}(\Xi_{cc}^{++},   \pi^+,        \Lambda_c^+,           D^{*0},       D^{+},        p   )
  + {\cal M}(\Xi_{cc}^{++},   \pi^+,        \Sigma_c^+,            D^{*0},       D^{+},        p   )
  + {\cal M}(\Xi_{cc}^{++},   \rho^+,       \Lambda_c^+,           D^{0},        D^{+},        p   ) \nonumber\\
&&+ {\cal M}(\Xi_{cc}^{++},   \rho^+,       \Sigma_c^+,            D^{0},        D^{+},        p   )
  + {\cal M}(\Xi_{cc}^{++},   \rho^0,       \Sigma_c^{++},         D^{+},        D^{+},        p   )
  + {\cal M}(\Xi_{cc}^{++},   \omega,       \Sigma_c^{++},         D^{+},        D^{+},        p   )  \nonumber\\
&&+ {\cal M}(\Xi_{cc}^{++},   \pi^0,        \Sigma_c^{++},         D^{*+},       D^{+},        p   )
  + {\cal M}(\Xi_{cc}^{++},   \eta_1,       \Sigma_c^{++},         D^{*+},       D^{+},        p   )
  + {\cal M}(\Xi_{cc}^{++},   \eta_8,       \Sigma_c^{++},         D^{*+},       D^{+},        p   )  \nonumber\\
&&+ {\cal M}(\Xi_{cc}^{++},   \pi^+,        \Lambda_c^+,           n,            p,            D^+ )
  + {\cal M}(\Xi_{cc}^{++},   \pi^+,        \Sigma_c^+,            n,            p,            D^+ )
  + {\cal M}(\Xi_{cc}^{++},   \rho^+,       \Lambda_c^+,           n,            p,            D^+ )  \nonumber\\
&&+ {\cal M}(\Xi_{cc}^{++},   \rho^+,       \Sigma_c^+,            n,            p,            D^+ )
  + {\cal M}(\Xi_{cc}^{++},   \rho^0,       \Sigma_c^{++},         p,            p,            D^+ )
  + {\cal M}(\Xi_{cc}^{++},   \pi^0,        \Sigma_c^{++},         p,            p,            D^+ )  \nonumber\\
&&+ {\cal M}(\Xi_{cc}^{++},   \eta_1,       \Sigma_c^{++},         p,            p,            D^+ )
  + {\cal M}(\Xi_{cc}^{++},   \eta_8,       \Sigma_c^{++},         p,            p,            D^+ )
  + {\cal M}(\Xi_{cc}^{++},    K^+,         \Xi_c^+,               \Sigma^0,     p,            D^+ )  \nonumber\\
&&+ {\cal M}(\Xi_{cc}^{++},    K^+,         \Xi_c^+,               \Lambda,      p,            D^+ )
  + {\cal M}(\Xi_{cc}^{++},    K^+,         \Xi_c^{\prime +},      \Sigma^0,     p,            D^+ )
  + {\cal M}(\Xi_{cc}^{++},    K^+,         \Xi_c^{\prime +},      \Lambda,      p,            D^+ )  \nonumber\\
&&+ {\cal M}(\Xi_{cc}^{++},    K^{*+},      \Xi_c^{\prime +},      \Sigma^0,     p,            D^+ )
  + {\cal M}(\Xi_{cc}^{++},    K^{*+},      \Xi_c^{\prime +},      \Lambda,      p,            D^+ )
  + {\cal M}(\Xi_{cc}^{++},    K^{*+},      \Xi_c^+,               \Sigma^0,     p,            D^+ )  \nonumber\\
&&+ {\cal M}(\Xi_{cc}^{++},    K^{*+},      \Xi_c^+,               \Lambda,      p,            D^+ )
 ],
\end{eqnarray}

\begin{eqnarray}
{\cal A}(\Xi_{cc}^{++}\rightarrow p D^{*+})
&=&i [
    {\cal M}(\Xi_{cc}^{++},   \pi^+,        \Lambda_c^+,           D^{0},        D^{*+},        p   )
  + {\cal M}(\Xi_{cc}^{++},   \pi^+,        \Sigma_c^+,            D^{0},        D^{*+},        p   )
  + {\cal M}(\Xi_{cc}^{++},   \rho^+,       \Lambda_c^+,           D^{*0},       D^{*+},        p   )  \nonumber\\
&&+ {\cal M}(\Xi_{cc}^{++},   \rho^+,       \Sigma_c^+,            D^{*0},       D^{*+},        p   )
  + {\cal M}(\Xi_{cc}^{++},   \rho^0,       \Sigma_c^{++},         D^{*+},       D^{*+},        p   )
  + {\cal M}(\Xi_{cc}^{++},   \omega,       \Sigma_c^{++},         D^{*+},       D^{*+},        p   )   \nonumber\\
&&+ {\cal M}(\Xi_{cc}^{++},   \pi^0,        \Sigma_c^{++},         D^{+},        D^{*+},        p   )
  + {\cal M}(\Xi_{cc}^{++},   \eta_1,       \Sigma_c^{++},         D^{+},        D^{*+},        p   )
  + {\cal M}(\Xi_{cc}^{++},   \eta_8,       \Sigma_c^{++},         D^{+},        D^{*+},        p   )   \nonumber\\
&&+ {\cal M}(\Xi_{cc}^{++},   \pi^+,        \Lambda_c^+,           n,            p,            D^{*+} )
  + {\cal M}(\Xi_{cc}^{++},   \pi^+,        \Sigma_c^+,            n,            p,            D^{*+} )
  + {\cal M}(\Xi_{cc}^{++},   \rho^+,       \Lambda_c^+,           n,            p,            D^{*+} )  \nonumber\\
&&+ {\cal M}(\Xi_{cc}^{++},   \rho^+,       \Sigma_c^+,            n,            p,            D^{*+} )
  + {\cal M}(\Xi_{cc}^{++},   \rho^0,       \Sigma_c^{++},         p,            p,            D^{*+} )
  + {\cal M}(\Xi_{cc}^{++},   \pi^0,        \Sigma_c^{++},         p,            p,            D^{*+} )  \nonumber\\
&&+ {\cal M}(\Xi_{cc}^{++},   \eta_1,       \Sigma_c^{++},         p,            p,            D^{*+} )
  + {\cal M}(\Xi_{cc}^{++},   \eta_8,       \Sigma_c^{++},         p,            p,            D^{*+} )
  + {\cal M}(\Xi_{cc}^{++},    K^+,         \Xi_c^+,               \Sigma^0,     p,            D^{*+} )  \nonumber\\
&&+ {\cal M}(\Xi_{cc}^{++},    K^+,         \Xi_c^+,               \Lambda,      p,            D^{*+} )
  + {\cal M}(\Xi_{cc}^{++},    K^+,         \Xi_c^{\prime +},      \Sigma^0,     p,            D^{*+} )
  + {\cal M}(\Xi_{cc}^{++},    K^+,         \Xi_c^{\prime +},      \Lambda,      p,            D^{*+} )  \nonumber\\
&&+ {\cal M}(\Xi_{cc}^{++},    K^{*+},      \Xi_c^{\prime +},      \Sigma^0,     p,            D^{*+} )
  + {\cal M}(\Xi_{cc}^{++},    K^{*+},      \Xi_c^{\prime +},      \Lambda,      p,            D^{*+} )
  + {\cal M}(\Xi_{cc}^{++},    K^{*+},      \Xi_c^+,               \Sigma^0,     p,            D^{*+} )  \nonumber\\
&&+ {\cal M}(\Xi_{cc}^{++},    K^{*+},      \Xi_c^+,               \Lambda,      p,            D^{*+} )
   ],
\end{eqnarray}

\begin{eqnarray}
{\cal A}(\Xi_{cc}^{++}\rightarrow p D^{+}_{s})
&=&i [
    {\cal M}(\Xi_{cc}^{++},   K^+,        \Lambda_c^+,           D^{*0},       D_s^{+},        p       )
  + {\cal M}(\Xi_{cc}^{++},   K^+,        \Sigma_c^+,            D^{*0},       D_s^{+},        p       )
  + {\cal M}(\Xi_{cc}^{++},   K^{*+},     \Lambda_c^+,           D^{0},        D_s^{+},        p       )  \nonumber\\
&&+ {\cal M}(\Xi_{cc}^{++},   K^{*+},     \Sigma_c^+,            D^{0},        D_s^{+},        p       )
  + {\cal M}(\Xi_{cc}^{++},   K^+,        \Lambda_c^+,          \Sigma^0,      p,              D_s^{+} )
  + {\cal M}(\Xi_{cc}^{++},   K^+,        \Lambda_c^+,          \Lambda,       p,              D_s^{+} )  \nonumber\\
&&+ {\cal M}(\Xi_{cc}^{++},   K^+,        \Sigma_c^+,           \Sigma^0,      p,              D_s^{+} )
  + {\cal M}(\Xi_{cc}^{++},   K^+,        \Sigma_c^+,           \Lambda,       p,              D_s^{+} )
  + {\cal M}(\Xi_{cc}^{++},   K^{*+},     \Lambda_c^+,          \Sigma^0,      p,              D_s^{+} )  \nonumber\\
&&+ {\cal M}(\Xi_{cc}^{++},   K^{*+},     \Lambda_c^+,          \Lambda,       p,              D_s^{+} )
  + {\cal M}(\Xi_{cc}^{++},   K^{*+},     \Sigma_c^+,           \Sigma^0,      p,              D_s^{+} )
  + {\cal M}(\Xi_{cc}^{++},   K^{*+},     \Sigma_c^+,           \Lambda,       p,              D_s^{+} )  \nonumber\\
&&+ {\cal M}(\Xi_{cc}^{++},   K^0,        \Sigma_c^{++},         D^{*+},       D_s^{+},        p       )
  + {\cal M}(\Xi_{cc}^{++},   K^{*0},     \Sigma_c^{++},         D^{+},        D_s^{+},        p       )
  + {\cal M}(\Xi_{cc}^{++},   K^0,        \Sigma_c^{++},        \Sigma^+,      p,              D_s^{+} )  \nonumber\\
&&+ {\cal M}(\Xi_{cc}^{++},   K^{*0},     \Sigma_c^{++},        \Sigma^+,      p,              D_s^{+} )
 ],
\end{eqnarray}

\begin{eqnarray}
{\cal A}(\Xi_{cc}^{++}\rightarrow p D^{*+}_{s}
)&=&i [
    {\cal M}(\Xi_{cc}^{++},   K^+,        \Lambda_c^+,           D^{0},        D_s^{*+},       p        )
  + {\cal M}(\Xi_{cc}^{++},   K^+,        \Sigma_c^+,            D^{0},        D_s^{*+},       p        )
  + {\cal M}(\Xi_{cc}^{++},   K^{*+},     \Lambda_c^+,           D^{*0},       D_s^{*+},       p        )  \nonumber\\
&&+ {\cal M}(\Xi_{cc}^{++},   K^{*+},     \Sigma_c^+,            D^{*0},       D_s^{*+},       p        )
  + {\cal M}(\Xi_{cc}^{++},   K^+,        \Lambda_c^+,          \Sigma^0,      p,              D_s^{*+} )
  + {\cal M}(\Xi_{cc}^{++},   K^+,        \Lambda_c^+,          \Lambda,       p,              D_s^{*+} )  \nonumber\\
&&+ {\cal M}(\Xi_{cc}^{++},   K^+,        \Sigma_c^+,           \Sigma^0,      p,              D_s^{*+} )
  + {\cal M}(\Xi_{cc}^{++},   K^+,        \Sigma_c^+,           \Lambda,       p,              D_s^{*+} )
  + {\cal M}(\Xi_{cc}^{++},   K^{*+},     \Lambda_c^+,          \Sigma^0,      p,              D_s^{*+} )  \nonumber\\
&&+ {\cal M}(\Xi_{cc}^{++},   K^{*+},     \Lambda_c^+,          \Lambda,       p,              D_s^{*+} )
  + {\cal M}(\Xi_{cc}^{++},   K^{*+},     \Sigma_c^+,           \Sigma^0,      p,              D_s^{*+} )
  + {\cal M}(\Xi_{cc}^{++},   K^{*+},     \Sigma_c^+,           \Lambda,       p,              D_s^{*+} )  \nonumber\\
&&+ {\cal M}(\Xi_{cc}^{++},   K^0,        \Sigma_c^{++},         D^{+},        D_s^{*+},       p        )
  + {\cal M}(\Xi_{cc}^{++},   K^{*0},     \Sigma_c^{++},         D^{*+},       D_s^{*+},       p        )
  + {\cal M}(\Xi_{cc}^{++},   K^0,        \Sigma_c^{++},        \Sigma^+,      p,              D_s^{*+} )  \nonumber\\
&&+ {\cal M}(\Xi_{cc}^{++},   K^{*0},     \Sigma_c^{++},        \Sigma^+,      p,              D_s^{*+} )
    ],
\end{eqnarray}

\begin{eqnarray}
{\cal A}(\Xi_{cc}^{+}\rightarrow\Sigma^{0} D^{+})
&=&i [
    {\cal M}(\Xi_{cc}^{+},   \pi^{+},      \Xi_c^{ 0},       D^{*0},        D^{+},      \Sigma^{0}  )
  + {\cal M}(\Xi_{cc}^{+},   \rho^{+},     \Xi_c^{ 0},       D^{0},         D^{+},      \Sigma^{0}  )
  + {\cal M}(\Xi_{cc}^{+},   \pi^{+},      \Xi_c^{ 0},      \Sigma^{-},    \Sigma^{0},   D^{+}      )    \nonumber\\
&&+ {\cal M}(\Xi_{cc}^{+},   \rho^{+},     \Xi_c^{ 0},      \Sigma^{-},    \Sigma^{0},   D^{+}      )
  + {\cal M}(\Xi_{cc}^{+},   \pi^{+},      \Xi_c^{\prime 0}, D^{*0},        D^{+},      \Sigma^{0}  )
  + {\cal M}(\Xi_{cc}^{+},   \rho^{+},     \Xi_c^{\prime 0}, D^{0},         D^{+},      \Sigma^{0}  )    \nonumber\\
&&+ {\cal M}(\Xi_{cc}^{+},   \pi^{+},      \Xi_c^{\prime 0},\Sigma^{-},    \Sigma^{0},   D^{+}      )
  + {\cal M}(\Xi_{cc}^{+},   \rho^{+},     \Xi_c^{\prime 0},\Sigma^{-},    \Sigma^{0},   D^{+}      )
  + {\cal M}(\Xi_{cc}^{+},   \bar K^{0},   \Sigma_c^{+},     D_{s}^{*+},    D^{+},      \Sigma^{0}  )    \nonumber\\
&&+ {\cal M}(\Xi_{cc}^{+},   \bar K^{*0},  \Sigma_c^{+},     D_{s}^{+},     D^{+},      \Sigma^{0}  )
  + {\cal M}(\Xi_{cc}^{+},   \bar K^{0},   \Lambda_c^{+},    D_{s}^{*+},    D^{+},      \Sigma^{0}  )
  + {\cal M}(\Xi_{cc}^{+},   \bar K^{*0},  \Lambda_c^{+},    D_{s}^{+},     D^{+},      \Sigma^{0}  )    \nonumber\\
&&+ {\cal M}(\Xi_{cc}^{+},   \bar K^{0},   \Lambda_c^{+},    n,            \Sigma^{0},   D^{+}      )
  + {\cal M}(\Xi_{cc}^{+},   \bar K^{0},   \Sigma_c^{+},     n,            \Sigma^{0},   D^{+}      )
  + {\cal M}(\Xi_{cc}^{+},   \bar K^{*0},  \Lambda_c^{+},    n,            \Sigma^{0},   D^{+}      )    \nonumber\\
&&+ {\cal M}(\Xi_{cc}^{+},   \bar K^{*0},  \Sigma_c^{+},     n,            \Sigma^{0},   D^{+}      )
 ],
\end{eqnarray}
\begin{eqnarray}
{\cal A}(\Xi_{cc}^{+}\rightarrow\Lambda D^{+})
&=&i [
    {\cal M}(\Xi_{cc}^{+},   \pi^{+},      \Xi_c^{ 0},       D^{*0},        D^{+},      \Lambda  )
  + {\cal M}(\Xi_{cc}^{+},   \rho^{+},     \Xi_c^{ 0},       D^{0},         D^{+},      \Lambda  )
  + {\cal M}(\Xi_{cc}^{+},   \pi^{+},      \Xi_c^{ 0},      \Sigma^{-},    \Lambda,      D^{+}   )    \nonumber\\
&&+ {\cal M}(\Xi_{cc}^{+},   \rho^{+},     \Xi_c^{ 0},      \Sigma^{-},    \Lambda,      D^{+}   )
  + {\cal M}(\Xi_{cc}^{+},   \bar K^{0},   \Sigma_c^{+},     D_{s}^{*+},    D^{+},      \Lambda  )
  + {\cal M}(\Xi_{cc}^{+},   \bar K^{*0},  \Sigma_c^{+},     D_{s}^{+},     D^{+},      \Lambda  )    \nonumber\\
&&+ {\cal M}(\Xi_{cc}^{+},   \bar K^{0},   \Lambda_c^{+},    D_{s}^{*+},    D^{+},      \Lambda  )
  + {\cal M}(\Xi_{cc}^{+},   \bar K^{*0},  \Lambda_c^{+},    D_{s}^{+},     D^{+},      \Lambda  )
  + {\cal M}(\Xi_{cc}^{+},   \bar K^{0},   \Lambda_c^{+},    n,            \Lambda,      D^{+}   )    \nonumber\\
&&+ {\cal M}(\Xi_{cc}^{+},   \bar K^{0},   \Sigma_c^{+},     n,            \Lambda,      D^{+}   )
  + {\cal M}(\Xi_{cc}^{+},   \bar K^{*0},  \Lambda_c^{+},    n,            \Lambda,      D^{+}   )
  + {\cal M}(\Xi_{cc}^{+},   \bar K^{*0},  \Sigma_c^{+},     n,            \Lambda,      D^{+}   )    \nonumber\\
&&+ {\cal M}(\Xi_{cc}^{+},   \pi^{+},      \Xi_c^{\prime 0}, D^{*0},        D^{+},      \Lambda  )
  + {\cal M}(\Xi_{cc}^{+},   \rho^{+},     \Xi_c^{\prime 0}, D^{0},         D^{+},      \Lambda  )
  + {\cal M}(\Xi_{cc}^{+},   \pi^{+},      \Xi_c^{\prime 0},\Sigma^{-},    \Lambda,      D^{+}   )    \nonumber\\
&&+ {\cal M}(\Xi_{cc}^{+},   \rho^{+},     \Xi_c^{\prime 0},\Sigma^{-},    \Lambda,      D^{+}   )
 ],
\end{eqnarray}
\begin{eqnarray}
{\cal A}(\Xi_{cc}^{+}\rightarrow\Sigma^{0} D^{*+}
)&=&i [
    {\cal M}(\Xi_{cc}^{+},   \pi^{+},      \Xi_c^{ 0},       D^{0},        D^{*+},      \Sigma^{0}  )
  + {\cal M}(\Xi_{cc}^{+},   \rho^{+},     \Xi_c^{ 0},       D^{*0},       D^{*+},      \Sigma^{0}  )
  + {\cal M}(\Xi_{cc}^{+},   \pi^{+},      \Xi_c^{ 0},      \Sigma^{-},    \Sigma^{0},   D^{*+}     )    \nonumber\\
&&+ {\cal M}(\Xi_{cc}^{+},   \rho^{+},     \Xi_c^{ 0},      \Sigma^{-},    \Sigma^{0},   D^{*+}     )
  + {\cal M}(\Xi_{cc}^{+},   \bar K^{0},   \Sigma_c^{+},     D_{s}^{+},    D^{*+},      \Sigma^{0}  )
  + {\cal M}(\Xi_{cc}^{+},   \bar K^{*0},  \Sigma_c^{+},     D_{s}^{*+},   D^{*+},      \Sigma^{0}  )    \nonumber\\
&&+ {\cal M}(\Xi_{cc}^{+},   \bar K^{0},   \Lambda_c^{+},    D_{s}^{+},    D^{*+},      \Sigma^{0}  )
  + {\cal M}(\Xi_{cc}^{+},   \bar K^{*0},  \Lambda_c^{+},    D_{s}^{*+},   D^{*+},      \Sigma^{0}  )
  + {\cal M}(\Xi_{cc}^{+},   \bar K^{0},   \Lambda_c^{+},    n,            \Sigma^{0},   D^{*+}     )    \nonumber\\
&&+ {\cal M}(\Xi_{cc}^{+},   \bar K^{0},   \Sigma_c^{+},     n,            \Sigma^{0},   D^{*+}     )
  + {\cal M}(\Xi_{cc}^{+},   \bar K^{*0},  \Lambda_c^{+},    n,            \Sigma^{0},   D^{*+}     )
  + {\cal M}(\Xi_{cc}^{+},   \bar K^{*0},  \Sigma_c^{+},     n,            \Sigma^{0},   D^{*+}     )    \nonumber\\
&&+ {\cal M}(\Xi_{cc}^{+},   \pi^{+},      \Xi_c^{\prime 0}, D^{0},        D^{*+},      \Sigma^{0}  )
  + {\cal M}(\Xi_{cc}^{+},   \rho^{+},     \Xi_c^{\prime 0}, D^{*0},       D^{*+},      \Sigma^{0}  )
  + {\cal M}(\Xi_{cc}^{+},   \pi^{+},      \Xi_c^{\prime 0},\Sigma^{-},    \Sigma^{0},   D^{*+}     )    \nonumber\\
&&+ {\cal M}(\Xi_{cc}^{+},   \rho^{+},     \Xi_c^{\prime 0},\Sigma^{-},    \Sigma^{0},   D^{*+}     ) ],
\end{eqnarray}
\begin{eqnarray}
{\cal A}(\Xi_{cc}^{+}\rightarrow\Lambda D^{*+})
&=&i [
    {\cal M}(\Xi_{cc}^{+},   \pi^{+},      \Xi_c^{ 0},       D^{0},         D^{*+},      \Lambda  )
  + {\cal M}(\Xi_{cc}^{+},   \rho^{+},     \Xi_c^{ 0},       D^{*0},        D^{*+},      \Lambda  )
  + {\cal M}(\Xi_{cc}^{+},   \pi^{+},      \Xi_c^{ 0},      \Sigma^{-},    \Lambda,      D^{*+}   )    \nonumber\\
&&+ {\cal M}(\Xi_{cc}^{+},   \rho^{+},     \Xi_c^{ 0},      \Sigma^{-},    \Lambda,      D^{*+}   )
  + {\cal M}(\Xi_{cc}^{+},   \bar K^{0},   \Sigma_c^{+},     D_{s}^{+},    D^{*+},      \Lambda   )
  + {\cal M}(\Xi_{cc}^{+},   \bar K^{*0},  \Sigma_c^{+},     D_{s}^{*+},   D^{*+},      \Lambda   )    \nonumber\\
&&+ {\cal M}(\Xi_{cc}^{+},   \bar K^{0},   \Lambda_c^{+},    D_{s}^{+},    D^{*+},      \Lambda   )
  + {\cal M}(\Xi_{cc}^{+},   \bar K^{*0},  \Lambda_c^{+},    D_{s}^{*+},   D^{*+},      \Lambda   )
  + {\cal M}(\Xi_{cc}^{+},   \bar K^{0},   \Lambda_c^{+},    n,            \Lambda,      D^{*+}   )    \nonumber\\
&&+ {\cal M}(\Xi_{cc}^{+},   \bar K^{0},   \Sigma_c^{+},     n,            \Lambda,      D^{*+}   )
  + {\cal M}(\Xi_{cc}^{+},   \bar K^{*0},  \Lambda_c^{+},    n,            \Lambda,      D^{*+}   )
  + {\cal M}(\Xi_{cc}^{+},   \bar K^{*0},  \Sigma_c^{+},     n,            \Lambda,      D^{*+}   )    \nonumber\\
&&+ {\cal M}(\Xi_{cc}^{+},   \pi^{+},      \Xi_c^{\prime 0}, D^{0},         D^{*+},      \Lambda  )
  + {\cal M}(\Xi_{cc}^{+},   \rho^{+},     \Xi_c^{\prime 0}, D^{*0},        D^{*+},      \Lambda  )
  + {\cal M}(\Xi_{cc}^{+},   \pi^{+},      \Xi_c^{\prime 0},\Sigma^{-},    \Lambda,      D^{*+}   )    \nonumber\\
&&+ {\cal M}(\Xi_{cc}^{+},   \rho^{+},     \Xi_c^{\prime 0},\Sigma^{-},    \Lambda,      D^{*+}   ) ],
\end{eqnarray}

\begin{eqnarray}
{\cal A}( \Xi_{cc}^{+}\rightarrow\Lambda D_{s}^{+}
)&=&i [
    {\cal M}(\Xi_{cc}^{+},   K^{+},      \Xi_c^{ 0},       D^{*0},         D_s^{+},      \Lambda  )
  + {\cal M}(\Xi_{cc}^{+},   K^{*+},     \Xi_c^{ 0},       D^{0},          D_s^{+},      \Lambda  )
  + {\cal M}(\Xi_{cc}^{+},   K^{+},      \Xi_c^{ 0},      \Xi^{-},        \Lambda,     D_s^{+}    )  \nonumber\\
&&+ {\cal M}(\Xi_{cc}^{+},   K^{*+},     \Xi_c^{ 0},      \Xi^{-},        \Lambda,     D_s^{+}    )
  + {\cal M}(\Xi_{cc}^{+},   \phi,       \Lambda_{c}^{+},  D_s^{+},        D_s^{+},      \Lambda  )
  + {\cal M}(\Xi_{cc}^{+},   \phi,       \Sigma_{c}^{+},   D_s^{+},        D_s^{+},      \Lambda  )  \nonumber\\
&&+ {\cal M}(\Xi_{cc}^{+},   \eta_1,     \Lambda_{c}^{+},  D_s^{*+},       D_s^{+},      \Lambda  )
  + {\cal M}(\Xi_{cc}^{+},   \eta_1,     \Sigma_{c}^{+},   D_s^{*+},       D_s^{+},      \Lambda  )
  + {\cal M}(\Xi_{cc}^{+},   \eta_8,     \Lambda_{c}^{+},  D_s^{*+},       D_s^{+},      \Lambda  )  \nonumber\\
&&+ {\cal M}(\Xi_{cc}^{+},   \eta_8,     \Sigma_{c}^{+},   D_s^{*+},       D_s^{+},      \Lambda  )
  + {\cal M}(\Xi_{cc}^{+},   \phi,       \Lambda_{c}^{+},  \Lambda,       \Lambda,     D_s^{+}    )
  + {\cal M}(\Xi_{cc}^{+},   \eta_1,     \Lambda_{c}^{+},  \Lambda,       \Lambda,     D_s^{+}    )  \nonumber\\
&&+ {\cal M}(\Xi_{cc}^{+},   \eta_8,     \Lambda_{c}^{+},  \Lambda,       \Lambda,     D_s^{+}    )
  + {\cal M}(\Xi_{cc}^{+},   \phi,       \Sigma_{c}^{+},   \Lambda,       \Lambda,     D_s^{+}    )
  + {\cal M}(\Xi_{cc}^{+},   \eta_1,     \Sigma_{c}^{+},   \Lambda,       \Lambda,     D_s^{+}    )  \nonumber\\
&&+ {\cal M}(\Xi_{cc}^{+},   \eta_8,     \Sigma_{c}^{+},   \Lambda,       \Lambda,     D_s^{+}    )
  + {\cal M}(\Xi_{cc}^{+},   \pi^+,      \Sigma_{c}^{0},   \Sigma^{-},    \Lambda,     D_s^{+}    )
  + {\cal M}(\Xi_{cc}^{+},   \rho^+,     \Sigma_{c}^{0},   \Sigma^{-},    \Lambda,     D_s^{+}    )  \nonumber\\
&&+ {\cal M}(\Xi_{cc}^{+},   \rho^0,     \Lambda_{c}^{+},  \Sigma^{0},    \Lambda,     D_s^{+}    )
  + {\cal M}(\Xi_{cc}^{+},   \rho^0,     \Sigma_{c}^{+},   \Sigma^{0},    \Lambda,     D_s^{+}    )
  + {\cal M}(\Xi_{cc}^{+},   \omega,     \Lambda_{c}^{+},  \Lambda,       \Lambda,     D_s^{+}    )  \nonumber\\
&&+ {\cal M}(\Xi_{cc}^{+},   \omega,     \Sigma_{c}^{+},   \Lambda,       \Lambda,     D_s^{+}    )
  + {\cal M}(\Xi_{cc}^{+},   \pi^0,      \Lambda_{c}^{+},  \Sigma^{0},    \Lambda,     D_s^{+}    )
  + {\cal M}(\Xi_{cc}^{+},   \pi^0,      \Sigma_{c}^{+},   \Sigma^{0},    \Lambda,     D_s^{+}    )  \nonumber\\
&&+ {\cal M}(\Xi_{cc}^{+},   \eta_1,     \Lambda_{c}^{+},  \Lambda,       \Lambda,     D_s^{+}    )
  + {\cal M}(\Xi_{cc}^{+},   \eta_8,     \Lambda_{c}^{+},  \Lambda,       \Lambda,     D_s^{+}    )
  + {\cal M}(\Xi_{cc}^{+},   \eta_1,     \Sigma_{c}^{+},   \Lambda,       \Lambda,     D_s^{+}    )  \nonumber\\
&&+ {\cal M}(\Xi_{cc}^{+},   \eta_8,     \Sigma_{c}^{+},   \Lambda,       \Lambda,     D_s^{+}    )
  + {\cal M}(\Xi_{cc}^{+},   K^{+},      \Xi_c^{\prime 0}, D^{*0},         D_s^{+},      \Lambda  )
  + {\cal M}(\Xi_{cc}^{+},   K^{*+},     \Xi_c^{\prime 0}, D^{0},          D_s^{+},      \Lambda  )  \nonumber\\
&&+ {\cal M}(\Xi_{cc}^{+},   K^{+},      \Xi_c^{\prime 0},\Xi^{-},        \Lambda,     D_s^{+}    )
  + {\cal M}(\Xi_{cc}^{+},   K^{*+},     \Xi_c^{\prime 0},\Xi^{-},        \Lambda,     D_s^{+}    )
  ],
\end{eqnarray}
\begin{eqnarray}
{\cal A}(\Xi_{cc}^{+}\rightarrow\Sigma^{0}  D_{s}^{+})
&=&i [
    {\cal M}(\Xi_{cc}^{+},   K^{+},      \Xi_c^{ 0},       D^{*0},         D_s^{+},      \Sigma^{0}  )
  + {\cal M}(\Xi_{cc}^{+},   K^{*+},     \Xi_c^{ 0},       D^{0},          D_s^{+},      \Sigma^{0}  )
  + {\cal M}(\Xi_{cc}^{+},   K^{+},      \Xi_c^{ 0},      \Xi^{-},        \Sigma^{0},     D_s^{+}    )  \nonumber\\
&&+ {\cal M}(\Xi_{cc}^{+},   K^{*+},     \Xi_c^{ 0},      \Xi^{-},        \Sigma^{0},     D_s^{+}    )
  + {\cal M}(\Xi_{cc}^{+},   \phi,       \Lambda_{c}^{+},  D_s^{+},        D_s^{+},      \Sigma^{0}  )
  + {\cal M}(\Xi_{cc}^{+},   \phi,       \Sigma_{c}^{+},   D_s^{+},        D_s^{+},      \Sigma^{0}  )  \nonumber\\
&&+ {\cal M}(\Xi_{cc}^{+},   \eta_1,     \Lambda_{c}^{+},  D_s^{*+},       D_s^{+},      \Sigma^{0}  )
  + {\cal M}(\Xi_{cc}^{+},   \eta_1,     \Sigma_{c}^{+},   D_s^{*+},       D_s^{+},      \Sigma^{0}  )
  + {\cal M}(\Xi_{cc}^{+},   \eta_8,     \Lambda_{c}^{+},  D_s^{*+},       D_s^{+},      \Sigma^{0}  )  \nonumber\\
&&+ {\cal M}(\Xi_{cc}^{+},   \eta_8,     \Sigma_{c}^{+},   D_s^{*+},       D_s^{+},      \Sigma^{0}  )
  + {\cal M}(\Xi_{cc}^{+},   \phi,       \Lambda_{c}^{+},  \Sigma^{0},    \Sigma^{0},     D_s^{+}    )
  + {\cal M}(\Xi_{cc}^{+},   \eta_1,     \Lambda_{c}^{+},  \Sigma^{0},    \Sigma^{0},     D_s^{+}    )  \nonumber\\
&&+ {\cal M}(\Xi_{cc}^{+},   \eta_8,     \Lambda_{c}^{+},  \Sigma^{0},    \Sigma^{0},     D_s^{+}    )
  + {\cal M}(\Xi_{cc}^{+},   \phi,       \Sigma_{c}^{+},   \Sigma^{0},    \Sigma^{0},     D_s^{+}    )
  + {\cal M}(\Xi_{cc}^{+},   \eta_1,     \Sigma_{c}^{+},   \Sigma^{0},    \Sigma^{0},     D_s^{+}    )  \nonumber\\
&&+ {\cal M}(\Xi_{cc}^{+},   \eta_8,     \Sigma_{c}^{+},   \Sigma^{0},    \Sigma^{0},     D_s^{+}    )
  + {\cal M}(\Xi_{cc}^{+},   \pi^+,      \Sigma_{c}^{0},   \Sigma^{-},    \Sigma^{0},     D_s^{+}    )
  + {\cal M}(\Xi_{cc}^{+},   \rho^+,     \Sigma_{c}^{0},   \Sigma^{-},    \Sigma^{0},     D_s^{+}    )  \nonumber\\
&&+ {\cal M}(\Xi_{cc}^{+},   \rho^0,     \Lambda_{c}^{+},  \Lambda,       \Sigma^{0},     D_s^{+}    )
  + {\cal M}(\Xi_{cc}^{+},   \rho^0,     \Sigma_{c}^{+},   \Lambda,       \Sigma^{0},     D_s^{+}    )
  + {\cal M}(\Xi_{cc}^{+},   \omega,     \Lambda_{c}^{+},  \Sigma^{0},    \Sigma^{0},     D_s^{+}    )  \nonumber\\
&&+ {\cal M}(\Xi_{cc}^{+},   \omega,     \Sigma_{c}^{+},   \Sigma^{0},    \Sigma^{0},     D_s^{+}    )
  + {\cal M}(\Xi_{cc}^{+},   \pi^0,      \Lambda_{c}^{+},  \Lambda,       \Sigma^{0},     D_s^{+}    )
  + {\cal M}(\Xi_{cc}^{+},   \pi^0,      \Sigma_{c}^{+},   \Lambda,       \Sigma^{0},     D_s^{+}    )  \nonumber\\
&&+ {\cal M}(\Xi_{cc}^{+},   \eta_1,     \Lambda_{c}^{+},  \Sigma^{0},    \Sigma^{0},     D_s^{+}    )
  + {\cal M}(\Xi_{cc}^{+},   \eta_8,     \Lambda_{c}^{+},  \Sigma^{0},    \Sigma^{0},     D_s^{+}    )
  + {\cal M}(\Xi_{cc}^{+},   \eta_1,     \Sigma_{c}^{+},   \Sigma^{0},    \Sigma^{0},     D_s^{+}    )  \nonumber\\
&&+ {\cal M}(\Xi_{cc}^{+},   \eta_8,     \Sigma_{c}^{+},   \Sigma^{0},    \Sigma^{0},     D_s^{+}    )
  + {\cal M}(\Xi_{cc}^{+},   K^{+},      \Xi_c^{\prime 0}, D^{*0},         D_s^{+},      \Sigma^{0}  )
  + {\cal M}(\Xi_{cc}^{+},   K^{*+},     \Xi_c^{\prime 0}, D^{0},          D_s^{+},      \Sigma^{0}  )  \nonumber\\
&&+ {\cal M}(\Xi_{cc}^{+},   K^{+},      \Xi_c^{\prime 0},\Xi^{-},        \Sigma^{0},     D_s^{+}    )
  + {\cal M}(\Xi_{cc}^{+},   K^{*+},     \Xi_c^{\prime 0},\Xi^{-},        \Sigma^{0},     D_s^{+}    ) ],
\end{eqnarray}
\begin{eqnarray}
{\cal A}(\Xi_{cc}^{+}\rightarrow\Sigma^{0} D_{s}^{*+})
&=&i [
    {\cal M}(\Xi_{cc}^{+},   K^{+},      \Xi_c^{ 0},       D^{0},         D_s^{*+},      \Sigma^{0}  )
  + {\cal M}(\Xi_{cc}^{+},   K^{*+},     \Xi_c^{ 0},       D^{*0},        D_s^{*+},      \Sigma^{0}   )
  + {\cal M}(\Xi_{cc}^{+},   K^{+},      \Xi_c^{ 0},      \Xi^{-},        \Sigma^{0},     D_s^{*+}    )  \nonumber\\
&&+ {\cal M}(\Xi_{cc}^{+},   K^{*+},     \Xi_c^{ 0},      \Xi^{-},        \Sigma^{0},     D_s^{*+}    )
  + {\cal M}(\Xi_{cc}^{+},   \phi,       \Lambda_{c}^{+},  D_s^{*+},      D_s^{*+},      \Sigma^{0}   )
  + {\cal M}(\Xi_{cc}^{+},   \phi,       \Sigma_{c}^{+},   D_s^{*+},      D_s^{*+},      \Sigma^{0}   )  \nonumber\\
&&+ {\cal M}(\Xi_{cc}^{+},   \eta_1,     \Lambda_{c}^{+},  D_s^{+},       D_s^{*+},      \Sigma^{0}   )
  + {\cal M}(\Xi_{cc}^{+},   \eta_1,     \Sigma_{c}^{+},   D_s^{+},       D_s^{*+},      \Sigma^{0}   )
  + {\cal M}(\Xi_{cc}^{+},   \eta_8,     \Lambda_{c}^{+},  D_s^{+},       D_s^{*+},      \Sigma^{0}   )  \nonumber\\
&&+ {\cal M}(\Xi_{cc}^{+},   \eta_8,     \Sigma_{c}^{+},   D_s^{+},       D_s^{*+},      \Sigma^{0}   )
  + {\cal M}(\Xi_{cc}^{+},   \phi,       \Lambda_{c}^{+},  \Sigma^{0},    \Sigma^{0},     D_s^{*+}    )
  + {\cal M}(\Xi_{cc}^{+},   \eta_1,     \Lambda_{c}^{+},  \Sigma^{0},    \Sigma^{0},     D_s^{*+}    )  \nonumber\\
&&+ {\cal M}(\Xi_{cc}^{+},   \eta_8,     \Lambda_{c}^{+},  \Sigma^{0},    \Sigma^{0},     D_s^{*+}    )
  + {\cal M}(\Xi_{cc}^{+},   \phi,       \Sigma_{c}^{+},   \Sigma^{0},    \Sigma^{0},     D_s^{*+}    )
  + {\cal M}(\Xi_{cc}^{+},   \eta_1,     \Sigma_{c}^{+},   \Sigma^{0},    \Sigma^{0},     D_s^{*+}    )  \nonumber\\
&&+ {\cal M}(\Xi_{cc}^{+},   \eta_8,     \Sigma_{c}^{+},   \Sigma^{0},    \Sigma^{0},     D_s^{*+}    )
  + {\cal M}(\Xi_{cc}^{+},   \pi^+,      \Sigma_{c}^{0},   \Sigma^{-},    \Sigma^{0},     D_s^{*+}    )
  + {\cal M}(\Xi_{cc}^{+},   \rho^+,     \Sigma_{c}^{0},   \Sigma^{-},    \Sigma^{0},     D_s^{*+}    )  \nonumber\\
&&+ {\cal M}(\Xi_{cc}^{+},   \rho^0,     \Lambda_{c}^{+},  \Lambda,       \Sigma^{0},     D_s^{*+}    )
  + {\cal M}(\Xi_{cc}^{+},   \rho^0,     \Sigma_{c}^{+},   \Lambda,       \Sigma^{0},     D_s^{*+}    )
  + {\cal M}(\Xi_{cc}^{+},   \omega,     \Lambda_{c}^{+},  \Sigma^{0},    \Sigma^{0},     D_s^{*+}    )  \nonumber\\
&&+ {\cal M}(\Xi_{cc}^{+},   \omega,     \Sigma_{c}^{+},   \Sigma^{0},    \Sigma^{0},     D_s^{*+}    )
  + {\cal M}(\Xi_{cc}^{+},   \pi^0,      \Lambda_{c}^{+},  \Lambda,       \Sigma^{0},     D_s^{*+}    )
  + {\cal M}(\Xi_{cc}^{+},   \pi^0,      \Sigma_{c}^{+},   \Lambda,       \Sigma^{0},     D_s^{*+}    )  \nonumber\\
&&+ {\cal M}(\Xi_{cc}^{+},   \eta_1,     \Lambda_{c}^{+},  \Sigma^{0},    \Sigma^{0},     D_s^{*+}    )
  + {\cal M}(\Xi_{cc}^{+},   \eta_8,     \Lambda_{c}^{+},  \Sigma^{0},    \Sigma^{0},     D_s^{*+}    )
  + {\cal M}(\Xi_{cc}^{+},   \eta_1,     \Sigma_{c}^{+},   \Sigma^{0},    \Sigma^{0},     D_s^{*+}    )  \nonumber\\
&&+ {\cal M}(\Xi_{cc}^{+},   \eta_8,     \Sigma_{c}^{+},   \Sigma^{0},    \Sigma^{0},     D_s^{*+}    )
  + {\cal M}(\Xi_{cc}^{+},   K^{+},      \Xi_c^{\prime 0}, D^{0},         D_s^{*+},      \Sigma^{0}   )
  + {\cal M}(\Xi_{cc}^{+},   K^{*+},     \Xi_c^{\prime 0}, D^{*0},        D_s^{*+},      \Sigma^{0}   )  \nonumber\\
&&+ {\cal M}(\Xi_{cc}^{+},   K^{+},      \Xi_c^{\prime 0}, \Xi^{-},       \Sigma^{0},     D_s^{*+}    )
  + {\cal M}(\Xi_{cc}^{+},   K^{*+},     \Xi_c^{\prime 0}, \Xi^{-},       \Sigma^{0},     D_s^{*+}    )
 ],
\end{eqnarray}
\begin{eqnarray}
{\cal A}(\Xi_{cc}^{+}\rightarrow\Lambda  D_{s}^{*+})
&=&i [
    {\cal M}(\Xi_{cc}^{+},   K^{+},      \Xi_c^{ 0},       D^{0},         D_s^{*+},   \Lambda      )
  + {\cal M}(\Xi_{cc}^{+},   K^{*+},     \Xi_c^{ 0},       D^{*0},        D_s^{*+},   \Lambda      )
  + {\cal M}(\Xi_{cc}^{+},   K^{+},      \Xi_c^{ 0},       \Xi^{-},       \Lambda,     D_s^{*+}    )  \nonumber\\
&&+ {\cal M}(\Xi_{cc}^{+},   K^{*+},     \Xi_c^{ 0},       \Xi^{-},       \Lambda,     D_s^{*+}    )
  + {\cal M}(\Xi_{cc}^{+},   \phi,       \Lambda_{c}^{+},  D_s^{*+},      D_s^{*+},    \Lambda     )
  + {\cal M}(\Xi_{cc}^{+},   \phi,       \Sigma_{c}^{+},   D_s^{*+},      D_s^{*+},    \Lambda     )  \nonumber\\
&&+ {\cal M}(\Xi_{cc}^{+},   \eta_1,     \Lambda_{c}^{+},  D_s^{+},       D_s^{*+},    \Lambda     )
  + {\cal M}(\Xi_{cc}^{+},   \eta_1,     \Sigma_{c}^{+},   D_s^{+},       D_s^{*+},    \Lambda     )
  + {\cal M}(\Xi_{cc}^{+},   \eta_8,     \Lambda_{c}^{+},  D_s^{+},       D_s^{*+},    \Lambda     )  \nonumber\\
&&+ {\cal M}(\Xi_{cc}^{+},   \eta_8,     \Sigma_{c}^{+},   D_s^{+},       D_s^{*+},    \Lambda     )
  + {\cal M}(\Xi_{cc}^{+},   \phi,       \Lambda_{c}^{+},  \Lambda,      \Lambda,      D_s^{*+}    )
  + {\cal M}(\Xi_{cc}^{+},   \eta_1,     \Lambda_{c}^{+},  \Lambda,      \Lambda,      D_s^{*+}    )  \nonumber\\
&&+ {\cal M}(\Xi_{cc}^{+},   \eta_8,     \Lambda_{c}^{+},  \Lambda,      \Lambda,      D_s^{*+}    )
  + {\cal M}(\Xi_{cc}^{+},   \phi,       \Sigma_{c}^{+},   \Lambda,      \Lambda,      D_s^{*+}    )
  + {\cal M}(\Xi_{cc}^{+},   \eta_1,     \Sigma_{c}^{+},   \Lambda,      \Lambda,      D_s^{*+}    )  \nonumber\\
&&+ {\cal M}(\Xi_{cc}^{+},   \eta_8,     \Sigma_{c}^{+},   \Lambda,      \Lambda,      D_s^{*+}    )
  + {\cal M}(\Xi_{cc}^{+},   \pi^+,      \Sigma_{c}^{0},   \Sigma^{-},   \Lambda,      D_s^{*+}    )
  + {\cal M}(\Xi_{cc}^{+},   \rho^+,     \Sigma_{c}^{0},   \Sigma^{-},   \Lambda,      D_s^{*+}    )  \nonumber\\
&&+ {\cal M}(\Xi_{cc}^{+},   \rho^0,     \Lambda_{c}^{+},  \Sigma^{0},   \Lambda,      D_s^{*+}    )
  + {\cal M}(\Xi_{cc}^{+},   \rho^0,     \Sigma_{c}^{+},   \Sigma^{0},   \Lambda,      D_s^{*+}    )
  + {\cal M}(\Xi_{cc}^{+},   \omega,     \Lambda_{c}^{+},  \Lambda,      \Lambda,      D_s^{*+}    )  \nonumber\\
&&+ {\cal M}(\Xi_{cc}^{+},   \omega,     \Sigma_{c}^{+},   \Lambda,      \Lambda,      D_s^{*+}    )
  + {\cal M}(\Xi_{cc}^{+},   \pi^0,      \Lambda_{c}^{+},  \Sigma^{0},   \Lambda,      D_s^{*+}    )
  + {\cal M}(\Xi_{cc}^{+},   \pi^0,      \Sigma_{c}^{+},   \Sigma^{0},   \Lambda,      D_s^{*+}    )  \nonumber\\
&&+ {\cal M}(\Xi_{cc}^{+},   \eta_1,     \Lambda_{c}^{+},  \Lambda,      \Lambda,      D_s^{*+}    )
  + {\cal M}(\Xi_{cc}^{+},   \eta_8,     \Lambda_{c}^{+},  \Lambda,      \Lambda,      D_s^{*+}    )
  + {\cal M}(\Xi_{cc}^{+},   \eta_1,     \Sigma_{c}^{+},   \Lambda,      \Lambda,      D_s^{*+}    )  \nonumber\\
&&+ {\cal M}(\Xi_{cc}^{+},   \eta_8,     \Sigma_{c}^{+},   \Lambda,      \Lambda,      D_s^{*+}    )
  + {\cal M}(\Xi_{cc}^{+},   K^{+},      \Xi_c^{\prime 0}, D^{0},         D_s^{*+},   \Lambda      )
  + {\cal M}(\Xi_{cc}^{+},   K^{*+},     \Xi_c^{\prime 0}, D^{*0},        D_s^{*+},   \Lambda      )  \nonumber\\
&&+ {\cal M}(\Xi_{cc}^{+},   K^{+},      \Xi_c^{\prime 0}, \Xi^{-},       \Lambda,     D_s^{*+}    )
  + {\cal M}(\Xi_{cc}^{+},   K^{*+},     \Xi_c^{\prime 0}, \Xi^{-},       \Lambda,     D_s^{*+}    )
 ],
\end{eqnarray}

\begin{eqnarray}
{\cal A}(\Xi_{cc}^{+}\rightarrow n D_{s}^{+})
&=&i [
    {\cal M}(\Xi_{cc}^{+},   K^{+},      \Sigma_c^{0},        D^{*0},        D_s^{+},      n           )
  + {\cal M}(\Xi_{cc}^{+},   K^{*+},     \Sigma_c^{0},        D^{0},         D_s^{+},      n           )
  + {\cal M}(\Xi_{cc}^{+},   K^{+},      \Sigma_c^{0},       \Sigma^{-},     n,            D_s^{+}     )  \nonumber\\
&&+ {\cal M}(\Xi_{cc}^{+},   K^{*+},     \Sigma_c^{0},       \Sigma^{-},     n,            D_s^{+}     )
  + {\cal M}(\Xi_{cc}^{+},   K^{0},      \Lambda_{c}^{+},     D^{*+},        D_s^{+},      n           )
  + {\cal M}(\Xi_{cc}^{+},   K^{0},      \Sigma_c^{+},        D^{*+},        D_s^{+},      n           )  \nonumber\\
&&+ {\cal M}(\Xi_{cc}^{+},   K^{*0},     \Lambda_{c}^{+},     D^{+},         D_s^{+},      n           )
  + {\cal M}(\Xi_{cc}^{+},   K^{*0},     \Sigma_c^{+},        D^{+},         D_s^{+},      n           )
  + {\cal M}(\Xi_{cc}^{+},   K^{0},      \Lambda_{c}^{+},    \Sigma^{0},     n,            D_s^{+}     )  \nonumber\\
&&+ {\cal M}(\Xi_{cc}^{+},   K^{0},      \Lambda_{c}^{+},    \Lambda,        n,            D_s^{+}     )
  + {\cal M}(\Xi_{cc}^{+},   K^{0},      \Sigma_c^{+},       \Sigma^{0},     n,            D_s^{+}     )
  + {\cal M}(\Xi_{cc}^{+},   K^{0},      \Sigma_c^{+},       \Lambda,        n,            D_s^{+}     )  \nonumber\\
&&+ {\cal M}(\Xi_{cc}^{+},   K^{*0},     \Lambda_{c}^{+},    \Sigma^{0},     n,            D_s^{+}     )
  + {\cal M}(\Xi_{cc}^{+},   K^{*0},     \Lambda_{c}^{+},    \Lambda,        n,            D_s^{+}     )
  + {\cal M}(\Xi_{cc}^{+},   K^{*0},     \Sigma_c^{+},       \Sigma^{0},     n,            D_s^{+}     )  \nonumber\\
&&+ {\cal M}(\Xi_{cc}^{+},   K^{*0},     \Sigma_c^{+},       \Lambda,        n,            D_s^{+}     )
  ],
\end{eqnarray}
\begin{eqnarray}
{\cal A}(\Xi_{cc}^{+}\rightarrow n D_{s}^{*+})
&=&i [
    {\cal M}(\Xi_{cc}^{+},   K^{+},      \Sigma_c^{0},        D^{0},         D_s^{*+},     n           )
  + {\cal M}(\Xi_{cc}^{+},   K^{*+},     \Sigma_c^{0},        D^{*0},        D_s^{*+},     n           )
  + {\cal M}(\Xi_{cc}^{+},   K^{+},      \Sigma_c^{0},       \Sigma^{-},     n,            D_s^{*+}    )  \nonumber\\
&&+ {\cal M}(\Xi_{cc}^{+},   K^{*+},     \Sigma_c^{0},       \Sigma^{-},     n,            D_s^{*+}    )
  + {\cal M}(\Xi_{cc}^{+},   K^{0},      \Lambda_{c}^{+},     D^{+},         D_s^{*+},     n           )
  + {\cal M}(\Xi_{cc}^{+},   K^{0},      \Sigma_c^{+},        D^{+},         D_s^{*+},     n           )  \nonumber\\
&&+ {\cal M}(\Xi_{cc}^{+},   K^{*0},     \Lambda_{c}^{+},     D^{*+},        D_s^{*+},     n           )
  + {\cal M}(\Xi_{cc}^{+},   K^{*0},     \Sigma_c^{+},        D^{*+},        D_s^{*+},     n           )
  + {\cal M}(\Xi_{cc}^{+},   K^{0},      \Lambda_{c}^{+},    \Sigma^{0},     n,            D_s^{*+}    )  \nonumber\\
&&+ {\cal M}(\Xi_{cc}^{+},   K^{0},      \Lambda_{c}^{+},    \Lambda,        n,            D_s^{*+}    )
  + {\cal M}(\Xi_{cc}^{+},   K^{0},      \Sigma_c^{+},       \Sigma^{0},     n,            D_s^{*+}    )
  + {\cal M}(\Xi_{cc}^{+},   K^{0},      \Sigma_c^{+},       \Lambda,        n,            D_s^{*+}    )  \nonumber\\
&&+ {\cal M}(\Xi_{cc}^{+},   K^{*0},     \Lambda_{c}^{+},    \Sigma^{0},     n,            D_s^{*+}    )
  + {\cal M}(\Xi_{cc}^{+},   K^{*0},     \Lambda_{c}^{+},    \Lambda,        n,            D_s^{*+}    )
  + {\cal M}(\Xi_{cc}^{+},   K^{*0},     \Sigma_c^{+},       \Sigma^{0},     n,            D_s^{*+}    )  \nonumber\\
&&+ {\cal M}(\Xi_{cc}^{+},   K^{*0},     \Sigma_c^{+},       \Lambda,        n,            D_s^{*+}    )
  ],
\end{eqnarray}

\begin{eqnarray}
{\cal A}(\Xi_{cc}^{+}\rightarrow n D^{+})
&=&i [
    {\cal M}(\Xi_{cc}^{+},   \pi^{+},       \Sigma_c^{ 0},       D^{*0},         D^{+},        n      )
  + {\cal M}(\Xi_{cc}^{+},   \rho^{+},      \Sigma_c^{ 0},       D^{0},          D^{+},        n      )
  + {\cal M}(\Xi_{cc}^{+},   \rho^{0},      \Lambda_{c}^{+},     D^{+},          D^{+},        n      )  \nonumber\\
&&+ {\cal M}(\Xi_{cc}^{+},   \rho^{0},      \Sigma_c^{ +},       D^{+},          D^{+},        n      )
  + {\cal M}(\Xi_{cc}^{+},   \omega,        \Lambda_{c}^{+},     D^{+},          D^{+},        n      )
  + {\cal M}(\Xi_{cc}^{+},   \omega,        \Sigma_c^{ +},       D^{+},          D^{+},        n      )  \nonumber\\
&&+ {\cal M}(\Xi_{cc}^{+},   \pi^{0},       \Lambda_{c}^{+},     D^{*+},         D^{+},        n      )
  + {\cal M}(\Xi_{cc}^{+},   \pi^{0},       \Sigma_c^{ +},       D^{*+},         D^{+},        n      )
  + {\cal M}(\Xi_{cc}^{+},   \eta_1,        \Lambda_{c}^{+},     D^{*+},         D^{+},        n      )  \nonumber\\
&&+ {\cal M}(\Xi_{cc}^{+},   \eta_8,        \Lambda_{c}^{+},     D^{*+},         D^{+},        n      )
  + {\cal M}(\Xi_{cc}^{+},   \eta_1,        \Sigma_c^{ +},       D^{*+},         D^{+},        n      )
  + {\cal M}(\Xi_{cc}^{+},   \eta_8,        \Sigma_c^{ +},       D^{*+},         D^{+},        n      )  \nonumber\\
&&+ {\cal M}(\Xi_{cc}^{+},   \rho^{0},      \Lambda_{c}^{+},     n,              n,            D^{+}  )
  + {\cal M}(\Xi_{cc}^{+},   \rho^{0},      \Sigma_c^{ +},       n,              n,            D^{+}  )
  + {\cal M}(\Xi_{cc}^{+},   \pi^{0},       \Lambda_{c}^{+},     n,              n,            D^{+}  )  \nonumber\\
&&+ {\cal M}(\Xi_{cc}^{+},   \pi^{0},       \Sigma_c^{ +},       n,              n,            D^{+}  )
  + {\cal M}(\Xi_{cc}^{+},   \eta_1,        \Lambda_{c}^{+},     n,              n,            D^{+}  )
  + {\cal M}(\Xi_{cc}^{+},   \eta_8,        \Lambda_{c}^{+},     n,              n,            D^{+}  )  \nonumber\\
&&+ {\cal M}(\Xi_{cc}^{+},   \eta_1,        \Sigma_c^{ +},       n,              n,            D^{+}  )
  + {\cal M}(\Xi_{cc}^{+},   \eta_8,        \Sigma_c^{ +},       n,              n,            D^{+}  )
  + {\cal M}(\Xi_{cc}^{+},   K^+,           \Xi_c^{ 0},         \Sigma^-,        n,            D^{+}  ) \nonumber\\
&&+ {\cal M}(\Xi_{cc}^{+},   K^{*+},        \Xi_c^{ 0},         \Sigma^-,        n,            D^{+}  )
  + {\cal M}(\Xi_{cc}^{+},   K^+,           \Xi_c^{\prime 0},   \Sigma^-,        n,            D^{+}  )
  + {\cal M}(\Xi_{cc}^{+},   K^{*+},        \Xi_c^{\prime 0},   \Sigma^-,        n,            D^{+}  )
],
\end{eqnarray}
\begin{eqnarray}
{\cal A}(\Xi_{cc}^{+}\rightarrow n D^{*+})
&=&i [
    {\cal M}(\Xi_{cc}^{+},   \pi^{+},       \Sigma_c^{ 0},       D^{0},         D^{*+},       n      )
  + {\cal M}(\Xi_{cc}^{+},   \rho^{+},      \Sigma_c^{ 0},       D^{*0},        D^{*+},       n      )
  + {\cal M}(\Xi_{cc}^{+},   \rho^{0},      \Lambda_{c}^{+},     D^{*+},        D^{*+},       n      )  \nonumber\\
&&+ {\cal M}(\Xi_{cc}^{+},   \rho^{0},      \Sigma_c^{ +},       D^{*+},        D^{*+},       n      )
  + {\cal M}(\Xi_{cc}^{+},   \omega,        \Lambda_{c}^{+},     D^{*+},        D^{*+},       n      )
  + {\cal M}(\Xi_{cc}^{+},   \omega,        \Sigma_c^{ +},       D^{*+},        D^{*+},       n      )  \nonumber\\
&&+ {\cal M}(\Xi_{cc}^{+},   \pi^{0},       \Lambda_{c}^{+},     D^{+},         D^{*+},       n      )
  + {\cal M}(\Xi_{cc}^{+},   \pi^{0},       \Sigma_c^{ +},       D^{+},         D^{*+},       n      )
  + {\cal M}(\Xi_{cc}^{+},   \eta_1,        \Lambda_{c}^{+},     D^{+},         D^{*+},       n      )  \nonumber\\
&&+ {\cal M}(\Xi_{cc}^{+},   \eta_8,        \Lambda_{c}^{+},     D^{+},         D^{*+},       n      )
  + {\cal M}(\Xi_{cc}^{+},   \eta_1,        \Sigma_c^{ +},       D^{+},         D^{*+},       n      )
  + {\cal M}(\Xi_{cc}^{+},   \eta_8,        \Sigma_c^{ +},       D^{+},         D^{*+},       n      )  \nonumber\\
&&+ {\cal M}(\Xi_{cc}^{+},   \rho^{0},      \Lambda_{c}^{+},     n,             n,            D^{*+}  )
  + {\cal M}(\Xi_{cc}^{+},   \rho^{0},      \Sigma_c^{ +},       n,             n,            D^{*+}  )
  + {\cal M}(\Xi_{cc}^{+},   \pi^{0},       \Lambda_{c}^{+},     n,             n,            D^{*+}  )  \nonumber\\
&&+ {\cal M}(\Xi_{cc}^{+},   \pi^{0},       \Sigma_c^{ +},       n,             n,            D^{*+}  )
  + {\cal M}(\Xi_{cc}^{+},   \eta_1,        \Lambda_{c}^{+},     n,             n,            D^{*+}  )
  + {\cal M}(\Xi_{cc}^{+},   \eta_8,        \Lambda_{c}^{+},     n,             n,            D^{*+}  )  \nonumber\\
&&+ {\cal M}(\Xi_{cc}^{+},   \eta_1,        \Sigma_c^{ +},       n,             n,            D^{*+}  )
  + {\cal M}(\Xi_{cc}^{+},   \eta_8,        \Sigma_c^{ +},       n,             n,            D^{*+}  )
  + {\cal M}(\Xi_{cc}^{+},   K^+,           \Xi_c^{ 0},         \Sigma^-,       n,            D^{*+}  )\nonumber\\
&&+ {\cal M}(\Xi_{cc}^{+},   K^+,           \Xi_c^{\prime 0},   \Sigma^-,       n,            D^{*+}  )
  + {\cal M}(\Xi_{cc}^{+},   K^{*+},        \Xi_c^{\prime 0},   \Sigma^-,       n,            D^{*+}  )
  + {\cal M}(\Xi_{cc}^{+},   K^{*+},        \Xi_c^{ 0},         \Sigma^-,       n,            D^{*+}  )
 ],
\end{eqnarray}

\begin{eqnarray}
{\cal A}(\Xi_{cc}^{+}\rightarrow p D^{0})
&=&i [
    {\cal M}(\Xi_{cc}^{+},   K^{+},      \Xi_c^{ 0},       \Sigma^{0},        p,             D^{0}   )
  + {\cal M}(\Xi_{cc}^{+},   K^{+},      \Xi_c^{ 0},       \Lambda,           p,             D^{0}   )
  + {\cal M}(\Xi_{cc}^{+},   K^{+},      \Xi_c^{\prime 0}, \Sigma^{0},        p,             D^{0}   )  \nonumber\\
&&+ {\cal M}(\Xi_{cc}^{+},   K^{+},      \Xi_c^{\prime 0}, \Lambda,           p,             D^{0}   )
  + {\cal M}(\Xi_{cc}^{+},   K^{*+},     \Xi_c^{ 0},       \Sigma^{0},        p,             D^{0}   )
  + {\cal M}(\Xi_{cc}^{+},   K^{*+},     \Xi_c^{ 0},       \Lambda,           p,             D^{0}   )  \nonumber\\
&&+ {\cal M}(\Xi_{cc}^{+},   K^{*+},     \Xi_c^{\prime 0}, \Sigma^{0},        p,             D^{0}   )
  + {\cal M}(\Xi_{cc}^{+},   K^{*+},     \Xi_c^{\prime 0}, \Lambda,           p,             D^{0}   )
  + {\cal M}(\Xi_{cc}^{+},  \pi^{+},     \Sigma_c^{ 0},     n,                p,             D^{0}   )  \nonumber\\
&&+ {\cal M}(\Xi_{cc}^{+},  \rho^+,      \Sigma_c^{ 0},     n,                p,             D^{0}   )
  + {\cal M}(\Xi_{cc}^{+},  \rho^0,      \Sigma_c^{ +},     p,                p,             D^{0}   )
  + {\cal M}(\Xi_{cc}^{+},  \pi^{0},     \Sigma_c^{ +},     p,                p,             D^{0}   )  \nonumber\\
&&+ {\cal M}(\Xi_{cc}^{+},  \eta_1,      \Sigma_c^{ +},     p,                p,             D^{0}   )
  + {\cal M}(\Xi_{cc}^{+},  \eta_8,      \Sigma_c^{ +},     p,                p,             D^{0}   )
  + {\cal M}(\Xi_{cc}^{+},  \rho^0,      \Lambda_c^{ +},    p,                p,             D^{0}   )  \nonumber\\
&&+ {\cal M}(\Xi_{cc}^{+},  \pi^{0},     \Lambda_c^{ +},    p,                p,             D^{0}   )
  + {\cal M}(\Xi_{cc}^{+},  \eta_1,      \Lambda_c^{ +},    p,                p,             D^{0}   )
  + {\cal M}(\Xi_{cc}^{+},  \eta_8,      \Lambda_c^{ +},    p,                p,             D^{0}   )
],
\end{eqnarray}
\begin{eqnarray}
{\cal A}(\Xi_{cc}^{+}\rightarrow p D^{*0})
&=&i [
    {\cal M}(\Xi_{cc}^{+},   K^{+},      \Xi_c^{ 0},       \Sigma^{0},        p,             D^{*0}   )
  + {\cal M}(\Xi_{cc}^{+},   K^{+},      \Xi_c^{ 0},       \Lambda,           p,             D^{*0}   )
  + {\cal M}(\Xi_{cc}^{+},   K^{+},      \Xi_c^{\prime 0}, \Sigma^{0},        p,             D^{*0}   )  \nonumber\\
&&+ {\cal M}(\Xi_{cc}^{+},   K^{+},      \Xi_c^{\prime 0}, \Lambda,           p,             D^{*0}   )
  + {\cal M}(\Xi_{cc}^{+},   K^{*+},     \Xi_c^{ 0},       \Sigma^{0},        p,             D^{*0}   )
  + {\cal M}(\Xi_{cc}^{+},   K^{*+},     \Xi_c^{ 0},       \Lambda,           p,             D^{*0}   )  \nonumber\\
&&+ {\cal M}(\Xi_{cc}^{+},   K^{*+},     \Xi_c^{\prime 0}, \Sigma^{0},        p,             D^{*0}   )
  + {\cal M}(\Xi_{cc}^{+},   K^{*+},     \Xi_c^{\prime 0}, \Lambda,           p,             D^{*0}   )
  + {\cal M}(\Xi_{cc}^{+},  \pi^{+},     \Sigma_c^{ 0},     n,                p,             D^{*0}   )  \nonumber\\
&&+ {\cal M}(\Xi_{cc}^{+},  \rho^+,      \Sigma_c^{ 0},     n,                p,             D^{*0}   )
  + {\cal M}(\Xi_{cc}^{+},  \rho^0,      \Sigma_c^{ +},     p,                p,             D^{*0}   )
  + {\cal M}(\Xi_{cc}^{+},  \pi^{0},     \Sigma_c^{ +},     p,                p,             D^{*0}   )  \nonumber\\
&&+ {\cal M}(\Xi_{cc}^{+},  \eta_1,      \Sigma_c^{ +},     p,                p,             D^{*0}   )
  + {\cal M}(\Xi_{cc}^{+},  \eta_8,      \Sigma_c^{ +},     p,                p,             D^{*0}   )
  + {\cal M}(\Xi_{cc}^{+},  \rho^0,      \Lambda_c^{ +},    p,                p,             D^{*0}   )  \nonumber\\
&&+ {\cal M}(\Xi_{cc}^{+},  \pi^{0},     \Lambda_c^{ +},    p,                p,             D^{*0}   )
  + {\cal M}(\Xi_{cc}^{+},  \eta_1,      \Lambda_c^{ +},    p,                p,             D^{*0}   )
  + {\cal M}(\Xi_{cc}^{+},  \eta_8,      \Lambda_c^{ +},    p,                p,             D^{*0}   )
 ],
\end{eqnarray}

\begin{eqnarray}
{\cal A}(\Xi_{cc}^{+}\rightarrow\Sigma^{+} D^{0})
&=&i [
    {\cal M}(\Xi_{cc}^{+},   \pi^{+},      \Xi_c^{ 0},       \Sigma^{0},        \Sigma^{ +},      D^{0}  )
  + {\cal M}(\Xi_{cc}^{+},   \pi^{+},      \Xi_c^{ 0},       \Lambda,           \Sigma^{ +},      D^{0}  )
  + {\cal M}(\Xi_{cc}^{+},   \pi^{+},      \Xi_c^{\prime 0}, \Sigma^{0},        \Sigma^{ +},      D^{0}  )  \nonumber\\
&&+ {\cal M}(\Xi_{cc}^{+},   \pi^{+},      \Xi_c^{\prime 0}, \Lambda,           \Sigma^{ +},      D^{0}  )
  + {\cal M}(\Xi_{cc}^{+},   \rho^+,       \Xi_c^{ 0},       \Sigma^{0},        \Sigma^{ +},      D^{0}  )
  + {\cal M}(\Xi_{cc}^{+},   \rho^+,       \Xi_c^{ 0},       \Lambda,           \Sigma^{ +},      D^{0}  )  \nonumber\\
&&+ {\cal M}(\Xi_{cc}^{+},   \rho^+,       \Xi_c^{\prime 0}, \Sigma^{0},        \Sigma^{ +},      D^{0}  )
  + {\cal M}(\Xi_{cc}^{+},   \rho^+,       \Xi_c^{\prime 0}, \Lambda,           \Sigma^{ +},      D^{0}  )
  + {\cal M}(\Xi_{cc}^{+},   \bar K^{0},   \Sigma_c^{ +},     p,                \Sigma^{ +},      D^{0}  )  \nonumber\\
&&+ {\cal M}(\Xi_{cc}^{+},   \bar K^{*0},  \Sigma_c^{ +},     p,                \Sigma^{ +},      D^{0}  )
  + {\cal M}(\Xi_{cc}^{+},   \bar K^{0},   \Lambda_c^{ +},    p,                \Sigma^{ +},      D^{0}  )
  + {\cal M}(\Xi_{cc}^{+},   \bar K^{*0},  \Lambda_c^{ +},    p,                \Sigma^{ +},      D^{0}  )
 ],
\end{eqnarray}
\begin{eqnarray}
{\cal A}(\Xi_{cc}^{+}\rightarrow\Sigma^{+} D^{*0})
&=&i [
    {\cal M}(\Xi_{cc}^{+},   \pi^{+},      \Xi_c^{ 0},       \Sigma^{0},        \Sigma^{ +},      D^{*0}  )
  + {\cal M}(\Xi_{cc}^{+},   \pi^{+},      \Xi_c^{ 0},       \Lambda,           \Sigma^{ +},      D^{*0}  )
  + {\cal M}(\Xi_{cc}^{+},   \pi^{+},      \Xi_c^{\prime 0}, \Sigma^{0},        \Sigma^{ +},      D^{*0}  )  \nonumber\\
&&+ {\cal M}(\Xi_{cc}^{+},   \pi^{+},      \Xi_c^{\prime 0}, \Lambda,           \Sigma^{ +},      D^{*0}  )
  + {\cal M}(\Xi_{cc}^{+},   \rho^+,       \Xi_c^{ 0},       \Sigma^{0},        \Sigma^{ +},      D^{*0}  )
  + {\cal M}(\Xi_{cc}^{+},   \rho^+,       \Xi_c^{ 0},       \Lambda,           \Sigma^{ +},      D^{*0}  )  \nonumber\\
&&+ {\cal M}(\Xi_{cc}^{+},   \rho^+,       \Xi_c^{\prime 0}, \Sigma^{0},        \Sigma^{ +},      D^{*0}  )
  + {\cal M}(\Xi_{cc}^{+},   \rho^+,       \Xi_c^{\prime 0}, \Lambda,           \Sigma^{ +},      D^{*0}  )
  + {\cal M}(\Xi_{cc}^{+},   \bar K^{0},   \Sigma_c^{ +},     p,                \Sigma^{ +},      D^{*0}  )  \nonumber\\
&&+ {\cal M}(\Xi_{cc}^{+},   \bar K^{*0},  \Sigma_c^{ +},     p,                \Sigma^{ +},      D^{*0}  )
  + {\cal M}(\Xi_{cc}^{+},   \bar K^{0},   \Lambda_c^{ +},    p,                \Sigma^{ +},      D^{*0}  )
  + {\cal M}(\Xi_{cc}^{+},   \bar K^{*0},  \Lambda_c^{ +},    p,                \Sigma^{ +},      D^{*0}  )
  ],
\end{eqnarray}

\begin{eqnarray}
{\cal A}(\Xi_{cc}^{+}\rightarrow\Xi^{0} D_{s}^{+})
&=&i [
    {\cal M}(\Xi_{cc}^{+},   \pi^{+},      \Xi_c^{ 0},       \Xi^{-},         \Xi^{0},      D_s^+  )
  + {\cal M}(\Xi_{cc}^{+},   \rho^+,       \Xi_c^{ 0},       \Xi^{-},         \Xi^{0},      D_s^+  )
  + {\cal M}(\Xi_{cc}^{+},   \bar K^{0},   \Sigma_c^{ +},    \Sigma^{0},      \Xi^{0},      D_s^+  )  \nonumber\\
&&+ {\cal M}(\Xi_{cc}^{+},   \bar K^{0},   \Sigma_c^{ +},    \Lambda,         \Xi^{0},      D_s^+  )
  + {\cal M}(\Xi_{cc}^{+},   \bar K^{*0},  \Sigma_c^{ +},    \Sigma^{0},      \Xi^{0},      D_s^+  )
  + {\cal M}(\Xi_{cc}^{+},   \bar K^{*0},  \Sigma_c^{ +},    \Lambda,         \Xi^{0},      D_s^+  )  \nonumber\\
&&+ {\cal M}(\Xi_{cc}^{+},   \pi^{+},      \Xi_c^{\prime 0}, \Xi^{-},         \Xi^{0},      D_s^+  )
  + {\cal M}(\Xi_{cc}^{+},   \rho^+,       \Xi_c^{\prime 0}, \Xi^{-},         \Xi^{0},      D_s^+  )
  + {\cal M}(\Xi_{cc}^{+},   \bar K^{0},   \Lambda_c^{ +},   \Sigma^{0},      \Xi^{0},      D_s^+  )  \nonumber\\
&&+ {\cal M}(\Xi_{cc}^{+},   \bar K^{0},   \Lambda_c^{ +},   \Lambda,         \Xi^{0},      D_s^+  )
  + {\cal M}(\Xi_{cc}^{+},   \bar K^{*0},  \Lambda_c^{ +},   \Sigma^{0},      \Xi^{0},      D_s^+  )
  + {\cal M}(\Xi_{cc}^{+},   \bar K^{*0},  \Lambda_c^{ +},   \Lambda,         \Xi^{0},      D_s^+  )
],
\end{eqnarray}
\begin{eqnarray}
{\cal A}(\Xi_{cc}^{+}\rightarrow\Xi^{0} D_{s}^{*+})
&=&i [
    {\cal M}(\Xi_{cc}^{+},   \pi^{+},      \Xi_c^{ 0},       \Xi^{-},         \Xi^{0},      D_s^{*+}  )
  + {\cal M}(\Xi_{cc}^{+},   \rho^+,       \Xi_c^{ 0},       \Xi^{-},         \Xi^{0},      D_s^{*+}  )
  + {\cal M}(\Xi_{cc}^{+},   \bar K^{0},   \Sigma_c^{ +},    \Sigma^{0},      \Xi^{0},      D_s^{*+}  )  \nonumber\\
&&+ {\cal M}(\Xi_{cc}^{+},   \bar K^{0},   \Sigma_c^{ +},    \Lambda,         \Xi^{0},      D_s^{*+}  )
  + {\cal M}(\Xi_{cc}^{+},   \bar K^{*0},  \Sigma_c^{ +},    \Sigma^{0},      \Xi^{0},      D_s^{*+}  )
  + {\cal M}(\Xi_{cc}^{+},   \bar K^{*0},  \Sigma_c^{ +},    \Lambda,         \Xi^{0},      D_s^{*+}  )  \nonumber\\
&&+ {\cal M}(\Xi_{cc}^{+},   \pi^{+},      \Xi_c^{\prime 0}, \Xi^{-},         \Xi^{0},      D_s^{*+}  )
  + {\cal M}(\Xi_{cc}^{+},   \rho^+,       \Xi_c^{\prime 0}, \Xi^{-},         \Xi^{0},      D_s^{*+}  )
  + {\cal M}(\Xi_{cc}^{+},   \bar K^{0},   \Lambda_c^{ +},   \Sigma^{0},      \Xi^{0},      D_s^{*+}  )  \nonumber\\
&&+ {\cal M}(\Xi_{cc}^{+},   \bar K^{0},   \Lambda_c^{ +},   \Lambda,         \Xi^{0},      D_s^{*+}  )
  + {\cal M}(\Xi_{cc}^{+},   \bar K^{*0},  \Lambda_c^{ +},   \Sigma^{0},      \Xi^{0},      D_s^{*+}  )
  + {\cal M}(\Xi_{cc}^{+},   \bar K^{*0},  \Lambda_c^{ +},   \Lambda,         \Xi^{0},      D_s^{*+}  )],
\end{eqnarray}

\begin{eqnarray}
{\cal A}(\Omega_{cc}^{+}\rightarrow\Xi^{0} D^{+})
&=&i [
    {\cal M}(\Omega_{cc}^{+},   \pi^{+},    \Omega_{c}^{0},    D^{*0},       D^{+},     \Xi^{0}   )
  + {\cal M}(\Omega_{cc}^{+},   \rho^+,     \Omega_{c}^{0},    D^{0},        D^{+},     \Xi^{0}   )
  + {\cal M}(\Omega_{cc}^{+},   \pi^{+},    \Omega_{c}^{0},   \Xi^-,        \Xi^{0},     D^{+}    )  \nonumber\\
&&+ {\cal M}(\Omega_{cc}^{+},   \rho^+,     \Omega_{c}^{0},   \Xi^-,        \Xi^{0},     D^{+}    )
  + {\cal M}(\Omega_{cc}^{+},   \bar K^{0}, \Xi_c^{ +},       \Sigma^{ 0},  \Xi^{0},     D^{+}    )
  + {\cal M}(\Omega_{cc}^{+},   \bar K^{0}, \Xi_c^{ +},       \Lambda,      \Xi^{0},     D^{+}    )  \nonumber\\
&&+ {\cal M}(\Omega_{cc}^{+},   \bar K^{0}, \Xi_c^{\prime +}, \Sigma^{ 0},  \Xi^{0},     D^{+}    )
  + {\cal M}(\Omega_{cc}^{+},   \bar K^{0}, \Xi_c^{\prime +}, \Lambda,      \Xi^{0},     D^{+}    )
  + {\cal M}(\Omega_{cc}^{+},   \bar K^{*0},\Xi_c^{ +},       \Sigma^{ 0},  \Xi^{0},     D^{+}    )  \nonumber\\
&&+ {\cal M}(\Omega_{cc}^{+},   \bar K^{*0},\Xi_c^{ +},       \Lambda,      \Xi^{0},     D^{+}    )
  + {\cal M}(\Omega_{cc}^{+},   \bar K^{*0},\Xi_c^{\prime +}, \Sigma^{ 0},  \Xi^{0},     D^{+}    )
  + {\cal M}(\Omega_{cc}^{+},   \bar K^{*0},\Xi_c^{\prime +}, \Lambda,      \Xi^{0},     D^{+}    )  \nonumber\\
&&+ {\cal M}(\Omega_{cc}^{+},   \bar K^{0}, \Xi_c^{ +},        D_s^{*+},     D^{+},     \Xi^{0}   )
  + {\cal M}(\Omega_{cc}^{+},   \bar K^{0}, \Xi_c^{\prime +},  D_s^{*+},     D^{+},     \Xi^{0}   )
  + {\cal M}(\Omega_{cc}^{+},   \bar K^{*0},\Xi_c^{ +},        D_s^{+},      D^{+},     \Xi^{0}   )  \nonumber\\
&&+ {\cal M}(\Omega_{cc}^{+},   \bar K^{*0},\Xi_c^{\prime +},  D_s^{+},      D^{+},     \Xi^{0}   )
],
\end{eqnarray}
\begin{eqnarray}
{\cal A}(\Omega_{cc}^{+}\rightarrow\Xi^{0} D^{*+})
&=&i [
    {\cal M}(\Omega_{cc}^{+},   \pi^{+},    \Omega_{c}^{0},    D^{0},       D^{*+},     \Xi^{0}    )
  + {\cal M}(\Omega_{cc}^{+},   \rho^+,     \Omega_{c}^{0},    D^{*0},      D^{*+},     \Xi^{0}    )
  + {\cal M}(\Omega_{cc}^{+},   \pi^{+},    \Omega_{c}^{0},   \Xi^-,        \Xi^{0},     D^{*+}    )  \nonumber\\
&&+ {\cal M}(\Omega_{cc}^{+},   \rho^+,     \Omega_{c}^{0},   \Xi^-,        \Xi^{0},     D^{*+}    )
  + {\cal M}(\Omega_{cc}^{+},   \bar K^{0}, \Xi_c^{ +},       \Sigma^{ 0},  \Xi^{0},     D^{*+}    )
  + {\cal M}(\Omega_{cc}^{+},   \bar K^{0}, \Xi_c^{ +},       \Lambda,      \Xi^{0},     D^{*+}    )  \nonumber\\
&&+ {\cal M}(\Omega_{cc}^{+},   \bar K^{0}, \Xi_c^{\prime +}, \Sigma^{ 0},  \Xi^{0},     D^{*+}    )
  + {\cal M}(\Omega_{cc}^{+},   \bar K^{0}, \Xi_c^{\prime +}, \Lambda,      \Xi^{0},     D^{*+}    )
  + {\cal M}(\Omega_{cc}^{+},   \bar K^{*0},\Xi_c^{ +},       \Sigma^{ 0},  \Xi^{0},     D^{*+}    )  \nonumber\\
&&+ {\cal M}(\Omega_{cc}^{+},   \bar K^{*0},\Xi_c^{ +},       \Lambda,      \Xi^{0},     D^{*+}    )
  + {\cal M}(\Omega_{cc}^{+},   \bar K^{*0},\Xi_c^{\prime +}, \Sigma^{ 0},  \Xi^{0},     D^{*+}    )
  + {\cal M}(\Omega_{cc}^{+},   \bar K^{*0},\Xi_c^{\prime +}, \Lambda,      \Xi^{0},     D^{*+}    )  \nonumber\\
&&+ {\cal M}(\Omega_{cc}^{+},   \bar K^{0}, \Xi_c^{ +},        D_s^{+},     D^{*+},     \Xi^{0}    )
  + {\cal M}(\Omega_{cc}^{+},   \bar K^{0}, \Xi_c^{\prime +},  D_s^{+},     D^{*+},     \Xi^{0}    )
  + {\cal M}(\Omega_{cc}^{+},   \bar K^{*0},\Xi_c^{ +},        D_s^{*+},    D^{*+},     \Xi^{0}    )  \nonumber\\
&&+ {\cal M}(\Omega_{cc}^{+},   \bar K^{*0},\Xi_c^{\prime +},  D_s^{*+},    D^{*+},     \Xi^{0}    )
 ],
\end{eqnarray}

\begin{eqnarray}
{\cal A}(\Omega_{cc}^{+}\rightarrow\Xi^{0} D_{s}^{+})
&=&i[
    {\cal M}(\Omega_{cc}^{+},  K^{+},   \Omega_{c}^{0},    D^{*0},    D_s^{+},     \Xi^{0}   )
  + {\cal M}(\Omega_{cc}^{+},  K^{*+},  \Omega_{c}^{0},    D^{0},     D_s^{+},     \Xi^{0}   )
  + {\cal M}(\Omega_{cc}^{+}, \phi,     \Xi_c^+,           D_s^{+},   D_s^{+},     \Xi^{0}   )  \nonumber\\
&&+ {\cal M}(\Omega_{cc}^{+}, \phi,     \Xi_c^{\prime +},  D_s^{+},   D_s^{+},     \Xi^{0}   )
  + {\cal M}(\Omega_{cc}^{+}, \eta_1,   \Xi_c^{ +},        D_s^{*+},  D_s^{+},     \Xi^{0}   )
  + {\cal M}(\Omega_{cc}^{+}, \eta_1,   \Xi_c^{\prime +},  D_s^{*+},  D_s^{+},     \Xi^{0}   )  \nonumber\\
&&+ {\cal M}(\Omega_{cc}^{+}, \eta_8,   \Xi_c^{ +},        D_s^{*+},  D_s^{+},     \Xi^{0}   )
  + {\cal M}(\Omega_{cc}^{+}, \eta_8,   \Xi_c^{\prime +},  D_s^{*+},  D_s^{+},     \Xi^{0}   )
  + {\cal M}(\Omega_{cc}^{+}, \phi,     \Xi_c^+,          \Xi^0,     \Xi^{0},       D_s^{+}  )  \nonumber\\
&&+ {\cal M}(\Omega_{cc}^{+}, \eta_1,   \Xi_c^+,          \Xi^0,     \Xi^{0},       D_s^{+}  )
  + {\cal M}(\Omega_{cc}^{+}, \eta_8,   \Xi_c^+,          \Xi^0,     \Xi^{0},       D_s^{+}  )
  + {\cal M}(\Omega_{cc}^{+}, \phi,     \Xi_c^{\prime +}, \Xi^0,     \Xi^{0},       D_s^{+}  ) \nonumber\\
&&+ {\cal M}(\Omega_{cc}^{+}, \eta_1,   \Xi_c^{\prime +}, \Xi^0,     \Xi^{0},       D_s^{+}  )
  + {\cal M}(\Omega_{cc}^{+}, \eta_8,   \Xi_c^{\prime +}, \Xi^0,     \Xi^{0},       D_s^{+}  )
  + {\cal M}(\Omega_{cc}^{+}, \pi^+,    \Xi_c^0,          \Xi^-,     \Xi^{0},       D_s^{+}  ) \nonumber\\
&&+ {\cal M}(\Omega_{cc}^{+}, \rho^+,   \Xi_c^0,          \Xi^-,     \Xi^{0},       D_s^{+}  )
  + {\cal M}(\Omega_{cc}^{+}, \pi^+,    \Xi_c^{\prime 0}, \Xi^-,     \Xi^{0},       D_s^{+}  )
  + {\cal M}(\Omega_{cc}^{+}, \rho^+,   \Xi_c^{\prime 0}, \Xi^-,     \Xi^{0},       D_s^{+}  ) ],
\end{eqnarray}
\begin{eqnarray}
{\cal A}(\Omega_{cc}^{+}\rightarrow\Xi^{0} D_{s}^{*+})
&=&i [
    {\cal M}(\Omega_{cc}^{+},  K^{+},   \Omega_{c}^{0},    D^{0},     D_s^{*+},     \Xi^{0}   )
  + {\cal M}(\Omega_{cc}^{+},  K^{*+},  \Omega_{c}^{0},    D^{*0},    D_s^{*+},     \Xi^{0}   )
  + {\cal M}(\Omega_{cc}^{+}, \phi,     \Xi_c^+,           D_s^{*+},  D_s^{*+},     \Xi^{0}   )  \nonumber\\
&&+ {\cal M}(\Omega_{cc}^{+}, \phi,     \Xi_c^{\prime +},  D_s^{*+},  D_s^{*+},     \Xi^{0}   )
  + {\cal M}(\Omega_{cc}^{+}, \eta_1,   \Xi_c^{ +},        D_s^{+},   D_s^{*+},     \Xi^{0}   )
  + {\cal M}(\Omega_{cc}^{+}, \eta_1,   \Xi_c^{\prime +},  D_s^{+},   D_s^{*+},     \Xi^{0}   )  \nonumber\\
&&+ {\cal M}(\Omega_{cc}^{+}, \eta_8,   \Xi_c^{ +},        D_s^{+},   D_s^{*+},     \Xi^{0}   )
  + {\cal M}(\Omega_{cc}^{+}, \eta_8,   \Xi_c^{\prime +},  D_s^{+},   D_s^{*+},     \Xi^{0}   )
  + {\cal M}(\Omega_{cc}^{+}, \phi,     \Xi_c^+,          \Xi^0,     \Xi^{0},       D_s^{*+}  )  \nonumber\\
&&+ {\cal M}(\Omega_{cc}^{+}, \eta_1,   \Xi_c^+,          \Xi^0,     \Xi^{0},       D_s^{*+}  )
  + {\cal M}(\Omega_{cc}^{+}, \eta_8,   \Xi_c^+,          \Xi^0,     \Xi^{0},       D_s^{*+}  )
  + {\cal M}(\Omega_{cc}^{+}, \phi,     \Xi_c^{\prime +}, \Xi^0,     \Xi^{0},       D_s^{*+}  )  \nonumber\\
&&+ {\cal M}(\Omega_{cc}^{+}, \eta_1,   \Xi_c^{\prime +}, \Xi^0,     \Xi^{0},       D_s^{*+}  )
  + {\cal M}(\Omega_{cc}^{+}, \eta_8,   \Xi_c^{\prime +}, \Xi^0,     \Xi^{0},       D_s^{*+}  )
  + {\cal M}(\Omega_{cc}^{+}, \pi^+,    \Xi_c^0,          \Xi^-,     \Xi^{0},       D_s^{*+}  )  \nonumber\\
&&+ {\cal M}(\Omega_{cc}^{+}, \rho^+,   \Xi_c^0,          \Xi^-,     \Xi^{0},       D_s^{*+}  )
  + {\cal M}(\Omega_{cc}^{+}, \pi^+,    \Xi_c^{\prime 0}, \Xi^-,     \Xi^{0},       D_s^{*+}  )
  + {\cal M}(\Omega_{cc}^{+}, \rho^+,   \Xi_c^{\prime 0}, \Xi^-,     \Xi^{0},       D_s^{*+}  )
 ],
\end{eqnarray}

\begin{eqnarray}
{\cal A}(\Omega_{cc}^{+}\rightarrow\Sigma^{0} D_{s}^{+})
&=&i [
    {\cal M}(\Omega_{cc}^{+},   K^{+},    \Xi_c^0,           D^{*0},      D_s^{+},      \Sigma^{0}  )
  + {\cal M}(\Omega_{cc}^{+},   K^{+},    \Xi_c^{\prime 0},  D^{*0},      D_s^{+},      \Sigma^{0}  )
  + {\cal M}(\Omega_{cc}^{+},   K^{*+},   \Xi_c^0,           D^{0},       D_s^{+},      \Sigma^{0}  )\nonumber\\
&&+ {\cal M}(\Omega_{cc}^{+},   K^{*+},   \Xi_c^{\prime 0},  D^{0},       D_s^{+},      \Sigma^{0}  )
  + {\cal M}(\Omega_{cc}^{+},   K^{+},    \Xi_c^0,           \Xi^{-},     \Sigma^{0},    D_s^{+}    )
  + {\cal M}(\Omega_{cc}^{+},   K^{+},    \Xi_c^{\prime 0},  \Xi^{-},     \Sigma^{0},    D_s^{+}    )\nonumber\\
&&+ {\cal M}(\Omega_{cc}^{+},   K^{*+},   \Xi_c^0,           \Xi^{-},     \Sigma^{0},    D_s^{+}    )
  + {\cal M}(\Omega_{cc}^{+},   K^{*+},   \Xi_c^{\prime 0},  \Xi^{-},     \Sigma^{0},    D_s^{+}    )
  + {\cal M}(\Omega_{cc}^{+},   K^{0},    \Xi_c^+,           \Xi^{0},     \Sigma^{0},    D_s^{+}    )\nonumber\\
&&+ {\cal M}(\Omega_{cc}^{+},   K^{0},    \Xi_c^{\prime +},  \Xi^{0},     \Sigma^{0},    D_s^{+}    )
  + {\cal M}(\Omega_{cc}^{+},   K^{*0},   \Xi_c^+,           \Xi^{0},     \Sigma^{0},    D_s^{+}    )
  + {\cal M}(\Omega_{cc}^{+},   K^{*0},   \Xi_c^{\prime +},  \Xi^{0},     \Sigma^{0},    D_s^{+}    )\nonumber\\
&&+ {\cal M}(\Omega_{cc}^{+},   K^{0},    \Xi_c^+,           D^{*+},      D_s^{+},      \Sigma^{0}  )
  + {\cal M}(\Omega_{cc}^{+},   K^{0},    \Xi_c^{\prime +},  D^{*+},      D_s^{+},      \Sigma^{0}  )
  + {\cal M}(\Omega_{cc}^{+},   K^{*0},   \Xi_c^+,           D^{+},       D_s^{+},      \Sigma^{0}  )\nonumber\\
&&+ {\cal M}(\Omega_{cc}^{+},   K^{*0},   \Xi_c^{\prime +},  D^{+},       D_s^{+},      \Sigma^{0}  )
 ],
\end{eqnarray}
\begin{eqnarray}
{\cal A}(\Omega_{cc}^{+}\rightarrow\Lambda D_{s}^{+})
&=&i [
    {\cal M}(\Omega_{cc}^{+},   K^{+},    \Xi_c^0,           D^{*0},      D_s^{+},      \Lambda     )
  + {\cal M}(\Omega_{cc}^{+},   K^{+},    \Xi_c^{\prime 0},  D^{*0},      D_s^{+},      \Lambda     )
  + {\cal M}(\Omega_{cc}^{+},   K^{*+},   \Xi_c^0,           D^{0},       D_s^{+},      \Lambda     )\nonumber\\
&&+ {\cal M}(\Omega_{cc}^{+},   K^{*+},   \Xi_c^{\prime 0},  D^{0},       D_s^{+},      \Lambda     )
  + {\cal M}(\Omega_{cc}^{+},   K^{+},    \Xi_c^0,           \Xi^{-},     \Lambda,       D_s^{+}    )
  + {\cal M}(\Omega_{cc}^{+},   K^{+},    \Xi_c^{\prime 0},  \Xi^{-},     \Lambda,       D_s^{+}    )\nonumber\\
&&+ {\cal M}(\Omega_{cc}^{+},   K^{*+},   \Xi_c^0,           \Xi^{-},     \Lambda,       D_s^{+}    )
  + {\cal M}(\Omega_{cc}^{+},   K^{*+},   \Xi_c^{\prime 0},  \Xi^{-},     \Lambda,       D_s^{+}    )
  + {\cal M}(\Omega_{cc}^{+},   K^{0},    \Xi_c^+,           \Xi^{0},     \Lambda,       D_s^{+}    )\nonumber\\
&&+ {\cal M}(\Omega_{cc}^{+},   K^{0},    \Xi_c^{\prime +},  \Xi^{0},     \Lambda,       D_s^{+}    )
  + {\cal M}(\Omega_{cc}^{+},   K^{*0},   \Xi_c^+,           \Xi^{0},     \Lambda,       D_s^{+}    )
  + {\cal M}(\Omega_{cc}^{+},   K^{*0},   \Xi_c^{\prime +},  \Xi^{0},     \Lambda,       D_s^{+}    )\nonumber\\
&&+ {\cal M}(\Omega_{cc}^{+},   K^{0},    \Xi_c^+,           D^{*+},      D_s^{+},      \Lambda     )
  + {\cal M}(\Omega_{cc}^{+},   K^{0},    \Xi_c^{\prime +},  D^{*+},      D_s^{+},      \Lambda     )
  + {\cal M}(\Omega_{cc}^{+},   K^{*0},   \Xi_c^+,           D^{+},       D_s^{+},      \Lambda     )\nonumber\\
&&+ {\cal M}(\Omega_{cc}^{+},   K^{*0},   \Xi_c^{\prime +},  D^{+},       D_s^{+},      \Lambda     )
 ],
\end{eqnarray}
\begin{eqnarray}
{\cal A}(\Omega_{cc}^{+}\rightarrow\Sigma^{0} D_{s}^{*+})
&=&i [
    {\cal M}(\Omega_{cc}^{+},   K^{+},    \Xi_c^0,           D^{0},        D_s^{*+},      \Sigma^{0}  )
  + {\cal M}(\Omega_{cc}^{+},   K^{+},    \Xi_c^{\prime 0},  D^{0},        D_s^{*+},      \Sigma^{0}  )
  + {\cal M}(\Omega_{cc}^{+},   K^{*+},   \Xi_c^0,           D^{*0},       D_s^{*+},      \Sigma^{0}  )\nonumber\\
&&+ {\cal M}(\Omega_{cc}^{+},   K^{*+},   \Xi_c^{\prime 0},  D^{*0},       D_s^{*+},      \Sigma^{0}  )
  + {\cal M}(\Omega_{cc}^{+},   K^{+},    \Xi_c^0,           \Xi^{-},     \Sigma^{0},     D_s^{*+}    )
  + {\cal M}(\Omega_{cc}^{+},   K^{+},    \Xi_c^{\prime 0},  \Xi^{-},     \Sigma^{0},     D_s^{*+}    )\nonumber\\
&&+ {\cal M}(\Omega_{cc}^{+},   K^{*+},   \Xi_c^0,           \Xi^{-},     \Sigma^{0},     D_s^{*+}    )
  + {\cal M}(\Omega_{cc}^{+},   K^{*+},   \Xi_c^{\prime 0},  \Xi^{-},     \Sigma^{0},     D_s^{*+}    )
  + {\cal M}(\Omega_{cc}^{+},   K^{0},    \Xi_c^+,           \Xi^{0},     \Sigma^{0},     D_s^{*+}    )\nonumber\\
&&+ {\cal M}(\Omega_{cc}^{+},   K^{0},    \Xi_c^{\prime +},  \Xi^{0},     \Sigma^{0},     D_s^{*+}    )
  + {\cal M}(\Omega_{cc}^{+},   K^{*0},   \Xi_c^+,           \Xi^{0},     \Sigma^{0},     D_s^{*+}    )
  + {\cal M}(\Omega_{cc}^{+},   K^{*0},   \Xi_c^{\prime +},  \Xi^{0},     \Sigma^{0},     D_s^{*+}    )\nonumber\\
&&+ {\cal M}(\Omega_{cc}^{+},   K^{0},    \Xi_c^+,           D^{+},        D_s^{*+},      \Sigma^{0}  )
  + {\cal M}(\Omega_{cc}^{+},   K^{0},    \Xi_c^{\prime +},  D^{+},        D_s^{*+},      \Sigma^{0}  )
  + {\cal M}(\Omega_{cc}^{+},   K^{*0},   \Xi_c^+,           D^{*+},       D_s^{*+},      \Sigma^{0}  )\nonumber\\
&&+ {\cal M}(\Omega_{cc}^{+},   K^{*0},   \Xi_c^{\prime +},  D^{*+},       D_s^{*+},      \Sigma^{0}  )
 ],
\end{eqnarray}
\begin{eqnarray}
{\cal A}(\Omega_{cc}^{+}\rightarrow\Lambda D_{s}^{*+})
&=&i [
    {\cal M}(\Omega_{cc}^{+},   K^{+},    \Xi_c^0,           D^{0},        D_s^{*+},      \Lambda     )
  + {\cal M}(\Omega_{cc}^{+},   K^{+},    \Xi_c^{\prime 0},  D^{0},        D_s^{*+},      \Lambda     )
  + {\cal M}(\Omega_{cc}^{+},   K^{*+},   \Xi_c^0,           D^{*0},       D_s^{*+},      \Lambda     )\nonumber\\
&&+ {\cal M}(\Omega_{cc}^{+},   K^{*+},   \Xi_c^{\prime 0},  D^{*0},       D_s^{*+},      \Lambda     )
  + {\cal M}(\Omega_{cc}^{+},   K^{+},    \Xi_c^0,           \Xi^{-},     \Lambda,        D_s^{*+}    )
  + {\cal M}(\Omega_{cc}^{+},   K^{+},    \Xi_c^{\prime 0},  \Xi^{-},     \Lambda,        D_s^{*+}    )\nonumber\\
&&+ {\cal M}(\Omega_{cc}^{+},   K^{*+},   \Xi_c^0,           \Xi^{-},     \Lambda,        D_s^{*+}    )
  + {\cal M}(\Omega_{cc}^{+},   K^{*+},   \Xi_c^{\prime 0},  \Xi^{-},     \Lambda,        D_s^{*+}    )
  + {\cal M}(\Omega_{cc}^{+},   K^{0},    \Xi_c^+,           \Xi^{0},     \Lambda,        D_s^{*+}    )\nonumber\\
&&+ {\cal M}(\Omega_{cc}^{+},   K^{0},    \Xi_c^{\prime +},  \Xi^{0},     \Lambda,        D_s^{*+}    )
  + {\cal M}(\Omega_{cc}^{+},   K^{*0},   \Xi_c^+,           \Xi^{0},     \Lambda,        D_s^{*+}    )
  + {\cal M}(\Omega_{cc}^{+},   K^{*0},   \Xi_c^{\prime +},  \Xi^{0},     \Lambda,        D_s^{*+}    )\nonumber\\
&&+ {\cal M}(\Omega_{cc}^{+},   K^{0},    \Xi_c^+,           D^{+},        D_s^{*+},      \Lambda     )
  + {\cal M}(\Omega_{cc}^{+},   K^{0},    \Xi_c^{\prime +},  D^{+},        D_s^{*+},      \Lambda     )
  + {\cal M}(\Omega_{cc}^{+},   K^{*0},   \Xi_c^+,           D^{*+},       D_s^{*+},      \Lambda     )\nonumber\\
&&+ {\cal M}(\Omega_{cc}^{+},   K^{*0},   \Xi_c^{\prime +},  D^{*+},       D_s^{*+},      \Lambda     )
  ],
\end{eqnarray}

\begin{eqnarray}
{\cal A}(\Omega_{cc}^{+}\rightarrow\Sigma^{0} D^{+})
&=&i [
    {\cal M}(\Omega_{cc}^{+},  \pi^{+},   \Xi_{c}^{0},            D^{*0},     D^{+},     \Sigma^{0}   )
  + {\cal M}(\Omega_{cc}^{+},  \pi^{+},   \Xi_c^{\prime 0},       D^{*0},     D^{+},     \Sigma^{0}   )
  + {\cal M}(\Omega_{cc}^{+},  \rho^+,    \Xi_{c}^{0},            D^{*0},     D^{+},     \Sigma^{0}   )\nonumber\\
&&+ {\cal M}(\Omega_{cc}^{+},  \rho^+,    \Xi_c^{\prime 0},       D^{*0},     D^{+},     \Sigma^{0}   )
  + {\cal M}(\Omega_{cc}^{+},  \pi^{+},   \Xi_{c}^{0},           \Sigma^-,   \Sigma^{0},  D^{+}       )
  + {\cal M}(\Omega_{cc}^{+},  \pi^{+},   \Xi_{c}^{\prime 0},    \Sigma^-,   \Sigma^{0},  D^{+}       )\nonumber\\
&&+ {\cal M}(\Omega_{cc}^{+},  \rho^+,    \Xi_{c}^{0},           \Sigma^-,   \Sigma^{0},  D^{+}       )
  + {\cal M}(\Omega_{cc}^{+},  \rho^+,    \Xi_{c}^{\prime 0},    \Sigma^-,   \Sigma^{0},  D^{+}       )
  + {\cal M}(\Omega_{cc}^{+},  \rho^0,    \Xi_{c}^{+},            D^{+},      D^{+},     \Sigma^{0}   )\nonumber\\
&&+ {\cal M}(\Omega_{cc}^{+},  \rho^0,    \Xi_{c}^{\prime +},     D^{+},      D^{+},     \Sigma^{0}   )
  + {\cal M}(\Omega_{cc}^{+},  \pi^{0},   \Xi_{c}^{+},            D^{*+},     D^{+},     \Sigma^{0}   )
  + {\cal M}(\Omega_{cc}^{+},  \pi^{0},   \Xi_{c}^{\prime +},     D^{*+},     D^{+},     \Sigma^{0}   )\nonumber\\
&&+ {\cal M}(\Omega_{cc}^{+},  \eta_1,    \Xi_{c}^{+},            D^{*+},     D^{+},     \Sigma^{0}   )
  + {\cal M}(\Omega_{cc}^{+},  \eta_8,    \Xi_{c}^{+},            D^{*+},     D^{+},     \Sigma^{0}   )
  + {\cal M}(\Omega_{cc}^{+},  \eta_1,    \Xi_{c}^{\prime +},     D^{*+},     D^{+},     \Sigma^{0}   )\nonumber\\
&&+ {\cal M}(\Omega_{cc}^{+},  \eta_8,    \Xi_{c}^{\prime +},     D^{*+},     D^{+},     \Sigma^{0}   )
  + {\cal M}(\Omega_{cc}^{+},  \omega,    \Xi_{c}^{+},            D^{+},      D^{+},     \Sigma^{0}   )
  + {\cal M}(\Omega_{cc}^{+},  \omega,    \Xi_{c}^{\prime +},     D^{+},      D^{+},     \Sigma^{0}   )\nonumber\\
&&+ {\cal M}(\Omega_{cc}^{+},  \rho^0,    \Xi_{c}^{+},           \Lambda,    \Sigma^{0},  D^{+}       )
  + {\cal M}(\Omega_{cc}^{+},  \rho^0,    \Xi_{c}^{\prime +},    \Lambda,    \Sigma^{0},  D^{+}       )
  + {\cal M}(\Omega_{cc}^{+},  \omega,    \Xi_{c}^{+},           \Sigma^{0}, \Sigma^{0},  D^{+}       )\nonumber\\
&&+ {\cal M}(\Omega_{cc}^{+},  \omega,    \Xi_{c}^{\prime +},    \Sigma^{0}, \Sigma^{0},  D^{+}       )
  + {\cal M}(\Omega_{cc}^{+},  \pi^{0},   \Xi_{c}^{+},           \Lambda,    \Sigma^{0},  D^{+}       )
  + {\cal M}(\Omega_{cc}^{+},  \pi^{0},   \Xi_{c}^{\prime +},    \Lambda,    \Sigma^{0},  D^{+}       )\nonumber\\
&&+ {\cal M}(\Omega_{cc}^{+},  \eta_1,    \Xi_{c}^{+},           \Sigma^{0}, \Sigma^{0},  D^{+}       )
  + {\cal M}(\Omega_{cc}^{+},  \eta_8,    \Xi_{c}^{+},           \Sigma^{0}, \Sigma^{0},  D^{+}       )
  + {\cal M}(\Omega_{cc}^{+},  \eta_1,    \Xi_{c}^{\prime +},    \Sigma^{0}, \Sigma^{0},  D^{+}       )\nonumber\\
&&+ {\cal M}(\Omega_{cc}^{+},  \eta_8,    \Xi_{c}^{\prime +},    \Sigma^{0}, \Sigma^{0},  D^{+}       )
  + {\cal M}(\Omega_{cc}^{+},   K^+,      \Omega_{c}^{0},        \Xi^-,      \Sigma^{0},  D^{+}       )
  + {\cal M}(\Omega_{cc}^{+},   K^{*+},   \Omega_{c}^{0},        \Xi^-,      \Sigma^{0},  D^{+}       )\nonumber\\
&&+ {\cal M}(\Omega_{cc}^{+},  \phi,      \Xi_{c}^{+},           \Sigma^{0}, \Sigma^{0},  D^{+}       )
  + {\cal M}(\Omega_{cc}^{+},  \eta_1,    \Xi_{c}^{+},           \Sigma^{0}, \Sigma^{0},  D^{+}       )
  + {\cal M}(\Omega_{cc}^{+},  \eta_8,    \Xi_{c}^{+},           \Sigma^{0}, \Sigma^{0},  D^{+}       )\nonumber\\
&&+ {\cal M}(\Omega_{cc}^{+},  \phi,      \Xi_{c}^{\prime +},    \Sigma^{0}, \Sigma^{0},  D^{+}       )
  + {\cal M}(\Omega_{cc}^{+},  \eta_1,    \Xi_{c}^{\prime +},    \Sigma^{0}, \Sigma^{0},  D^{+}       )
  + {\cal M}(\Omega_{cc}^{+},  \eta_8,    \Xi_{c}^{\prime +},    \Sigma^{0}, \Sigma^{0},  D^{+}       )
 ],
\end{eqnarray}
\begin{eqnarray}
{\cal A}(\Omega_{cc}^{+}\rightarrow\Lambda D^{+})
&=&i [
    {\cal M}(\Omega_{cc}^{+},  \pi^{+},   \Xi_{c}^{0},            D^{*0},     D^{+},     \Lambda     )
  + {\cal M}(\Omega_{cc}^{+},  \pi^{+},   \Xi_c^{\prime 0},       D^{*0},     D^{+},     \Lambda     )
  + {\cal M}(\Omega_{cc}^{+},  \rho^+,    \Xi_{c}^{0},            D^{*0},     D^{+},     \Lambda     )\nonumber\\
&&+ {\cal M}(\Omega_{cc}^{+},  \rho^+,    \Xi_c^{\prime 0},       D^{*0},     D^{+},     \Lambda     )
  + {\cal M}(\Omega_{cc}^{+},  \pi^{+},   \Xi_{c}^{0},           \Sigma^-,   \Lambda,    D^{+}       )
  + {\cal M}(\Omega_{cc}^{+},  \pi^{+},   \Xi_{c}^{\prime 0},    \Sigma^-,   \Lambda,    D^{+}       )\nonumber\\
&&+ {\cal M}(\Omega_{cc}^{+},  \rho^+,    \Xi_{c}^{0},           \Sigma^-,   \Lambda,    D^{+}       )
  + {\cal M}(\Omega_{cc}^{+},  \rho^+,    \Xi_{c}^{\prime 0},    \Sigma^-,   \Lambda,    D^{+}       )
  + {\cal M}(\Omega_{cc}^{+},  \rho^0,    \Xi_{c}^{+},            D^{+},      D^{+},     \Lambda     )\nonumber\\
&&+ {\cal M}(\Omega_{cc}^{+},  \rho^0,    \Xi_{c}^{\prime +},     D^{+},      D^{+},     \Lambda     )
  + {\cal M}(\Omega_{cc}^{+},  \pi^{0},   \Xi_{c}^{+},            D^{*+},     D^{+},     \Lambda     )
  + {\cal M}(\Omega_{cc}^{+},  \pi^{0},   \Xi_{c}^{\prime +},     D^{*+},     D^{+},     \Lambda     )\nonumber\\
&&+ {\cal M}(\Omega_{cc}^{+},  \eta_1,    \Xi_{c}^{+},            D^{*+},     D^{+},     \Lambda     )
  + {\cal M}(\Omega_{cc}^{+},  \eta_8,    \Xi_{c}^{+},            D^{*+},     D^{+},     \Lambda     )
  + {\cal M}(\Omega_{cc}^{+},  \eta_1,    \Xi_{c}^{\prime +},     D^{*+},     D^{+},     \Lambda     )\nonumber\\
&&+ {\cal M}(\Omega_{cc}^{+},  \eta_8,    \Xi_{c}^{\prime +},     D^{*+},     D^{+},     \Lambda     )
  + {\cal M}(\Omega_{cc}^{+},  \omega,    \Xi_{c}^{+},            D^{+},      D^{+},     \Lambda     )
  + {\cal M}(\Omega_{cc}^{+},  \omega,    \Xi_{c}^{\prime +},     D^{+},      D^{+},     \Lambda     )\nonumber\\
&&+ {\cal M}(\Omega_{cc}^{+},  \rho^0,    \Xi_{c}^{+},           \Lambda,    \Lambda,    D^{+}       )
  + {\cal M}(\Omega_{cc}^{+},  \rho^0,    \Xi_{c}^{\prime +},    \Lambda,    \Lambda,    D^{+}       )
  + {\cal M}(\Omega_{cc}^{+},  \omega,    \Xi_{c}^{+},           \Sigma^{0}, \Lambda,    D^{+}       )\nonumber\\
&&+ {\cal M}(\Omega_{cc}^{+},  \omega,    \Xi_{c}^{\prime +},    \Sigma^{0}, \Lambda,    D^{+}       )
  + {\cal M}(\Omega_{cc}^{+},  \pi^{0},   \Xi_{c}^{+},           \Lambda,    \Lambda,    D^{+}       )
  + {\cal M}(\Omega_{cc}^{+},  \pi^{0},   \Xi_{c}^{\prime +},    \Lambda,    \Lambda,    D^{+}       )\nonumber\\
&&+ {\cal M}(\Omega_{cc}^{+},  \eta_1,    \Xi_{c}^{+},           \Sigma^{0}, \Lambda,    D^{+}       )
  + {\cal M}(\Omega_{cc}^{+},  \eta_8,    \Xi_{c}^{+},           \Sigma^{0}, \Lambda,    D^{+}       )
  + {\cal M}(\Omega_{cc}^{+},  \eta_1,    \Xi_{c}^{\prime +},    \Sigma^{0}, \Lambda,    D^{+}       )\nonumber\\
&&+ {\cal M}(\Omega_{cc}^{+},  \eta_8,    \Xi_{c}^{\prime +},    \Sigma^{0}, \Lambda,    D^{+}       )
  + {\cal M}(\Omega_{cc}^{+},   K^+,      \Omega_{c}^{0},        \Xi^-,      \Lambda,    D^{+}       )
  + {\cal M}(\Omega_{cc}^{+},   K^{*+},   \Omega_{c}^{0},        \Xi^-,      \Lambda,    D^{+}       )\nonumber\\
&&+ {\cal M}(\Omega_{cc}^{+},  \phi,      \Xi_{c}^{+},           \Sigma^{0}, \Lambda,    D^{+}       )
  + {\cal M}(\Omega_{cc}^{+},  \eta_1,    \Xi_{c}^{+},           \Sigma^{0}, \Lambda,    D^{+}       )
  + {\cal M}(\Omega_{cc}^{+},  \eta_8,    \Xi_{c}^{+},           \Sigma^{0}, \Lambda,    D^{+}       )\nonumber\\
&&+ {\cal M}(\Omega_{cc}^{+},  \phi,      \Xi_{c}^{\prime +},    \Sigma^{0}, \Lambda,    D^{+}       )
  + {\cal M}(\Omega_{cc}^{+},  \eta_1,    \Xi_{c}^{\prime +},    \Sigma^{0}, \Lambda,    D^{+}       )
  + {\cal M}(\Omega_{cc}^{+},  \eta_8,    \Xi_{c}^{\prime +},    \Sigma^{0}, \Lambda,    D^{+}       )
 ],
\end{eqnarray}
\begin{eqnarray}
{\cal A}(\Omega_{cc}^{+}\rightarrow\Sigma^{0} D^{*+})
&=&i [
    {\cal M}(\Omega_{cc}^{+},  \pi^{+},   \Xi_{c}^{0},            D^{0},      D^{*+},     \Sigma^{0}   )
  + {\cal M}(\Omega_{cc}^{+},  \pi^{+},   \Xi_c^{\prime 0},       D^{0},      D^{*+},     \Sigma^{0}   )
  + {\cal M}(\Omega_{cc}^{+},  \rho^+,    \Xi_{c}^{0},            D^{*0},     D^{*+},     \Sigma^{0}   )\nonumber\\
&&+ {\cal M}(\Omega_{cc}^{+},  \rho^+,    \Xi_c^{\prime 0},       D^{*0},     D^{*+},     \Sigma^{0}   )
  + {\cal M}(\Omega_{cc}^{+},  \pi^{+},   \Xi_{c}^{0},           \Sigma^-,   \Sigma^{0},  D^{*+}       )
  + {\cal M}(\Omega_{cc}^{+},  \pi^{+},   \Xi_{c}^{\prime 0},    \Sigma^-,   \Sigma^{0},  D^{*+}       )\nonumber\\
&&+ {\cal M}(\Omega_{cc}^{+},  \rho^+,    \Xi_{c}^{0},           \Sigma^-,   \Sigma^{0},  D^{*+}       )
  + {\cal M}(\Omega_{cc}^{+},  \rho^+,    \Xi_{c}^{\prime 0},    \Sigma^-,   \Sigma^{0},  D^{*+}       )
  + {\cal M}(\Omega_{cc}^{+},  \rho^0,    \Xi_{c}^{+},            D^{*+},     D^{*+},     \Sigma^{0}   )\nonumber\\
&&+ {\cal M}(\Omega_{cc}^{+},  \rho^0,    \Xi_{c}^{\prime +},     D^{*+},     D^{*+},     \Sigma^{0}   )
  + {\cal M}(\Omega_{cc}^{+},  \pi^{0},   \Xi_{c}^{+},            D^{+},      D^{*+},     \Sigma^{0}   )
  + {\cal M}(\Omega_{cc}^{+},  \pi^{0},   \Xi_{c}^{\prime +},     D^{+},      D^{*+},     \Sigma^{0}   )\nonumber\\
&&+ {\cal M}(\Omega_{cc}^{+},  \eta_1,    \Xi_{c}^{+},            D^{+},      D^{*+},     \Sigma^{0}   )
  + {\cal M}(\Omega_{cc}^{+},  \eta_8,    \Xi_{c}^{+},            D^{+},      D^{*+},     \Sigma^{0}   )
  + {\cal M}(\Omega_{cc}^{+},  \eta_1,    \Xi_{c}^{\prime +},     D^{+},      D^{*+},     \Sigma^{0}   )\nonumber\\
&&+ {\cal M}(\Omega_{cc}^{+},  \eta_8,    \Xi_{c}^{\prime +},     D^{+},      D^{*+},     \Sigma^{0}   )
  + {\cal M}(\Omega_{cc}^{+},  \omega,    \Xi_{c}^{+},            D^{*+},     D^{*+},     \Sigma^{0}   )
  + {\cal M}(\Omega_{cc}^{+},  \omega,    \Xi_{c}^{\prime +},     D^{*+},     D^{*+},     \Sigma^{0}   )\nonumber\\
&&+ {\cal M}(\Omega_{cc}^{+},  \rho^0,    \Xi_{c}^{+},           \Lambda,    \Sigma^{0},  D^{*+}       )
  + {\cal M}(\Omega_{cc}^{+},  \rho^0,    \Xi_{c}^{\prime +},    \Lambda,    \Sigma^{0},  D^{*+}       )
  + {\cal M}(\Omega_{cc}^{+},  \omega,    \Xi_{c}^{+},           \Sigma^{0}, \Sigma^{0},  D^{*+}       )\nonumber\\
&&+ {\cal M}(\Omega_{cc}^{+},  \omega,    \Xi_{c}^{\prime +},    \Sigma^{0}, \Sigma^{0},  D^{*+}       )
  + {\cal M}(\Omega_{cc}^{+},  \pi^{0},   \Xi_{c}^{+},           \Lambda,    \Sigma^{0},  D^{*+}       )
  + {\cal M}(\Omega_{cc}^{+},  \pi^{0},   \Xi_{c}^{\prime +},    \Lambda,    \Sigma^{0},  D^{*+}       )\nonumber\\
&&+ {\cal M}(\Omega_{cc}^{+},  \eta_1,    \Xi_{c}^{+},           \Sigma^{0}, \Sigma^{0},  D^{*+}       )
  + {\cal M}(\Omega_{cc}^{+},  \eta_8,    \Xi_{c}^{+},           \Sigma^{0}, \Sigma^{0},  D^{*+}       )
  + {\cal M}(\Omega_{cc}^{+},  \eta_1,    \Xi_{c}^{\prime +},    \Sigma^{0}, \Sigma^{0},  D^{*+}       )\nonumber\\
&&+ {\cal M}(\Omega_{cc}^{+},  \eta_8,    \Xi_{c}^{\prime +},    \Sigma^{0}, \Sigma^{0},  D^{*+}       )
  + {\cal M}(\Omega_{cc}^{+},   K^+,      \Omega_{c}^{0},        \Xi^-,      \Sigma^{0},  D^{*+}       )
  + {\cal M}(\Omega_{cc}^{+},   K^{*+},   \Omega_{c}^{0},        \Xi^-,      \Sigma^{0},  D^{*+}       )\nonumber\\
&&+ {\cal M}(\Omega_{cc}^{+},  \phi,      \Xi_{c}^{+},           \Sigma^{0}, \Sigma^{0},  D^{*+}       )
  + {\cal M}(\Omega_{cc}^{+},  \eta_1,    \Xi_{c}^{+},           \Sigma^{0}, \Sigma^{0},  D^{*+}       )
  + {\cal M}(\Omega_{cc}^{+},  \eta_8,    \Xi_{c}^{+},           \Sigma^{0}, \Sigma^{0},  D^{*+}       )\nonumber\\
&&+ {\cal M}(\Omega_{cc}^{+},  \phi,      \Xi_{c}^{\prime +},    \Sigma^{0}, \Sigma^{0},  D^{*+}       )
  + {\cal M}(\Omega_{cc}^{+},  \eta_1,    \Xi_{c}^{\prime +},    \Sigma^{0}, \Sigma^{0},  D^{*+}       )
  + {\cal M}(\Omega_{cc}^{+},  \eta_8,    \Xi_{c}^{\prime +},    \Sigma^{0}, \Sigma^{0},  D^{*+}       )
 ],
\end{eqnarray}
\begin{eqnarray}
{\cal A}(\Omega_{cc}^{+}\rightarrow\Lambda D^{*+})
&=&i [
    {\cal M}(\Omega_{cc}^{+},  \pi^{+},   \Xi_{c}^{0},            D^{0},      D^{*+},     \Lambda     )
  + {\cal M}(\Omega_{cc}^{+},  \pi^{+},   \Xi_c^{\prime 0},       D^{0},      D^{*+},     \Lambda     )
  + {\cal M}(\Omega_{cc}^{+},  \rho^+,    \Xi_{c}^{0},            D^{*0},     D^{*+},     \Lambda     )\nonumber\\
&&+ {\cal M}(\Omega_{cc}^{+},  \rho^+,    \Xi_c^{\prime 0},       D^{*0},     D^{*+},     \Lambda     )
  + {\cal M}(\Omega_{cc}^{+},  \pi^{+},   \Xi_{c}^{0},           \Sigma^-,   \Lambda,    D^{*+}       )
  + {\cal M}(\Omega_{cc}^{+},  \pi^{+},   \Xi_{c}^{\prime 0},    \Sigma^-,   \Lambda,    D^{*+}       )\nonumber\\
&&+ {\cal M}(\Omega_{cc}^{+},  \rho^+,    \Xi_{c}^{0},           \Sigma^-,   \Lambda,    D^{*+}       )
  + {\cal M}(\Omega_{cc}^{+},  \rho^+,    \Xi_{c}^{\prime 0},    \Sigma^-,   \Lambda,    D^{*+}       )
  + {\cal M}(\Omega_{cc}^{+},  \rho^0,    \Xi_{c}^{+},            D^{*+},     D^{*+},     \Lambda     )\nonumber\\
&&+ {\cal M}(\Omega_{cc}^{+},  \rho^0,    \Xi_{c}^{\prime +},     D^{*+},     D^{*+},     \Lambda     )
  + {\cal M}(\Omega_{cc}^{+},  \pi^{0},   \Xi_{c}^{+},            D^{+},      D^{*+},     \Lambda     )
  + {\cal M}(\Omega_{cc}^{+},  \pi^{0},   \Xi_{c}^{\prime +},     D^{+},      D^{*+},     \Lambda     )\nonumber\\
&&+ {\cal M}(\Omega_{cc}^{+},  \eta_1,    \Xi_{c}^{+},            D^{+},      D^{*+},     \Lambda     )
  + {\cal M}(\Omega_{cc}^{+},  \eta_8,    \Xi_{c}^{+},            D^{+},      D^{*+},     \Lambda     )
  + {\cal M}(\Omega_{cc}^{+},  \eta_1,    \Xi_{c}^{\prime +},     D^{+},      D^{*+},     \Lambda     )\nonumber\\
&&+ {\cal M}(\Omega_{cc}^{+},  \eta_8,    \Xi_{c}^{\prime +},     D^{+},      D^{*+},     \Lambda     )
  + {\cal M}(\Omega_{cc}^{+},  \omega,    \Xi_{c}^{+},            D^{*+},     D^{*+},     \Lambda     )
  + {\cal M}(\Omega_{cc}^{+},  \omega,    \Xi_{c}^{\prime +},     D^{*+},     D^{*+},     \Lambda     )\nonumber\\
&&+ {\cal M}(\Omega_{cc}^{+},  \rho^0,    \Xi_{c}^{+},           \Lambda,    \Lambda,    D^{*+}       )
  + {\cal M}(\Omega_{cc}^{+},  \rho^0,    \Xi_{c}^{\prime +},    \Lambda,    \Lambda,    D^{*+}       )
  + {\cal M}(\Omega_{cc}^{+},  \omega,    \Xi_{c}^{+},           \Sigma^{0}, \Lambda,    D^{*+}       )\nonumber\\
&&+ {\cal M}(\Omega_{cc}^{+},  \omega,    \Xi_{c}^{\prime +},    \Sigma^{0}, \Lambda,    D^{*+}       )
  + {\cal M}(\Omega_{cc}^{+},  \pi^{0},   \Xi_{c}^{+},           \Lambda,    \Lambda,    D^{*+}       )
  + {\cal M}(\Omega_{cc}^{+},  \pi^{0},   \Xi_{c}^{\prime +},    \Lambda,    \Lambda,    D^{*+}       )\nonumber\\
&&+ {\cal M}(\Omega_{cc}^{+},  \eta_1,    \Xi_{c}^{+},           \Sigma^{0}, \Lambda,    D^{*+}       )
  + {\cal M}(\Omega_{cc}^{+},  \eta_8,    \Xi_{c}^{+},           \Sigma^{0}, \Lambda,    D^{*+}       )
  + {\cal M}(\Omega_{cc}^{+},  \eta_1,    \Xi_{c}^{\prime +},    \Sigma^{0}, \Lambda,    D^{*+}       )\nonumber\\
&&+ {\cal M}(\Omega_{cc}^{+},  \eta_8,    \Xi_{c}^{\prime +},    \Sigma^{0}, \Lambda,    D^{*+}       )
  + {\cal M}(\Omega_{cc}^{+},   K^+,      \Omega_{c}^{0},        \Xi^-,      \Lambda,    D^{*+}       )
  + {\cal M}(\Omega_{cc}^{+},   K^{*+},   \Omega_{c}^{0},        \Xi^-,      \Lambda,    D^{*+}       )\nonumber\\
&&+ {\cal M}(\Omega_{cc}^{+},  \phi,      \Xi_{c}^{+},           \Sigma^{0}, \Lambda,    D^{*+}       )
  + {\cal M}(\Omega_{cc}^{+},  \eta_1,    \Xi_{c}^{+},           \Sigma^{0}, \Lambda,    D^{*+}       )
  + {\cal M}(\Omega_{cc}^{+},  \eta_8,    \Xi_{c}^{+},           \Sigma^{0}, \Lambda,    D^{*+}       )\nonumber\\
&&+ {\cal M}(\Omega_{cc}^{+},  \phi,      \Xi_{c}^{\prime +},    \Sigma^{0}, \Lambda,    D^{*+}       )
  + {\cal M}(\Omega_{cc}^{+},  \eta_1,    \Xi_{c}^{\prime +},    \Sigma^{0}, \Lambda,    D^{*+}       )
  + {\cal M}(\Omega_{cc}^{+},  \eta_8,    \Xi_{c}^{\prime +},    \Sigma^{0}, \Lambda,    D^{*+}       )
 ],
\end{eqnarray}

\begin{eqnarray}
{\cal A}(\Omega_{cc}^{+}\rightarrow\Sigma^{+} D^{0})
&=&i [
    {\cal M}(\Omega_{cc}^{+},  K^{+},   \Omega_{c}^{0},     \Xi^{0},     \Sigma^{+},    D^{0}    )
  + {\cal M}(\Omega_{cc}^{+},  K^{*+},  \Omega_{c}^{0},     \Xi^{0},     \Sigma^{+},    D^{0}    )
  + {\cal M}(\Omega_{cc}^{+}, \pi^+,    \Xi_{c}^{0},        \Sigma^{0},  \Sigma^{+},    D^{0}    )\nonumber\\
&&+ {\cal M}(\Omega_{cc}^{+}, \pi^+,    \Xi_{c}^{\prime 0}, \Sigma^{0},  \Sigma^{+},    D^{0}    )
  + {\cal M}(\Omega_{cc}^{+}, \pi^+,    \Xi_{c}^{0},        \Lambda,     \Sigma^{+},    D^{0}    )
  + {\cal M}(\Omega_{cc}^{+}, \pi^+,    \Xi_{c}^{\prime 0}, \Lambda,     \Sigma^{+},    D^{0}    )\nonumber\\
&&+ {\cal M}(\Omega_{cc}^{+}, \rho^+,   \Xi_{c}^{0},        \Sigma^{0},  \Sigma^{+},    D^{0}    )
  + {\cal M}(\Omega_{cc}^{+}, \rho^+,   \Xi_{c}^{\prime 0}, \Sigma^{0},  \Sigma^{+},    D^{0}    )
  + {\cal M}(\Omega_{cc}^{+}, \rho^+,   \Xi_{c}^{0},        \Lambda,     \Sigma^{+},    D^{0}    )\nonumber\\
&&+ {\cal M}(\Omega_{cc}^{+}, \rho^+,   \Xi_{c}^{\prime 0}, \Lambda,     \Sigma^{+},    D^{0}    )
  + {\cal M}(\Omega_{cc}^{+}, \phi,     \Xi_{c}^{+},        \Sigma^{+},  \Sigma^{+},    D^{0}    )
  + {\cal M}(\Omega_{cc}^{+}, \eta_1,   \Xi_{c}^{+},        \Sigma^{+},  \Sigma^{+},    D^{0}    )\nonumber\\
&&+ {\cal M}(\Omega_{cc}^{+}, \eta_8,   \Xi_{c}^{+},        \Sigma^{+},  \Sigma^{+},    D^{0}    )
  + {\cal M}(\Omega_{cc}^{+}, \phi,     \Xi_{c}^{\prime 0}, \Sigma^{+},  \Sigma^{+},    D^{0}    )
  + {\cal M}(\Omega_{cc}^{+}, \eta_1,   \Xi_{c}^{\prime 0}, \Sigma^{+},  \Sigma^{+},    D^{0}    )\nonumber\\
&&+ {\cal M}(\Omega_{cc}^{+}, \eta_8,   \Xi_{c}^{\prime 0}, \Sigma^{+},  \Sigma^{+},    D^{0}    )
  ],
\end{eqnarray}
\begin{eqnarray}
{\cal A}(\Omega_{cc}^{+}\rightarrow\Sigma^{+} D^{*0})
&=&i [
    {\cal M}(\Omega_{cc}^{+},  K^{+},   \Omega_{c}^{0},     \Xi^{0},     \Sigma^{+},    D^{*0}    )
  + {\cal M}(\Omega_{cc}^{+},  K^{*+},  \Omega_{c}^{0},     \Xi^{0},     \Sigma^{+},    D^{*0}    )
  + {\cal M}(\Omega_{cc}^{+}, \pi^+,    \Xi_{c}^{0},        \Sigma^{0},  \Sigma^{+},    D^{*0}    )\nonumber\\
&&+ {\cal M}(\Omega_{cc}^{+}, \pi^+,    \Xi_{c}^{\prime 0}, \Sigma^{0},  \Sigma^{+},    D^{*0}    )
  + {\cal M}(\Omega_{cc}^{+}, \pi^+,    \Xi_{c}^{0},        \Lambda,     \Sigma^{+},    D^{*0}    )
  + {\cal M}(\Omega_{cc}^{+}, \pi^+,    \Xi_{c}^{\prime 0}, \Lambda,     \Sigma^{+},    D^{*0}    )\nonumber\\
&&+ {\cal M}(\Omega_{cc}^{+}, \rho^+,   \Xi_{c}^{0},        \Sigma^{0},  \Sigma^{+},    D^{*0}    )
  + {\cal M}(\Omega_{cc}^{+}, \rho^+,   \Xi_{c}^{\prime 0}, \Sigma^{0},  \Sigma^{+},    D^{*0}    )
  + {\cal M}(\Omega_{cc}^{+}, \rho^+,   \Xi_{c}^{0},        \Lambda,     \Sigma^{+},    D^{*0}    )\nonumber\\
&&+ {\cal M}(\Omega_{cc}^{+}, \rho^+,   \Xi_{c}^{\prime 0}, \Lambda,     \Sigma^{+},    D^{*0}    )
  + {\cal M}(\Omega_{cc}^{+}, \phi,     \Xi_{c}^{+},        \Sigma^{+},  \Sigma^{+},    D^{*0}    )
  + {\cal M}(\Omega_{cc}^{+}, \eta_1,   \Xi_{c}^{+},        \Sigma^{+},  \Sigma^{+},    D^{*0}    )\nonumber\\
&&+ {\cal M}(\Omega_{cc}^{+}, \eta_8,   \Xi_{c}^{+},        \Sigma^{+},  \Sigma^{+},    D^{*0}    )
  + {\cal M}(\Omega_{cc}^{+}, \phi,     \Xi_{c}^{\prime 0}, \Sigma^{+},  \Sigma^{+},    D^{*0}    )
  + {\cal M}(\Omega_{cc}^{+}, \eta_1,   \Xi_{c}^{\prime 0}, \Sigma^{+},  \Sigma^{+},    D^{*0}    )\nonumber\\
&&+ {\cal M}(\Omega_{cc}^{+}, \eta_8,   \Xi_{c}^{\prime 0}, \Sigma^{+},  \Sigma^{+},    D^{*0}    )
 ],
\end{eqnarray}

\begin{eqnarray}
{\cal A}(\Omega_{cc}^{+}\rightarrow p D^{0})
&=&i [
    {\cal M}(\Omega_{cc}^{+},  K^{+},   \Xi_{c}^{0},        \Sigma^{0},       p,          D^{0}   )
  + {\cal M}(\Omega_{cc}^{+},  K^{+},   \Xi_{c}^{\prime 0}, \Sigma^{0},       p,          D^{0}   )
  + {\cal M}(\Omega_{cc}^{+},  K^{+},   \Xi_{c}^{0},        \Lambda,          p,          D^{0}   )\nonumber\\
&&+ {\cal M}(\Omega_{cc}^{+},  K^{+},   \Xi_{c}^{\prime 0}, \Lambda,          p,          D^{0}   )
  + {\cal M}(\Omega_{cc}^{+},  K^{*+},  \Xi_{c}^{0},        \Sigma^{0},       p,          D^{0}   )
  + {\cal M}(\Omega_{cc}^{+},  K^{*+},  \Xi_{c}^{\prime 0}, \Sigma^{0},       p,          D^{0}   )\nonumber\\
&&+ {\cal M}(\Omega_{cc}^{+},  K^{*+},  \Xi_{c}^{0},        \Lambda,          p,          D^{0}   )
  + {\cal M}(\Omega_{cc}^{+},  K^{*+},  \Xi_{c}^{\prime 0}, \Lambda,          p,          D^{0}   )
  + {\cal M}(\Omega_{cc}^{+},  K^{0},   \Xi_{c}^{+},        \Sigma^{+},       p,          D^{0}   )\nonumber\\
&&+ {\cal M}(\Omega_{cc}^{+},  K^{0},   \Xi_{c}^{\prime +}, \Sigma^{+},       p,          D^{0}   )
  + {\cal M}(\Omega_{cc}^{+},  K^{*0},  \Xi_{c}^{+},        \Sigma^{+},       p,          D^{0}   )
  + {\cal M}(\Omega_{cc}^{+},  K^{*0},  \Xi_{c}^{\prime +}, \Sigma^{+},       p,          D^{0}   )
 ],
\end{eqnarray}
\begin{eqnarray}
{\cal A}(\Omega_{cc}^{+}\rightarrow p D^{*0})
&=&i [
    {\cal M}(\Omega_{cc}^{+},  K^{+},   \Xi_{c}^{0},        \Sigma^{0},       p,          D^{*0}   )
  + {\cal M}(\Omega_{cc}^{+},  K^{+},   \Xi_{c}^{\prime 0}, \Sigma^{0},       p,          D^{*0}   )
  + {\cal M}(\Omega_{cc}^{+},  K^{+},   \Xi_{c}^{0},        \Lambda,          p,          D^{*0}   )\nonumber\\
&&+ {\cal M}(\Omega_{cc}^{+},  K^{+},   \Xi_{c}^{\prime 0}, \Lambda,          p,          D^{*0}   )
  + {\cal M}(\Omega_{cc}^{+},  K^{*+},  \Xi_{c}^{0},        \Sigma^{0},       p,          D^{*0}   )
  + {\cal M}(\Omega_{cc}^{+},  K^{*+},  \Xi_{c}^{\prime 0}, \Sigma^{0},       p,          D^{*0}   )\nonumber\\
&&+ {\cal M}(\Omega_{cc}^{+},  K^{*+},  \Xi_{c}^{0},        \Lambda,          p,          D^{*0}   )
  + {\cal M}(\Omega_{cc}^{+},  K^{*+},  \Xi_{c}^{\prime 0}, \Lambda,          p,          D^{*0}   )
  + {\cal M}(\Omega_{cc}^{+},  K^{0},   \Xi_{c}^{+},        \Sigma^{+},       p,          D^{*0}   )\nonumber\\
&&+ {\cal M}(\Omega_{cc}^{+},  K^{0},   \Xi_{c}^{\prime +}, \Sigma^{+},       p,          D^{*0}   )
  + {\cal M}(\Omega_{cc}^{+},  K^{*0},  \Xi_{c}^{+},        \Sigma^{+},       p,          D^{*0}   )
  + {\cal M}(\Omega_{cc}^{+},  K^{*0},  \Xi_{c}^{\prime +}, \Sigma^{+},       p,          D^{*0}   )
 ],
\end{eqnarray}

\begin{eqnarray}
{\cal A}(\Omega_{cc}^{+}\rightarrow n D^{+})
&=&i [
    {\cal M}(\Omega_{cc}^{+},  K^{+},    \Xi_{c}^{0},           \Sigma^{-},      n,       D^{+}   )
  + {\cal M}(\Omega_{cc}^{+},  K^{+},    \Xi_{c}^{\prime 0},    \Sigma^{-},      n,       D^{+}   )
  + {\cal M}(\Omega_{cc}^{+},  K^{*+},   \Xi_{c}^{0},           \Sigma^{-},      n,       D^{+}   )\nonumber\\
&&+ {\cal M}(\Omega_{cc}^{+},  K^{*+},   \Xi_{c}^{\prime 0},    \Sigma^{-},      n,       D^{+}   )
  + {\cal M}(\Omega_{cc}^{+},  K^{0},    \Xi_{c}^{+},           \Sigma^{0},      n,       D^{+}   )
  + {\cal M}(\Omega_{cc}^{+},  K^{0},    \Xi_{c}^{+},           \Lambda,         n,       D^{+}   )\nonumber\\
&&+ {\cal M}(\Omega_{cc}^{+},  K^{0},    \Xi_{c}^{\prime +},    \Sigma^{0},      n,       D^{+}   )
  + {\cal M}(\Omega_{cc}^{+},  K^{0},    \Xi_{c}^{\prime +},    \Lambda,         n,       D^{+}   )
  + {\cal M}(\Omega_{cc}^{+},  K^{*0},   \Xi_{c}^{+},           \Sigma^{0},      n,       D^{+}   )\nonumber\\
&&+ {\cal M}(\Omega_{cc}^{+},  K^{*0},   \Xi_{c}^{+},           \Lambda,         n,       D^{+}   )
  + {\cal M}(\Omega_{cc}^{+},  K^{*0},   \Xi_{c}^{\prime +},    \Sigma^{0},      n,       D^{+}   )
  + {\cal M}(\Omega_{cc}^{+},  K^{*0},   \Xi_{c}^{\prime +},    \Lambda,         n,       D^{+}   )
 ],
\end{eqnarray}
\begin{eqnarray}
{\cal A}(\Omega_{cc}^{+}\rightarrow n D^{*+})
&=&i [
    {\cal M}(\Omega_{cc}^{+},  K^{+},    \Xi_{c}^{0},           \Sigma^{-},      n,       D^{*+}   )
  + {\cal M}(\Omega_{cc}^{+},  K^{+},    \Xi_{c}^{\prime 0},    \Sigma^{-},      n,       D^{*+}   )
  + {\cal M}(\Omega_{cc}^{+},  K^{*+},   \Xi_{c}^{0},           \Sigma^{-},      n,       D^{*+}   )\nonumber\\
&&+ {\cal M}(\Omega_{cc}^{+},  K^{*+},   \Xi_{c}^{\prime 0},    \Sigma^{-},      n,       D^{*+}   )
  + {\cal M}(\Omega_{cc}^{+},  K^{0},    \Xi_{c}^{+},           \Sigma^{0},      n,       D^{*+}   )
  + {\cal M}(\Omega_{cc}^{+},  K^{0},    \Xi_{c}^{+},           \Lambda,         n,       D^{*+}   )\nonumber\\
&&+ {\cal M}(\Omega_{cc}^{+},  K^{0},    \Xi_{c}^{\prime +},    \Sigma^{0},      n,       D^{*+}   )
  + {\cal M}(\Omega_{cc}^{+},  K^{0},    \Xi_{c}^{\prime +},    \Lambda,         n,       D^{*+}   )
  + {\cal M}(\Omega_{cc}^{+},  K^{*0},   \Xi_{c}^{+},           \Sigma^{0},      n,       D^{*+}   )\nonumber\\
&&+ {\cal M}(\Omega_{cc}^{+},  K^{*0},   \Xi_{c}^{+},           \Lambda,         n,       D^{*+}   )
  + {\cal M}(\Omega_{cc}^{+},  K^{*0},   \Xi_{c}^{\prime +},    \Sigma^{0},      n,       D^{*+}   )
  + {\cal M}(\Omega_{cc}^{+},  K^{*0},   \Xi_{c}^{\prime +},    \Lambda,         n,       D^{*+}   )
 ],
\end{eqnarray}
\begin{eqnarray}
{\cal A}(\Omega_{cc}^{+}\rightarrow n D^{*+})
&=&i [
    {\cal M}(\Omega_{cc}^{+},  K^{+},    \Xi_{c}^{0},           \Sigma^{-},      n,       D^{*+}   )
  + {\cal M}(\Omega_{cc}^{+},  K^{+},    \Xi_{c}^{\prime 0},    \Sigma^{-},      n,       D^{*+}   )
  + {\cal M}(\Omega_{cc}^{+},  K^{*+},   \Xi_{c}^{0},           \Sigma^{-},      n,       D^{*+}   )\nonumber\\
&&+ {\cal M}(\Omega_{cc}^{+},  K^{*+},   \Xi_{c}^{\prime 0},    \Sigma^{-},      n,       D^{*+}   )
  + {\cal M}(\Omega_{cc}^{+},  K^{0},    \Xi_{c}^{+},           \Sigma^{0},      n,       D^{*+}   )
  + {\cal M}(\Omega_{cc}^{+},  K^{0},    \Xi_{c}^{+},           \Lambda,         n,       D^{*+}   )\nonumber\\
&&+ {\cal M}(\Omega_{cc}^{+},  K^{0},    \Xi_{c}^{\prime +},    \Sigma^{0},      n,       D^{*+}   )
  + {\cal M}(\Omega_{cc}^{+},  K^{0},    \Xi_{c}^{\prime +},    \Lambda,         n,       D^{*+}   )
  + {\cal M}(\Omega_{cc}^{+},  K^{*0},   \Xi_{c}^{+},           \Sigma^{0},      n,       D^{*+}   )\nonumber\\
&&+ {\cal M}(\Omega_{cc}^{+},  K^{*0},   \Xi_{c}^{+},           \Lambda,         n,       D^{*+}   )
  + {\cal M}(\Omega_{cc}^{+},  K^{*0},   \Xi_{c}^{\prime +},    \Sigma^{0},      n,       D^{*+}   )
  + {\cal M}(\Omega_{cc}^{+},  K^{*0},   \Xi_{c}^{\prime +},    \Lambda,         n,       D^{*+}   )
 ].
\end{eqnarray}}
\section{Strong Coupling Constants}
\label{app:stcouplings}
In this section, we list all of the strong coupling constants used in our calculation. Some of these values are obtained from Refs. \cite{Cheng:FSIB,Aliev:2006xr,Aliev:2009ei,Khodjamirian:2011jp,Azizi:2014bua,Yu:2016pyo,Azizi:2015tya,Ballon-Bayona:2017bwk}. Those that cannot be found directly in the literature are calculated under the assumption of $SU(3)_F$ symmetry.

According to the $SU(3)_F$ multiplets to which the particles belong, the vertices in this study can be divided into ${\cal B}{\cal B}V$, ${\cal B}{\cal B}P$, $DD^*P$, $DDV$, $D^*D^*V$,  ${\cal B}_c{\cal B} D$ and ${\cal B}_c {\cal B} D^*$, where $P$ denotes a light pseudoscalar meson, $V$ represents a light vector meson, and ${\cal B}_c$ is a singly charmed baryon. These definitions clarify the meanings of our symbols for each vertex, the coupling constants of which are presented in Tables~\ref{tab:SCBBV}, \ref{tab:SCBBP}, \ref{tab:SCDDP}, and \ref{tab:SCLND}.


\begin{table}[!htbp]
  \centering
 \caption{Strong coupling constants of ${\cal B}{\cal B}V$ vertices}
 \label{tab:SCBBV}
  \begin{tabular}{|c|c|c|c|c|c|c|c|c|}  
   \hline %
   Vertex&$f_{1}$&$f_{2}$&Vertex&$f_{1}$&$f_{2}$  &Vertex&$f_{1}$&$f_{2}$\\
   \hline %
   $p\rightarrow\ n\rho^{+}$&-2.40&32.95&
   $\Lambda\rightarrow\Sigma^{-}\rho^{+}$&2.00&12.30&
   $\Sigma^{0}\rightarrow\Sigma^{+}\rho^{+}$&7.20&-25.00\\
   \hline %
   $\Sigma^{+}\rightarrow\Lambda\rho^{+}$&2.00&12.30&
   $\Sigma^{+}\rightarrow p\overline{K}^{*0}$&5.66&-1.70&
   $\Sigma^{+}\rightarrow\Xi^{0}K^{*+}$&-2.26&37.05\\
   \hline %
   $\Sigma^{+}\rightarrow\Sigma^{+}\phi$&-6.00&2.50&
   $ p\rightarrow\Sigma^{0}K^{*+}$&4.00&-1.20&
   $p\rightarrow\Lambda K^{*+}$&5.10&-28.00\\
   \hline %
   $p\rightarrow\Sigma^{+}K^{*0}$&5.66&-1.70&
   $\Sigma^{0}\rightarrow\Sigma^{-}\rho^{+}$&-7.20&25.00&
   $\Sigma^{0}\rightarrow n\overline{K}^{*0}$&-4.00&1.20\\
   \hline %
   $\Lambda\rightarrow n\overline{K}^{*0}$&5.10&-28.00&
   $\Sigma^{0}\rightarrow\Xi^{-}K^{*+}$&-1.60&26.20&
   $\Sigma^{0}\rightarrow\Sigma^{0}\phi$&-6.00&2.50\\
   \hline %
   $\Sigma^{0}\rightarrow\Sigma^{0}\omega$&4.30&-1.10&
   $\Lambda\rightarrow\Sigma^{0}\rho^{0}$&1.90&11.90&
   $p\rightarrow p\rho^{0}$&-2.50&22.20\\
   \hline %
   $\Lambda\rightarrow\Xi^{0}K^{*0}$&-6.00&17.10&
   $\Lambda\rightarrow\Xi^{-}K^{*+}$&-6.00&17.10&
   $\Lambda\rightarrow\Lambda\phi$&-5.30&24.60\\
   \hline %
   $n\rightarrow\Sigma^{0}K^{*0}$&-4.00&1.20&
   $\Lambda\rightarrow\Lambda\omega$&-7.10&8.70&
   $n\rightarrow\Sigma^{-}K^{*+}$&5.66&-1.70\\
   \hline %
   $n\rightarrow\Lambda K^{*0}$&5.10&-28.00&
   $\Xi^{-}\rightarrow\Sigma^{+}\overline{K}^{*0}$&-2.26&37.05&
   $n\rightarrow n\rho^{0}$&2.50&-22.20\\
   \hline %
   $\Xi^{0}\rightarrow\Lambda\overline{K}^{*0}$&4.45&-27.54&
   $\Xi^{0}\rightarrow\Xi^{-}\rho^{+}$&6.08&-1.56&
   $\Xi^{0}\rightarrow\Sigma^{0}\overline{K}^{*0}$&1.60&-26.20\\
   \hline %
   $\Xi^{0}\rightarrow\Xi^{0}\phi$&-9.50&32.30&
   $\Sigma^{0}\rightarrow\Xi^{0}K^{*0}$&1.60&-26.20&&&\\
   \hline %
     \end{tabular}
\end{table}

\begin{table}[!htbp]
  \centering
 \caption{Strong coupling constants of ${\cal B}{\cal B}P$ vertices}
 \label{tab:SCBBP}
  \begin{tabular}{|c|c|c|c|c|c|}  
   \hline %
   Vertex&g&Vertex&g&Vertex&g\\
   \hline %
   $p\rightarrow n\pi^{+}$&21.20&
   $\Lambda\rightarrow\Sigma^{-}\pi^{+}$&10.00&
   $\Sigma^{0}\rightarrow\Sigma^{+}\pi^{+}$&-10.70 \\
   \hline %
   $\Sigma^{+}\rightarrow\Lambda\pi^{+}$&10.00&
   $\Sigma^{+}\rightarrow p\overline{K}^{0}$&5.75&
   $\Sigma^{+}\rightarrow\Xi^{0}K^{+}$&19.80 \\
   \hline %
   $p\rightarrow\Sigma^{0}K^{+}$&4.25&
   $p\rightarrow\Lambda K^{+}$&-13.50&
   $p\rightarrow\Sigma^{+} K^{0}$&5.75 \\
   \hline %
   $\Sigma^{-}\rightarrow\Sigma^{0}\pi^{+}$&10.70&
   $\Sigma^{0}\rightarrow n\overline{K}^{0}$&-4.25&
   $\Xi^{-}\rightarrow\Sigma^{+}\overline{K}^{0}$&19.80\\
   \hline %
   $\Sigma^{0}\rightarrow\Xi^{-}K^{+}$&14.00&
   $\Lambda\rightarrow\Xi^{-}K^{+}$&4.25&
   $\Lambda\rightarrow n\overline{K}^{0}$&-13.50 \\
   \hline %
   $p\rightarrow p\eta_8$&4.25&
   $n\rightarrow\Sigma^{-} K^{+}$&4.70&
   $p\rightarrow p\pi^{0}$&14.90 \\
   \hline %
   $p\rightarrow p\eta_1$&14.14&
   $n\rightarrow n\pi^{0}$&-14.90&
   $n\rightarrow\Sigma^{0} K^{0}$&-4.25\\
   \hline %
   $n\rightarrow\Lambda K^{0}$&-13.50&
   $\Xi^{0}\rightarrow\Sigma^{0}\overline{K}^{0}$&-14.00&
   $n\rightarrow n\eta_8$&4.25\\
   \hline %
   $\Xi^{0}\rightarrow\Xi^{-}\pi^{+}$&4.70&
   $\Lambda\rightarrow\Xi^{0}K^{0}$&4.25&
   $n\rightarrow n\eta_1$&14.14\\
   \hline %
   $\Sigma^{0}\rightarrow\Xi^{0}K^{0}$&-14.00&
   $\Sigma^{0}\rightarrow\Sigma^{0}\eta_8$&10.00&
   $\Xi^{0}\rightarrow\Lambda\overline{K}^{0}$&4.25\\
   \hline %
   $\Lambda\rightarrow\Sigma^{0}\pi^{0}$&10.00&
   $\Sigma^{0}\rightarrow\Sigma^{0}\eta_1$&14.14&
   $\Xi^{0}\rightarrow\Xi^{0}\eta_8$&-13.50 \\
   \hline %
   $\Lambda\rightarrow\Lambda\eta_8$&-10.00&
   $\Sigma^{+}\rightarrow\Sigma^{+}\eta_8$&10.00&
   $\Xi^{0}\rightarrow\Xi^{0}\eta_1$&14.14 \\
   \hline %
   $\Lambda\rightarrow\Lambda\eta_1$&-14.14&
   $\Sigma^{+}\rightarrow\Sigma^{+}\eta_1$&14.14&&\\
   \hline %
   \end{tabular}
\end{table}

\begin{table}[!htbp]
  \centering
 \caption{Strong coupling constants of $D D^{*}P$, $D DV$ and $ D^{*} D^{*}V$ vertices}
 \label{tab:SCDDP}
  \begin{tabular}{|c|c|c|c|c|c|c|}  
   \hline %
   Vertex&g&Vertex&g&Vertex&g&f\\
   \hline %
   ${D}^{*}\rightarrow D\pi$&17.90&$D\rightarrow D\rho$&3.69&
   ${D}^{*}\rightarrow{D}^{*}\rho$&3.69&4.61 \\
   \hline %
     \end{tabular}
\end{table}

\begin{table}[!htbp]
  \centering
 \caption{Strong coupling constants of ${\cal B}_{c}{\cal B}D$ and ${\cal B}_{c}{\cal B}{ D}^{*}$ vertices}
 \label{tab:SCLND}
  \begin{tabular}{|c|c|c|c|c|}  
   \hline %
   vertex&g&vertex&g&f\\
   \hline %
   $\Lambda_{c}\rightarrow N{D}_{q}$&4.82&$\Lambda_{c}\rightarrow N{D}_{q}^{*}$&-5.80&3.60\\
   \hline %
   $\Sigma_{c}\rightarrow N{D}_{q}$&3.78&$\Sigma_{c}\rightarrow N{D}_{q}^{*}$&11.21&4.64\\
   \hline %
     \end{tabular}
\end{table}


 \end{document}